# Multidimensional Skills on LinkedIn Profiles: Measuring Human Capital and the Gender Skill Gap*

David Dorn, Florian Schoner, Moritz Seebacher, Lisa Simon, and Ludger Woessmann†

## Abstract

We measure human capital using the self-reported skill sets of nearly 9 million U.S. college graduates from professional profiles on LinkedIn. We aggregate skill strings into 48 clusters of general, occupation-specific, and managerial skills. Multidimensional skills can account for several important labor-market patterns. First, the number and composition of skills are systematically related to measures of human-capital investment such as education and work experience. The number of skills increases with experience, and the average age-skill profile closely resembles the well-established concave age-earnings profile. Second, workers who report more skills, especially specific and managerial ones, hold higher-paid jobs. Skill differences account for more earnings variation than detailed measures of education and experience. Third, we document a sizable gender gap in skills. While women and men report nearly equal numbers of skills shortly after college graduation, women's skill count increases more slowly with age subsequently. A simple quantitative exercise shows that women's slower skill accumulation can be fully accounted for by reduced work hours associated with motherhood. The resulting gender differences in skills rationalize a substantial proportion of the gender gap in job-based earnings.

---

* We gratefully acknowledge comments and suggestions from Joe Altonji, Rick Hanushek, seminar participants at UC Berkeley, MPI Bonn, Boston U, U Copenhagen, Harvard U, U Melbourne, Monash U, U Regensburg, NU Singapore, SOFI Stockholm, and SSE Stockholm, and conference participants at MIT Shaping the Future of Work Initiative, CESifo Economics of Education, European Association of Labour Economists, German Economic Association, IAB PhD workshop, ifo Big Data Seminar, and ifo Center for the Economics of Education. Severin Bayer, Martin Blasi, Alexander Krieg, Nicolas McPartlin, Lewin Nolden, Isabella Pau, and Pankraz Trauner provided excellent research assistance. Project funding by the Bavarian State Ministry of Economic Affairs, Regional Development and Energy for the project 'Education Policies to Empower People to Participate in the Social Market Economy in the Era of Digital Transformation' (Grant 0703/89372/2921) is gratefully acknowledged. Dorn acknowledges support by the University of Zurich's Research Priority Program 'Equality of Opportunity'. The contribution by Woessmann is part of the German Science Foundation project CRC TRR 190.

† Dorn: University of Zurich; david.dorn@econ.uzh.ch. Schoner: ifo Institute, University of Munich; schoner@ifo.de. Seebacher: ifo Institute, University of Munich; seebacher@ifo.de. Simon: Revelio Labs; lisa@reveliolabs.com. Woessmann: University of Munich, ifo Institute; Hoover Institution, Stanford University; woessmann@ifo.de.

## 1. Introduction

Theories of human capital posit that individuals can invest in the acquisition of valuable labor-market skills. Workers enhance their earnings potential both by spending time in formal education and through on-the-job learning (Becker 1964, Mincer 1974). While measures of investments into human capital production such as years of schooling or years of work experience are available in many large-scale, worker-level data sets, the skills that result from such investments are usually unobserved. Education and experience variables are thus often used to proxy for the skills that result from such investments (Card 1999, Heckman, Lochner and Todd 2006). However, task models of the labor market argue that workers with the same education level may possess a potentially large vector of distinct skills that qualifies them to perform different job tasks in the labor market (e.g., Autor, Levy, and Murnane 2003). Analyses of job advertisements confirm that employers search for a wide range of worker skills that may not be easily captured by traditional education and experience metrics, such as problem-solving ability or proficiency with specialized software (e.g., Deming and Kahn 2018). There is, however, little direct evidence on which groups of workers possess the skills that employers value, and how such skills are related to workers' human capital investments and subsequent attainment of highly paid jobs.

In this paper, we provide large-scale evidence on the self-reported skill sets of millions of U.S. college graduates, using data from individual profiles on LinkedIn, the leading social media platform for professional networking. We conduct three sets of analyses that provide new evidence on skill differentials across distinct groups of college graduates and assess the usefulness of LinkedIn skills as a measure for human capital. First, we investigate whether skills systematically relate to education and experience as predicted by classical models of human capital acquisition. Second, we analyze whether the size and composition of workers' skill sets helps to explain which workers hold highly paid jobs. Third, we document skill differences between women and men, and relate these gender skill gaps to motherhood and to gender differentials in earnings.

LinkedIn profiles are structured like a worker's resume and provide information on self-declared skills as well as educational and occupational trajectories. Our analysis sample draws on a database of scraped public profiles curated by the workforce intelligence company Revelio Labs and contains a cross-section of LinkedIn profiles with complete skill, education, and experience data for 8.85 million college graduates who held jobs in the United States in 2019, before the onset of the Covid-19 pandemic.

To analyze the self-reported skills, we employ a word association algorithm to aggregate the reported raw skill strings into 48 skill clusters. We distinguish three broad groups of skill clusters. A first group comprises general skills that are used in a wide range of occupations, such as ‘communication, problem solving’ or ‘customer satisfaction, customer retention’. A second group contains specific skills that are concentrated in a narrow set of occupations, such as ‘legal research, legal writing’ or ‘SQL, software development’. And a third group consists of managerial skills that are frequently observed in top executive jobs, such as ‘project management, budgets’ or ‘coaching, leadership development’. In our analysis, we discuss and address various measurement challenges emerging from the self-reporting of skills.

Our first set of results analyzes the relationship between skills and measures of experience and education. We document that the average age-skill profile has a concave shape which looks remarkably similar to the well-known average age-earnings profiles. Older workers report a larger number of skills than their younger peers, which is consistent with an acquisition of skills through on-the-job experience. That interpretation is supported by the observation that the number of skills increases more strongly with actual rather than potential years of experience. It is further corroborated by the finding that older workers indicate larger fractions of occupation-specific and managerial skills, suggesting that these skills are often acquired through on-the-job experience, while general skills that are not focused on specific occupations are primarily acquired during initial education. The composition of reported skills also varies with different types of college education. Most notably, workers with degrees in arts, humanities, or social sciences have considerably higher shares of general skills than graduates of science, medicine, or law programs (who have relatively more specific skills) and graduates of business programs (who have relatively more managerial skills).

We next study the relationship between skills and the attainment of highly paid jobs. We draw on salary data that Revelio Labs imputes primarily based on job characteristics such as job title and firm name. We show that workers with a larger number of skills, and those who report higher fractions of managerial and occupation-specific skills, tend to have higher-paid jobs. The skill variables from LinkedIn account for a larger proportion of job-based earnings variation than an augmented Mincer model with a highly detailed vector of education and experience variables, and they mediate the impact of these traditional human capital variables to a substantial extent. We obtain qualitatively similar results with a simple alternative earnings metric that relies on data from the Census Bureau's American Community Survey and assigns to each worker the average earnings in her occupation.

Through the lens of these relationships between education, experience, skills, and earnings, we analyze gender gaps between women and men. Young women in their twenties report only a slightly lower average number of skills as their male peers, and this difference is fully accounted for by gender-specific occupational choice. But women's age-skill profiles are considerably flatter than men's during their thirties and early forties, giving rise to a substantial gender skill gap. We provide evidence that this emerging skill gap is unlikely to be primarily driven by gender differences in overreporting, profile updating, or occupational choice. Instead, the evidence is consistent with an explanation where reduced female labor supply due to motherhood gives rise to slower skill accumulation on the job. Gender differences in self-reported skills can account for 60 percent of the gender gap in job-based earnings, a proportion that considerably exceeds the explanatory power of highly detailed education and experience variables.

Taken together, our results suggest that worker skills on LinkedIn profiles, despite being self-reported and potentially noisy, provide an informative new measurement of individual human capital that is richer and more detailed than most available measurements.

The remainder of this paper is organized as follows. Section 2 relates our analyses to several strands of the existing literature. Section 3 describes the LinkedIn data and discusses potential challenges for the measurement of skills. Section 4 defines variables and our analysis sample of college graduates who work in the US labor market. Section 5 tests whether skills relate to experience and education as predicted by basic theories of human capital acquisition, while section 6 assesses whether more skilled workers are employed in higher-paid jobs. Section 7 documents a gender skill gap and investigates its relation to motherhood and to the gender earnings gap. Section 8 concludes.

## 2. Related Literature

Our analysis contributes to the large literature on skills in the labor market. Seminal human capital models characterize the accumulation of skills in formal education and on the job (Becker 1964, Mincer 1974, Heckman, Lochner, and Todd 2006). In a framework of supply and demand for skills, the distribution of earnings by skill level emerges from a race between an evolving supply of worker skills and advances in production technology that shape the demand for skills (Tinbergen 1974, Katz and Murphy 1992, Goldin and Katz 2008, Autor 2014, Autor, Goldin, and Katz 2020). Task models posit the importance of multidimensional skills to perform a broad range of job tasks required in the labor market (e.g., Autor, Levy, and Murnane 2003, Acemoglu and Autor 2011, Autor and Dorn 2013, Acemoglu and Restrepo 2022). However, in all these literatures skills are usually not directly observed but instead proxied by measures of education and work experience.[1] Data from online postings of job vacancies, however, show that employers are interested in a wide range of skills beyond educational credentials (e.g., Deming and Kahn 2018, Hershbein and Kahn 2018, Deming and Noray 2020, Braxton and Taska 2023). But the job advertisement data do not directly indicate which workers hold these multidimensional skills. Our analysis complements this literature by measuring multidimensional skills in worker-level data.[2] In addition, we can observe a much richer set of education variables than most classical studies of human capital.

Prior analyses of individual skills usually observed subsets of skills and studied their relationship with earnings (see Deming and Silliman 2024 and Woessmann 2025 for recent reviews of the literature). Some investigate skills in various cognitive domains such as literacy, numeracy, and science (e.g., Murnane, Willett, and Levy 1995, Bowles, Gintis, and Osborne 2001, Hanushek and Woessmann 2008, Castex and Dechter 2014, Hanushek et al. 2015, Hermo et al. 2022). Others study social skills such as leadership and teamwork (e.g., Kuhn and Weinberger 2005, Weinberger 2014, Deming 2017, 2023) and other noncognitive aspects of skills (e.g., Heckman, Stixrud, and Urzua 2006, Lindqvist and Vestman 2011, Edin et al. 2022), partly in various combinations (e.g., Piopiunik et al. 2020, Lise and Postel-Vinay 2020, Guvenen et al. 2020). However, such studies tend to fall short of covering the wide range of skills seen in job advertisements or observed in our data. Prior use of LinkedIn data for labor-market analysis is limited (e.g., Wheeler et al. 2022, Conzelmann et al. 2022, Evsyukova, Rusche, and Mill 2025) and does not make use of the skill information contained in the professional profiles.

---

[1] In human capital models with a unidimensional skill and competitive labor markets (e.g., Mincer 1974), a worker's skill can be directly inferred from her earnings, and no separate measurement of skill is needed. In a setting with multidimensional skills (and possibly labor markets that are not perfectly competitive), it is no longer possible to simply back out a worker's skill set from her earnings.

[2] A measurement advantage in our data relative to job ads is that LinkedIn profiles comprise a dedicated field where users indicate skills. Job ads are instead often unstructured so that researcher discretion is necessary to define which elements of a job ad refer to skill requirements.

Our categorization of skills into general versus specific skills follows a long tradition in labor economics (Streeck 2011). This distinction, a central aspect of the original Becker (1964) model, is particularly relevant when analyzing job tasks (Sanders and Taber 2012) and life cycle dynamics (Hanushek et al. 2017). General skills are typically defined as being portable, in the sense that they can be productively used in many different jobs, whereas specific skills are valuable only in a limited set of jobs. Becker (1964) adopted a narrow definition of specificity, where specific skills are useful only at a single firm. We instead focus on the occupational specificity of skills, consistent with the empirical observation that experience acquired at one firm often remains beneficial when workers move to a similar occupation with comparable tasks at another firm (Gathmann and Schönberg 2010, Eggenberger, Rinawi, and Backes-Gellner 2018). A standard assumption in human capital models is that formal education provides primarily general skills, whereas specific skills are acquired relatively more through on-the-job experience (e.g., Acemoglu and Pischke 1999).

We additionally subdivide general skills into managerial and non-managerial skills. Managerial skills, which we define as the skills that are most concentrated among top executives, are general in the sense of having a high portability across occupations. However, the underrepresentation of recent college graduates in management jobs suggests that managerial skills are acquired more through work experience than through education, which makes them more comparable to specific skills in this regard. Our conceptualization of managerial skills is related to leadership skills (Kuhn and Weinberger 2005), people management skills (Hoffman and Tadelis 2021), and skills to coordinate, monitor, and motivate workers (Weidmann et al. 2024), but it is broader as it also includes skills related to tasks such as planning, budgeting, or business analysis that are not necessarily focused on people.

Human capital differences have long been considered as a potential explanation for differences in labor-market outcomes across different population segments, perhaps most notably the earnings gap between women and men (Goldin 2024). During the 1980s, about one quarter of the gender gap in full-time wages in the U.S. could be explained by women's lower average years of education and experience (Olivetti, Pan, and Petrongolo 2024). Over the following three decades, the male advantage in work experience shrank and women overtook men in terms of educational attainment. By 2018, basic education and experience variables can no longer rationalize a gender wage gap in favor of men and instead predict a modest gap in favor of women.[3] In the words of Goldin (2014, p. 1094), "as women have increased their productivity enhancing characteristics and as they 'look' more like men, the human capital part of the wage difference has been squeezed out."

While education and experience have lost explanatory power, the contribution of occupation and industry of employment to the gender wage gap has increased from 19 percent in 1980 to 54 percent in 2018 (Olivetti, Pan, and Petrongolo 2024). Blau and Kahn (2017, p. 797) note that these occupation-industry effects "may represent human capital, other labor-market skills, and commitment, on the one hand, or employer discrimination, on the other hand", which raises the possibility that the contribution of skill differentials to the gender wage gap may be considerably larger than implied by basic education and experience variables alone.[4] We contribute to this literature by documenting a gender gap in self-reported skills and by showing that the large and difficult-to-interpret occupation- or job-specific component of the gender earnings gap can to a substantial extent be rationalized by these skill differences.[5] We also relate to a literature on the role of motherhood in the emergence of gender earnings gaps over the life cycle (Bertrand, Goldin, and Katz 2010, Kleven, Landais, and Sogaard 2019). Our results show that motherhood is related to reduced female work hours and a slower skill accumulation of women relative to same-aged men.

## 3. LinkedIn Data and the Measurement of Skills

The primary source of our data consists of publicly accessible individual worker profiles obtained from the professional networking platform LinkedIn, which we introduce in section 3.1.

---

[3] Using survey data from the Panel Study of Income Dynamics, Olivetti, Pan, and Petrongolo (2024) estimate that education and experience explained 23 percent of the gender wage gap in 1980, 25 percent in 1989, 17 percent in 1998, 8 percent in 2010, and minus 8 percent in 2018.

[4] The differential occupational composition between women and men is partly a result of gender differences in college major choices (Bertrand 2011, Speer 2020). However, Sloane, Hurst, and Black (2021) note that the contribution of occupational effects to the gender wage gap declines only modestly once gender differences in college majors are taken into account.

[5] A few studies have been able to almost fully explain gender wage gaps in narrow subsets of highly educated workers. Azmat and Ferrer (2017) show that the gender earnings gap among young lawyers is substantially accounted for by performance differences, and Bertrand, Goldin, and Katz (2010) find that the gender earnings gap among MBA graduates largely derives from differences in prior training, career interruptions, and weekly work hours. It is however challenging to assess what portion of these effects result from gender differences in skills.

We observe a cross-section of these profiles in 2019 and discuss in section 3.2 the limitations imposed by the cross-sectional approach. The LinkedIn profiles offer detailed self-reported information on workers' skills. We discuss in section 3.3 how users report skills on their profiles and in how far self-reporting generates measurement challenges for our analysis.

### 3.1 LinkedIn Profiles

LinkedIn is the world's largest professional network and operates a social media platform that focuses on career development. Its more than one billion registered users worldwide (LinkedIn 2024) create online profiles that list CV information such as current and prior jobs, educational degrees, and skills. LinkedIn is widely used by workers and employers for job search and recruiting (Evsyukova, Rusche, and Mill 2025), and many universities and career consultants recommend creating and regularly updating profiles on the platform (e.g., Kratz 2021, Arizona State University Alumni 2021). In the United States, a majority of college graduates reports using LinkedIn (Auxier and Anderson 2021), and most corporate recruiters look for job candidates on the platform (Ryan 2020).

Most of the CV information on LinkedIn is part of a public profile that is visible to any user of the platform, while other information such as posts and shares is usually only visible to users that are logged in on LinkedIn or who have established a personal link with the profile holder on the platform. The LinkedIn records analyzed in our study are provided by Revelio Labs, a workforce intelligence company that builds an encompassing human-resource database from various sources. Revelio Labs scrapes the semi-structured public profiles on LinkedIn and then parses and curates the information into a structured database. They clean, standardize, and enrich the data using proprietary algorithms. Our raw data consist of a cross-section of LinkedIn records from the United States that were scraped between January and September 2019, before the massive labor-market disruptions caused by the Covid-19 pandemic.

### 3.2 Cross-Sectional Analysis

The cross-sectional structure of the data does not allow us to distinguish between age and cohort effects when analyzing skill differences across younger versus older workers. Our descriptive analysis is thus similar in spirit to Mincer's (1974) classical analysis of the relationship between education, potential experience, and earnings. Mincer's pioneering work has subsequently been complemented by longitudinal studies which establish that the originally observed cross-sectional patterns largely capture age effects and not just cohort effects (Thornton, Rodgers, and Brookshire 1997, Sandgren 2007) and by analyses with causal designs that establish a causal link between human capital investments and earnings (see Card 1999 and Heckman, Lochner, and Todd 2006 for reviews of the literature).

Consistent with human capital models that predict an accumulation of skills as workers age, and following a large empirical literature starting with Mincer (1974), we interpret the combined age-cohort effects in our data as age effects. This choice simplifies the language we can use to describe results but comes with the caveat that cohort effects can never be fully ruled out.[6]

### 3.3 Self-Reported Skills

Workers self-report skills on their LinkedIn profiles. The addition of skills to a user's profile is semi-structured. When the user starts typing a skill, LinkedIn provides suggestions of frequently used skill terms to facilitate comparability across user profiles. However, users are free to enter new skill terms that have not been previously used on the platform. We observe 1.8 million unique text strings for skills in our analysis sample, and the average profile that reports skills indicates 20 of them. Before describing our definition of skill variables and other relevant variables in section 4, we discuss potential measurement challenges that result from the self-reporting of skills and their implications for our empirical analysis.

***Relevance of LinkedIn Skills***. LinkedIn profiles are widely used for recruiting and job search, and skill information on users' profiles plays an important role in this process. In a survey of recruiting professionals who use LinkedIn, 94 percent of recruiters agree that it necessary to understand which skills workers possess to make informed talent decisions (Degraux 2023). According to LinkedIn, 48 percent of company recruiters on the platform explicitly use workers' skill data when they seek to fill vacancies (Anderson 2024), and searches by skill are considerably more likely than searches by years of experience (Degraux 2023). Whether or not a worker lists a particular skill thus affects the likelihood of being approached by a recruiter (Smith 2023). LinkedIn skills should therefore be of interest to labor economists because they have direct relevance for labor-market outcomes. To relate these skills to theories of human capital, we however must consider to what extent workers' self-reports of skills correspond to a theoretically grounded notion of skills.

[6] Cross-sectional analyses that relate earnings to potential or actual work experience remain widely used to measure human capital accumulation (e.g., Jedwab et al. 2023). An advantage of a cross-sectional analysis over a panel analysis is that the former does not confound age effects with year effects. Since 2023, LinkedIn phased in the feature that users can report skills separately for each position they held (Mason 2023). To the extent that this feature changed overall patterns of skill reporting, a panel analysis that follows individuals over time would tend to confound age effects with year-specific reporting effects.

***Definition of Skills***. A leading dictionary defines skills as "a learned power of doing something competently" (Merriam-Webster 2025), which is consistent with the premise of human capital theory that skills are produced through a learning process in school or on-the-job. In our empirical analysis, skills are however potentially more broadly defined as any concept that LinkedIn users report in the skill section of their profiles. While it is not guaranteed that users' understanding of skills always follows the dictionary definition, we show in section 4.1 that the most frequently reported skill strings indeed correspond to aptitudes that have plausibly been acquired in a learning process. We also note that there is substantial overlap between the keywords that researchers have used to extract skill requirements from job ads (e.g., Deming and Noray 2020) and the skills most frequently reported on LinkedIn profiles.

***Incomplete and Outdated LinkedIn Profiles***. LinkedIn data will underreport skills (and other relevant profile information) if users never created a complete profile or ceased to update their profile. We address these challenges as follows. First, our sample selection process described in section 4.4 removes apparently incomplete profiles, including those that report no skills or no educational information. Second, we present robustness tests in Appendix B for the subsample of workers who moved to a new employer within the last two years. The indication of a new employer ascertains that these workers modified their profile relatively recently, and the preceding job search arguably created an incentive to update and complete the entire profile information. While the subsample of recent movers should include user profiles with higher information quality, its focus on workers with recent job mobility makes it less representative of the overall population of college graduates in the US labor market, and we thus use it for robustness tests rather than for our main analysis. Appendix B shows that the qualitative results of our main analyses are confirmed in the sample of recent firm switchers, indicating that these results are unlikely to be driven by underreporting in outdated profiles.

***Underreporting of Basic Skills***. Workers typically use their LinkedIn profiles to highlight information that differentiates them from their peers while omitting widely shared qualifications.

For example, college graduates rarely provide a full record of their earlier education, such as elementary or high school, nor do they mention basic skills like reading. Similarly, users who list advanced competencies, such as proficiency in specialized data analysis software, may leave out more common abilities, such as using basic spreadsheet programs. If highly skilled individuals are more likely to underreport basic skills, such reporting bias may dampen an association between human capital investments and the number of reported skills, and between the number of skills and workers' earnings. This bias would work in the opposite direction of the results we find in our subsequent analysis.

***Idiosyncratic Reporting of Skill Detail and Proficiency***. LinkedIn users may apply different idiosyncratic standards to evaluate their own skill sets. First, users can decide on the level of detail of the skills they want to indicate. Whereas one user may report to be skilled in the 'MS Office' software suite, another user may indicate mastery of the individual programs 'MS Word' and 'MS Excel' as two separate skills. Second, users may differ in their assessment of how high one's level of proficiency should be to claim being skilled in domains such as 'project engineering', 'public speaking', or 'software development'.[7] Some users may even knowingly state skills that they do not possess to attract the attention of recruiters, even though career advisory services warn that such a strategy could backfire later in the recruiting process (UCLA Career Center 2024).

The observation that many recruiters use the LinkedIn skill data to search for suitable job candidates suggests that professional users of the platform consider the skill data to be generally informative about workers' actual qualifications. LinkedIn also seeks to enhance the quality of the reported skill information with a feature that allows people to endorse the skills reported by other users (typically co-workers) to whom they are connected on the platform. While this endorsement information is not part of the public profiles that can be scraped and is thus not contained in our data, the social control exerted by other users likely helps to improve the quality of the skill data that we have available.[8]

---

[7] Idiosyncratic reporting practices presumably introduce less noise in the measurement of other relevant variables on LinkedIn profiles such as education or work experience, since the question whether a worker has obtained an educational degree or held a job is less subject to individual interpretation. The reporting challenges that we discuss here for skills reported on LinkedIn profiles however also apply to the more established literature that studies the skill requirements reported in job advertisements, in which companies may apply different practices to report such skill demands.

[8] Users also have the option to report profiles with inaccurate information directly to LinkedIn, which will then review and potentially ban the profile.

For the purpose of our empirical analyses, the idiosyncratic reporting of skill detail and proficiency implies that the skill information contained in individual profiles provides a noisy measure of the skill sets that one would see if skills were assessed based on a common, objective standard. Noise in the measurement of individual-level skills will create attenuation bias when we regress job-based earnings on individuals' skills. For most parts of our subsequent analyses, we however aggregate the skill information of individuals to large groups of workers delineated by such variables as age, education level, or gender, and then focus on average skill differences across these groups. If the distribution of reporting noise is the same within each such group (e.g., if workers of different ages are equally likely to report a given set of skills using fewer or more skill strings), then the skill differences we measure across groups will be the same that would have been observed with a hypothetical objective measure of skills.[9] The group-level comparisons will be biased only if the groups have a different reporting behavior on average. While we do not see a strong reason to expect differential group-specific reporting for many of the worker groups we study, we will address in section 7.1 whether observed skill gaps between women and men might be the result of gendered reporting behavior.

## 4. Definition of Variables and Sample

We next describe how we organize the detailed skills reported on LinkedIn profiles into skill clusters and how we classify these clusters into the three broad domains of general non-managerial, general managerial, and specific skills (section 4.1). The following sections define other relevant variables for our analysis that are either directly observed in the profiles (section 4.2) or inferred from other profile information (section 4.3). Section 4.4 outlines our sample selection process and evaluates the representativeness of the final dataset.

[9] Suppose a worker $i$ of group $j$ has $s_{ij}$ skills according to an unobserved objective skill assessment, but she reports $\sigma_{ij} = s_{ij} + \varepsilon_{ij}$ skills on her LinkedIn profile, where $\varepsilon_{ij} \sim F(\mu_j, \theta_j)$ is a noise term drawn from a distribution function $F(.)$ with group-specific mean $\mu_j$ and shape parameter $\theta_j$. Due to the law of large numbers, the comparison of group means in reported skills between two groups *j*=1 and *j*=2 will tend towards $\bar{\sigma}_1 - \bar{\sigma}_2 \rightarrow \bar{s}_1 - \bar{s}_2 + (\mu_1 - \mu_2)$, and provides an unbiased measure of the objective skill differential $\bar{s}_1 - \bar{s}_2$ if $\mu_1 = \mu_2$.

## 4.1 Skill Clusters

***Derivation of Skill Clusters.*** We observe 1.8 million unique text strings for skills in our analysis sample of LinkedIn data, which retains profiles that report at least one skill (see section 4.4 below for more details on sample selection). The average such profile lists 20 skills, while the numbers of skills at the 10$^{th}$ and 90$^{th}$ percentiles of the distribution are 7 and 37, respectively.

To reduce the dimensionality of the skill data, we group skills that frequently appear together in the same user profiles into skill clusters. Revelio Labs cleans the skill terms and uses a Word2Vec word association algorithm to cluster the most common 10,500 skill strings (which account for more than 90 percent of all skill entries on U.S. LinkedIn profiles) into 50 clusters based on their co-occurrence patterns. After disregarding three clusters that report skill terms in languages other than English[10] and adding a residual cluster that aggregates all infrequent skill strings that are not among the 10,500 most common ones, we retain 48 skill clusters for our main analysis.

Table 1 lists the 48 skill clusters, which are named according to the two most frequent skill strings contained in the cluster. Appendix Table A1 provides further information on cluster composition by listing each cluster's five most frequent skill strings. We observe that the clustering algorithm usually groups terms that refer to either adjacent or overlapping skill concepts. For example, the five most frequent skill strings in the cluster 'accounting, financial reporting' are 'accounting', 'financial reporting', 'auditing', 'financial accounting', and 'accounts payable'.

The skill dimensions emerging from the clustering algorithm paint a rich picture of the multidimensional skills reported by workers. Many skill clusters refer to specific functions within firms such as 'recruiting, human resources', 'project management, budgets', or 'customer satisfaction, customer retention', or they group broadly applicable skills such as 'communication, problem solving' or 'Microsoft Office, customer service'. Other skill clusters instead indicate expertise related to a specific field of knowledge such as 'clinical research, medical devices' and 'legal research, legal writing', or show familiarity with software and technology applications such as 'Photoshop, Adobe Creative Suite' or 'mobile devices, mobile applications'.

---

[10] Three of the 50 skill clusters group a panoply of skills that were entered in Spanish, French, or Portuguese instead of English. Our analysis omits profiles with these skill clusters, since they cluster skills by the languages used to describe the skills rather than the content of the skills.

**Table 1: Skill clusters: Descriptive statistics**

| | Mean<br>(1) | Share positive<br>(2) | Specificity<br>(3) | Executive<br>(4) |
|---|---|---|---|---|
| **Specific skills** | | | | |
| Accounting, financial reporting | 0.409 | 0.094 | 0.604 | -0.005 |
| AutoCAD, SolidWorks | 0.166 | 0.062 | 0.733 | -0.031 |
| Biotechnology, molecular biology | 0.210 | 0.042 | 0.656 | -0.012 |
| Clinical research, medical devices | 0.222 | 0.070 | 0.650 | -0.013 |
| Engineering, project engineering | 0.289 | 0.086 | 0.618 | -0.011 |
| Healthcare, hospitals | 0.499 | 0.099 | 0.629 | -0.011 |
| Insurance, banking | 0.266 | 0.066 | 0.610 | 0.013 |
| Java enterprise edition, Jira | 0.104 | 0.033 | 0.780 | -0.025 |
| Java, Matlab | 0.397 | 0.130 | 0.624 | -0.067 |
| Legal research/writing | 0.337 | 0.049 | 0.732 | 0.030 |
| Mobile devices/applications | 0.047 | 0.028 | 0.611 | 0.002 |
| Real estate, investment properties | 0.191 | 0.039 | 0.644 | 0.046 |
| Revit, SketchUp | 0.168 | 0.051 | 0.640 | 0.001 |
| SQL, software development | 0.515 | 0.109 | 0.704 | -0.053 |
| Telecom., network security | 0.169 | 0.043 | 0.646 | -0.006 |
| Windows server, disaster recovery | 0.134 | 0.045 | 0.698 | -0.004 |
| **General managerial skills** | | | | |
| Analysis, financial analysis | 0.579 | 0.176 | 0.501 | 0.106 |
| Business analysis/process improv. | 0.351 | 0.128 | 0.515 | 0.068 |
| Coaching, leadership development | 0.477 | 0.147 | 0.455 | 0.071 |
| Marketing, social media marketing | 0.741 | 0.223 | 0.462 | 0.110 |
| Program mgmt., security clearance | 0.349 | 0.115 | 0.407 | 0.084 |
| Project management, budgets | 0.636 | 0.285 | 0.279 | 0.151 |
| Sales, strategic planning | 1.325 | 0.408 | 0.279 | 0.238 |
| **General non-managerial skills** | | | | |
| Access, software documentation | 0.131 | 0.080 | 0.515 | -0.038 |
| Communication, problem solving | 0.238 | 0.135 | 0.224 | -0.036 |
| Customer satisfaction/retention | 0.113 | 0.073 | 0.418 | -0.009 |
| Data analysis, databases | 0.363 | 0.162 | 0.398 | -0.059 |
| Editing, public relations | 0.525 | 0.174 | 0.467 | 0.036 |
| Energy, sustainability | 0.148 | 0.045 | 0.549 | 0.025 |
| English, Spanish | 0.166 | 0.108 | 0.214 | -0.014 |
| Excel, customer relations | 0.075 | 0.051 | 0.261 | 0.004 |
| Food, hospitality | 0.191 | 0.046 | 0.443 | 0.009 |
| HTML, JavaScript | 0.311 | 0.099 | 0.581 | -0.038 |

(continued on next page)

**Table 1 (continued)**

| | Mean (1) | Share positive (2) | Specificity (3) | Executive (4) |
|---|---|---|---|---|
| Info. technology, lean six sigma | 0.045 | 0.038 | 0.404 | 0.016 |
| Inventory/operations management | 0.270 | 0.110 | 0.476 | 0.050 |
| Microsoft Office, customer service | 2.861 | 0.682 | 0.127 | 0.020 |
| Other | 1.382 | 0.422 | 0.103 | -0.025 |
| Photoshop, Adobe CS | 0.586 | 0.154 | 0.435 | -0.052 |
| Process improv., cross-func. team lead. | 0.397 | 0.146 | 0.486 | 0.036 |
| Public speaking, research | 1.406 | 0.405 | 0.284 | 0.030 |
| Recruiting, human resources | 0.315 | 0.092 | 0.445 | 0.022 |
| Retail, forecasting | 0.322 | 0.113 | 0.471 | 0.018 |
| Security, emergency management | 0.156 | 0.048 | 0.488 | 0.008 |
| Social media/networking | 0.498 | 0.242 | 0.363 | -0.014 |
| Testing, quality assurance | 0.158 | 0.091 | 0.507 | -0.034 |
| U.s, software dev. life cycle[11] | 0.034 | 0.031 | 0.390 | -0.003 |
| Video production/editing | 0.379 | 0.080 | 0.523 | -0.009 |
| Windows, troubleshooting | 0.269 | 0.109 | 0.417 | -0.040 |

Notes: Column 1 reports the average number of raw skills reported in the cluster. Column 2 shows the share of profiles with at least one raw skill reported in the cluster. Column 3 indicates the Gini coefficient of skill concentration across occupations. Column 4 reports the difference between the fractions of profiles in the "top executives" occupation group and in all other occupation groups that report at least one skill in the cluster. Sample: U.S.-based workers with a college degree whose LinkedIn profiles contain required information. N = 8,850,314.

While LinkedIn users are in principle free to list any concept as a skill on their profiles, the list of most frequent skill strings by cluster in Appendix Table A1 suggests that users typically understand skills as learned abilities that may plausibly result from education, training, and on-the-job experience. The skills indicated by workers on LinkedIn also show considerable overlap with widely used skill terms in job ads.[12]

[11] The abbreviation 'U.s' in the cluster 'U.s, software development life cycle' likely stands for 'user stories', a feature of agile software development aimed to represent users' requirements. This interpretation is consistent with the other frequent strings in this cluster, such as 'relationship building' and 'waterfall methodologies', which also reflect customer-oriented software development skills.

[12] Data Appendix Table 1 of Deming and Noray (2020) lists 51 common skill strings in job ads which they group to the broader domains of social, cognitive, creativity, writing, management, finance, business systems, customer service, office software, technical support, data analysis, and specialized software skills. Nearly half of these skills (24 of 51), and at least one per domain, also appear among the frequently used skill terms on LinkedIn profiles shown in Appendix Table A1. We find little coverage in the LinkedIn data of the domain that Deming and Noray (2020)

***Classification of Skill Clusters by Occupational Specificity.*** We characterize the occupational specificity of each skill cluster based on the extent to which these skills are concentrated among workers of a few occupations rather than being widespread across occupations. For each LinkedIn profile, we measure whether it lists at least one skill of a given skill cluster and then compute each cluster's Gini coefficient across 336 six-digit occupations based on the Standard Occupational Classification (SOC). The resulting specificity measure for each skill cluster is reported in column 3 of Table 1, and we classify the top tercile of skill clusters with the highest Gini coefficients as *specific skills.* These specific skills include knowledge of specialized software for applications such as computer-aided product design ('AutoCAD, SolidWorks') or architectural planning ('Revit, SketchUp'), as well as skills that are specific to areas such as medicine/health ('clinical research, medical devices', 'healthcare, hospitals'), law ('legal research, legal writing'), finance ('insurance, banking', 'accounting, financial reporting'), and science/engineering ('biotechnology, molecular biology', 'engineering, project engineering').

We refer to the two terciles of skill clusters with lower Gini measures of occupational concentration as *general skills*. Many of these skills relate to basic business functions in firms, sometimes combined with the use of generic workplace software.

***Classification of General Skill Clusters into Managerial and Non-Managerial Skills.*** We observe that general skills comprise relatively basic skills such as 'Microsoft Office, customer service' or 'editing, public relations' that may qualify workers for entry-level jobs in firms, but also more advanced skills such as 'project management, budget' or 'coaching, leadership development' that would seem suitable for managerial positions. To reflect this distinction, which is relevant for some of our empirical results below, we further subdivide general skills into general managerial and general non-managerial skills. To this end, we compute for each skill cluster whether the corresponding skills are more concentrated in the occupation group 'top executives' (code 11-1000 of the 2018 SOC occupational classification) than among all other occupations.[13] The seven skill clusters that are most concentrated among top executives are classified as *general* describe as character, which comprises keywords such as 'detail-oriented', 'energetic', or 'self-starter' that may less clearly relate to the concept of skills as learned abilities. We also do not find keywords related to machine learning and artificial intelligence among the top skills in Appendix Table A1, perhaps because these technologies were not yet as broadly used in 2019.

[13] Specifically, we compute the fraction of profiles in the 'top executives' occupation group that report at least one skill from a given skill cluster and subtract from this the fraction of profiles from all other occupation groups that include skills from this cluster. The resulting statistic is indicated in column 4 of Table 1.

*managerial skills*, with the remainder of the general skills being *general non-managerial skills*. It is noteworthy that the skill clusters with the highest concentration among top executives all qualify as general skills as per our definition above. While these skills are particularly frequent among top executives, they are observed across a broad range of other occupations. This pattern likely results because managerial *occupations* capture only a small subset of the workers who perform managerial *functions* across many different occupations.

Our approach of classifying all skill clusters into the three categories specific skills, general managerial skills, and general non-managerial skills is both comprehensive and rule-based. It differs from other analyses that classified the skills of workers or the skill requirements of jobs as routine/non-routine (e.g., Autor, Levy and Murnane 2003, Spitz-Oener 2006, Autor and Handel 2013, Autor and Dorn 2013) or social/cognitive/non-cognitive (Deming 2017) based on only a carefully selected subset of available skill or task variables. The advantage of our approach is that we can include all skills that workers declare in our analysis since occupational specificity is a well-defined concept that can be applied to every skill. The rule-based classification approach reduces researcher discretion but can occasionally generate a classification that does not fit researcher intuition perfectly. For instance, the skill cluster 'sales, strategic planning' whose five most frequent skill strings are 'sales', 'strategic planning', 'team leadership', 'account management', and 'strategy' is classified as a managerial skill, even though 'sales' alone might intuitively be seen as a general non-managerial skill.

Table 1 provides descriptive statistics for the frequency of skills in our sample, as well as their overall occupational specificity and their concentration among top executives. Column 1 of the table indicates that the average LinkedIn profile in our sample comprises more than one skill string from each of the three skill clusters 'Microsoft Office, customer service', 'public speaking, research', and 'sales, strategic planning', while the four least frequent clusters contribute less than 0.1 skills each to the average profile.

There is a strong negative correlation (of -0.53) between skill frequency and skill specificity, which means that many specific skills are also relatively rare. While we classify a third of all skill clusters as specific skills, only one-fifth of the skills on the average LinkedIn profile are specific (4.1 out of 19.9 skills). Conversely, the seven general managerial skills are relatively frequent, accounting for slightly over one fifth of the skills on the average profile (4.6 out of 19.9 skills).

### 4.2 Additional Variables Observed in LinkedIn Profiles

***Education.*** LinkedIn profiles usually report educational degrees along with a field of study and the name of the educational institution that granted the degree. Our data cleaning seeks to standardize this information to obtain three variables: highest degree, field of study for the highest college degree, and college quality as proxied by a university ranking.

We distinguish six levels of education degrees: high school, associate, bachelor, master, professional degree, and doctorate. We parse the educational information from LinkedIn into these categories using information from several online sources on typical abbreviations of degrees (StudentNews Group 2024, Best Universities 2024, YourDictionary 2022) and subsequently hand-map the most frequent remaining categories. The most common non-mapped entries are either missing values or ambiguous terms such as 'certificate' or 'diploma'. If users report more than one college degree, we determine the highest degree based on the following descending ranking: doctorate, professional degree, master, bachelor, associate. If they report multiple degrees of the same type, we use the most recent one.

To classify the field of study of the highest degree, we start with the Classification of Instructional Programs (CIP) taxonomy from the National Center for Education Statistics (2024). As LinkedIn's field of study entries are semi-structured, most users report a field of study that corresponds to a CIP instructional program. For the largest non-mapped field-of-study entries, we again use a manual mapping. The most common field-of-study entries that remain unmapped are missing values, GPAs, or imprecise terms such as 'science'. We proxy for educational quality by merging information from the Times Higher Education (2019) U.S. College Ranking to each user profile based on the name of the college where the highest degree was obtained.

***Experience.*** LinkedIn profiles usually provide graduation dates for their educational degrees and start and end dates for each job spell. We use this information to calculate the actual work experience that every individual accumulated between graduation from the first undergraduate degree (associate or bachelor) and the scrape date in 2019.[14] We also compute potential experience as the entire time interval between graduation and scrape date.

[14] 93 percent of the LinkedIn profiles in our estimation sample report an undergraduate degree. In the case of profiles that report only a graduate or postgraduate degree, we compute experience since the start date of the first advanced degree (master, professional degree, doctorate). In cases where the end date of a job is missing, we assume that the employment is ongoing if no subsequent job is listed, or ended with the start date of the following job if a subsequent job is reported.

***Occupation and Location.*** LinkedIn entries for jobs typically contain information on occupation and place of work.[15] Revelio Labs aggregates the occupational information to 336 occupation groups according to the 2018 SOC classification and codes the place of work based on Metropolitan Statistical Areas (MSA) and U.S. states.

### 4.3 Variables Inferred from LinkedIn Profiles and External Sources

***Gender.*** LinkedIn profiles do not usually indicate workers' gender. Revelio Labs therefore infers gender from individuals' names. It uses an algorithm that computes the likelihood that a person with a given name is a woman and classifies profiles where this likelihood is greater than 50 percent as belonging to women. In practice, the predicted female probability is very close to zero or very close to one for a vast majority of all profiles. The imputation of gender is purely name-based and does not consider additional clues such as having a gender-typical education or occupation, thus avoiding potential stereotyping based on such variables.

***Age.*** To impute age, we utilize the date information of the first educational degree in a profile. If the profile indicates a high school degree, we assume the user finished high school at the typical graduation age of 18. On LinkedIn profiles that contain both the year of high school degree and the start years of subsequent studies, a median of zero years elapses between high school graduation and the beginning of undergraduate studies (associate, bachelor), and a median of six years elapses between high school graduation and the start of graduate studies (master, professional, doctorate). For LinkedIn profiles that do not indicate a high school graduation year, we thus assume an age of 18 at the start of the first undergraduate degree, or if that information is missing, an age of 24 at the start of graduate studies.[16]

***Job-Based and Occupation-Based Earnings.*** LinkedIn profiles do not report workers' earnings. Therefore, we are unable to observe individual-level earnings that may depend on many idiosyncratic and unobserved factors. However, we use two complementary strategies to infer whether individuals work in jobs or occupations that are typically highly paid.

[15] In case a profile reports multiple job spells at the time of the scrape date, we assign the occupation with the highest earnings.

[16] These estimated entry ages by educational level fall within the typical educational entry ages for the US that are reported by the OECD (2021).

Our primary such metric is based on Revelio Labs' proprietary salary model that predicts the annual salary for each job from the job title, company name, location, job tenure, and year of observation. This model is trained on salary information in publicly available visa application data, databases of self-reported salary data, and job postings data. It is noteworthy that this earnings imputation is largely based on characteristics of the job (job title, company, location) and considers only tenure as a worker-level earnings determinant. We refer to the earnings data provided by Revelio Labs as 'job-based earnings' to clarify that this measure is informative primarily for between-job but not for within-job earnings differentials across workers. This measure for instance allows us to analyze whether gender differences in skills help to statistically explain the differential representation of women and men across high-paid and low-paid jobs, but we cannot study whether a gender skill gap contributes to gender earnings gaps within a job type.[17]

We complement the job-based earnings from Revelio Labs with a measure of occupation-based earnings that we derive from the Census Bureau's American Community Survey (ACS). Using 95 four-digit SOC codes, we compute the average earnings of all employees of that occupation in the pooled 2018 and 2019 ACS data obtained from Ruggles et al. (2024).[18] The occupation-based earnings measure does not exploit the detail of the LinkedIn data as well as the job-based earnings from Revelio Labs, but it provides a useful complementary measure given its simplicity and transparency. Differentials in occupational wage levels have also been previously identified as an important component of the gender wage gap (e.g., Olivetti, Pan, and Petrongolo 2024). Appendix Figure A1 shows that there is a very high correlation between the occupation-based earnings from the ACS and the job-based earnings derived by Revelio Labs (correlation coefficient of 0.89).

### 4.4 Sample Selection and Representativeness

We focus on college graduates in the U.S. because LinkedIn is particularly widely used in that population (Auxier and Anderson 2021). Our estimation sample includes all LinkedIn profiles of U.S.-based workers with a college degree that contain information on skills and other basic variables that are critical for our analysis.

---

[17] The construction of Revelio Labs' job-based earnings measure does not consider either worker gender or skills. Therefore, it will assign the same job-typical earnings value to workers of the same job type regardless of their gender or skill sets.

[18] In unreported robustness analyses, we alternatively explored occupational earnings metrics that are based only on full-time, full-year employees, or only on male employees. Results for these measures are very similar to those that use earnings from all employees.

Appendix Table A2 documents the steps of our sampling procedure. In total, we observe 61.9 million profiles of workers who held a job in the U.S. in 2019. As a first step, we drop 6.4 million incomplete profiles for which Revelio Labs was not able to infer baseline data on gender, occupation, or location. We subsequently retain only the three fifths of individuals who indicate valid educational information. Among these, 73 percent report at least one college degree, which confirms the overrepresentation of college graduates on LinkedIn relative to their share of about 50 percent in the overall workforce (Bureau of Labor Statistics 2022). We further restrict the sample to profiles that indicate education dates from which we can infer age, those whose imputed age in 2019 is 23 to 64 years, and those where the start year of the first indicated work experience appears broadly consistent with age and graduation dates.[19] Finally, since the focus of our analysis is on skills, we restrict the analysis sample to the roughly half (49 percent) of the profiles remaining after the previous filtering steps that report at least one skill.

Our final analysis sample contains 8,850,314 profiles. While inclusion in this final sample requires that a profile comprises sufficiently complete information on education, experience, and skills, Appendix Table A2 indicates that the totality of the sample selection steps does not strongly alter the sample composition in terms of age or gender.

To assess the representativeness of the sample in relation to the U.S. college-educated labor force, we compare basic descriptive statistics of our estimation sample to representative worker data from the 2019 American Community Survey (ACS). The 8.85 million worker profiles with skills in our sample correspond to 13.6 percent of all U.S. workers with college degrees in the age range of 23 to 64 according to the ACS. The LinkedIn sample is somewhat younger (average age of 37.5 vs. 41.9 years) and more male (54 vs. 48 percent) than the ACS population. Panel A of Appendix Figure A2 shows that the LinkedIn sample also contains considerably more workers with graduate degrees (39 vs. 29 percent). These patterns are consistent with a slightly higher propensity of young people and males to use the platform, and a notable bias of the user base towards more highly educated individuals (Auxier and Anderson 2021).

[19] We consider work experience information to be inconsistent with educational timing if imputed age at the start of the first job is lower than 14, or if the duration between graduation from the highest degree and start of the first job is more than five years. While the omitted profiles may contain some valid profiles of individuals with unusual educational and professional career trajectories, we suspect that many of these profiles are either incomplete or contain erroneous data on either education or work years.

Occupation groups that tend to be concentrated in the public sector, such as education and health care, are underrepresented in the LinkedIn sample (Appendix Figure A2, Panel B), whereas occupation groups such as management, finance, IT, and media are overrepresented, perhaps because professional networking is seen as particularly important in these fields. The coverage of fields of study (Panel C) indicates a similar pattern with a relative underrepresentation of health and social science majors and an overrepresentation of business and STEM majors. The regional distribution of the LinkedIn observations across the nine U.S. census divisions maps the ACS population quite well (Panel D).

Overall, the LinkedIn sample seems broadly representative of the US workforce with college degrees. While our main results below give equal weight to each worker in the LinkedIn sample, we document in Appendix C that key results remain very similar when we reweight the LinkedIn observations based on population weights for demographic groups taken from the ACS.

## 5. Human Capital Investments and Skills

Classic human capital theory posits that workers acquire skills both through education and through work experience (e.g., Becker 1964, Mincer 1974). If LinkedIn users update their profiles with newly acquired skills over time, we would expect that workers who completed longer educational programs and those with longer periods of work experience report a larger number of skills. In addition, a well-known feature of the Mincer (1974) model is that the relationship between potential experience and earnings is concave, suggesting that skill accumulation is faster in the years that immediately follow college graduation.[20]

Another important feature of human capital theory is the distinction between skills accumulated by education and skills acquired through on-the-job experience. This distinction relates directly to the concepts of general versus specific skills: Many types of formal education seek to convey general skills that can be employed in a broad range of occupations, perhaps with the exception of graduate and professional programs that prepare students for particular occupational fields. Work experience instead fosters an accumulation of specific skills that may be specifically relevant for a worker's current occupation. Moreover, work experience is likely also important for the acquisition of managerial skills, which would explain why few workers attain managerial positions right after graduation from college.

---

[20] See Appendix Figure A3 for an age-earnings profile of US college graduates based on ACS data. The cross-sectional structure of our data prevents us from directly observing the skill accumulation of individual workers over time. Our empirical analysis therefore follows the classical approach of Mincer (1974) who tested his human capital theory by comparing workers with different educational attainment and work experience who were observed at the same time.

Based on these considerations, we formulate the following hypotheses that we test in our cross-sectional data:

(i) Workers with higher experience report
  a. a larger number of skills, with a concave relationship between experience and skills
  b. a larger fraction of specific and managerial skills

(ii) Workers with more advanced educational degrees report
  a. a larger number of skills
  b. a larger fraction of specific and managerial skills

We test the first two predictions in section 5.1 and the latter two predictions in section 5.2.

### 5.1 Skills by Age and Experience

***Number of Skills.*** One of the most widely documented empirical patterns in labor economics is a concave age-earnings profile that indicates a strong positive relationship between age and earnings at younger ages and a flat or even slightly declining relationship at older ages. In the Mincer (1974) model, this pattern results from on-the-job training that fosters skill acquisition especially at younger ages. If self-reported skills on LinkedIn profiles correspond to the skills considered in this model, we would expect to see a concave age-skill profile.

Intriguingly, the age-skill profile we derive from the LinkedIn data in Figure 1 indeed closely mirrors the well-known concave shape of age-earnings profiles. The figure depicts the average number of skills reported by college graduates for two-year age bins ranging from ages 23 to 64. The mean number of skills increases from 14.6 at age 23-24 to 22.3 at age 49-50 and then remains rather flat, declining only weakly at the very end of the age range.[21]

[21] Appendix B demonstrates that a concave shape of the age-skill profile similarly shows in a sample of workers who recently switched their employers, indicating that the concavity is not driven by lack of profile updating among older workers.

**Figure 1: Age-skill profile: Number of reported skills by age**

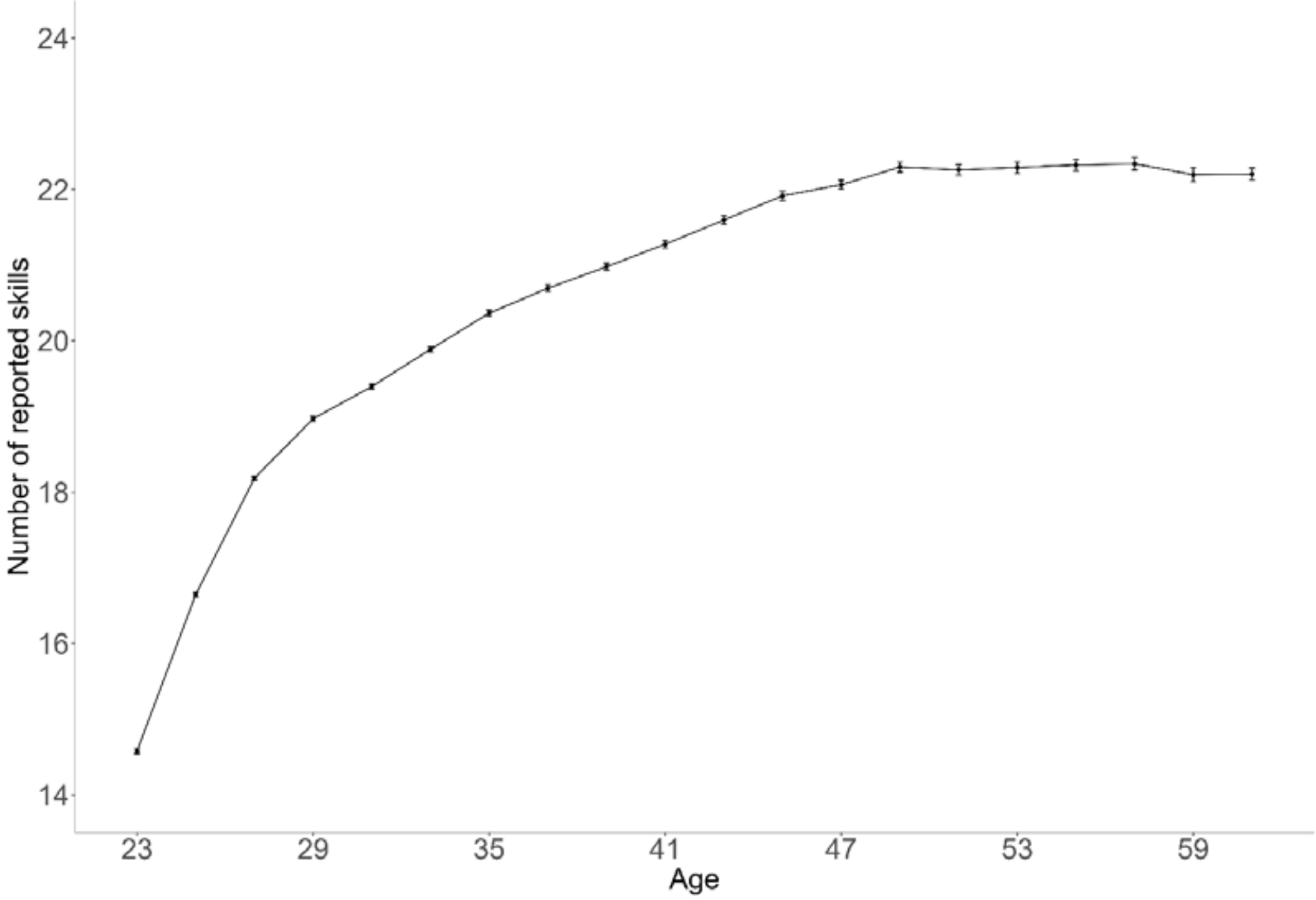

Notes: The figure plots the average number of reported skills for two-year age bins, except for the last bin that combines ages 61-64 due to lower sample size. The small error bars depict 99 percent confidence intervals for the skill averages.

Skill profiles have a similar concave shape when depicted by experience rather than age. Given that all individuals in the sample are college graduates, age is closely aligned with potential experience, i.e., the number of years of work experience that an individual could have accumulated had she been employed without interruption since graduation. Panel A of Figure 2 shows that the number of reported skills increases until about 25 years of potential experience and then flattens.

The LinkedIn profiles also allow us to compute years of actual work experience. For a given worker, actual work experience will be lower than potential experience if the worker did not hold a job for every month after graduation from college.[22] If skills are acquired primarily through work experience rather than through general life experience as a worker ages, then the number of reported skills should increase more with actual experience than with potential experience. Panel

[22] Our measure of actual work experience is likely an upper bound for the true experience value. Since LinkedIn profiles report the months but not exact dates of employment, we can only observe employment breaks that include at least a full calendar month. Moreover, employees may not report transitory absences from a job, such as parental or sickness leave.

A of Figure 2 confirms this prediction. For instance, the average worker with 20 years of actual work experience reports about one skill more than the average worker who has 20 years of potential work experience.

As we study workers who were employed in 2019, any difference between actual and potential experience is the result of temporary absences from work between college graduation and the start of the current job. For the median worker in our sample, the ratio of actual to potential experience is 0.93, and this small difference between actual and potential experience helps to explain why the relationships of skills with either potential or actual experience differ only modestly in Panel A of Figure 2.

In a complementary analysis, we single out individuals for whom actual experience lags considerably behind potential experience. We classify workers as having a 'low attachment' to the labor market if they have a ratio of actual to potential experience below 0.8, and cumulative non-employment spells of at least half a year. Everyone else is classified as high attachment. Panel B of Figure 2 shows that for a given age, low attachment workers with significant career breaks report substantially lower levels of skills. The gap is particularly pronounced during the thirties, when the difference between high- and low-attachment workers amounts to over three skills on average.

***Composition of Skills.*** We next test the hypothesis that younger workers whose human capital was formed primarily through schooling possess primarily general non-managerial skills, while older workers report relatively more specific and managerial skills that may have been acquired on the job. Figure 3 confirms these predictions. At age 23-24, 79.8 percent of workers' skills fall in the general non-managerial category, while this share decreases steadily to 47.5 percent at age 45-46 and remains roughly constant at older ages. The fraction of occupation-specific skills more than doubles from 10.5 percent at age 23-24 to 25.5 percent at age 39-40 and then levels off. The fraction of managerial skills has an even stronger positive age gradient and grows up to a higher age. It triples from 9.7 percent at age 23-24 to 28.9 percent at age 49-50 and remains at similar levels for higher ages.

**Figure 2: Number of reported skills by experience and labor-market attachment**

*Panel A: Number of reported skills by actual and potential work experience*

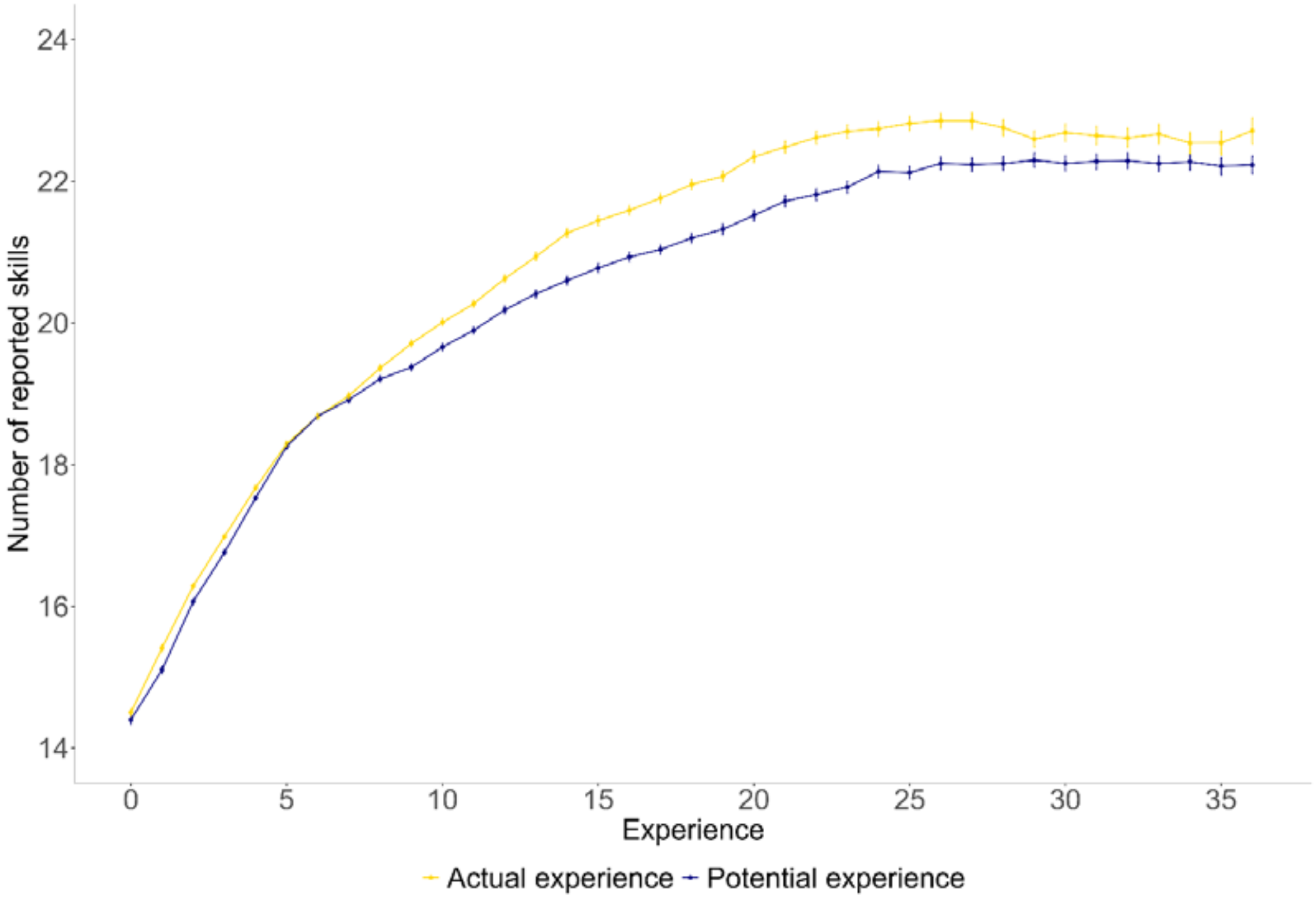

*Panel B: Number of reported skills by high and low labor-market attachment*

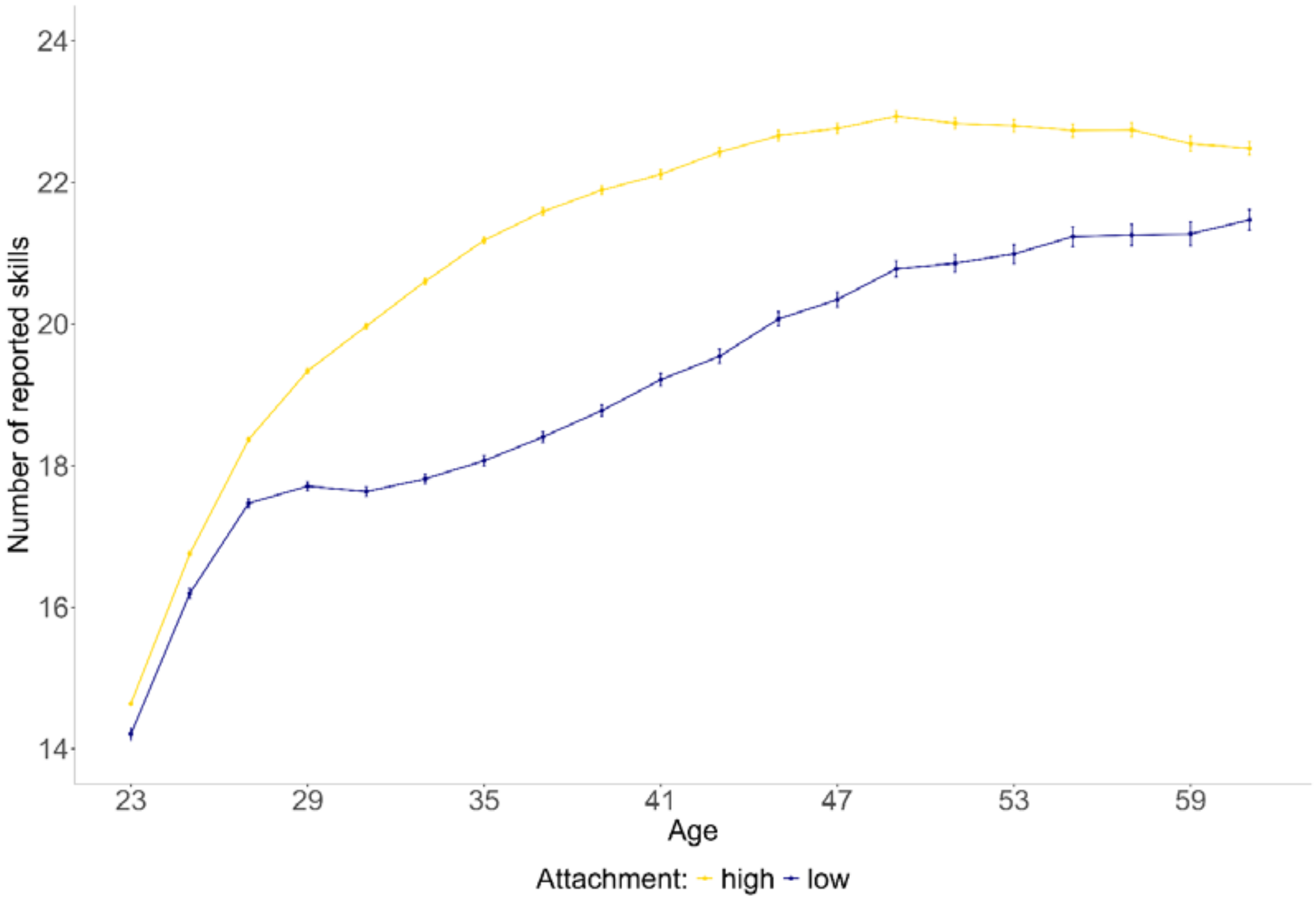

Notes: Panel A reports the average number of reported skills for workers with the indicated level of potential or actual work experience. Panel B reports the average number of reported skills for workers with low labor force attachment (defined as a ratio of actual to potential experience below 0.8 and cumulative non-employment spells of at least half a year) and for workers with high attachment (everyone else). Error bars depict 99 percent confidence intervals.

**Figure 3: Skill composition by age: Shares of general, managerial, and specific skills**

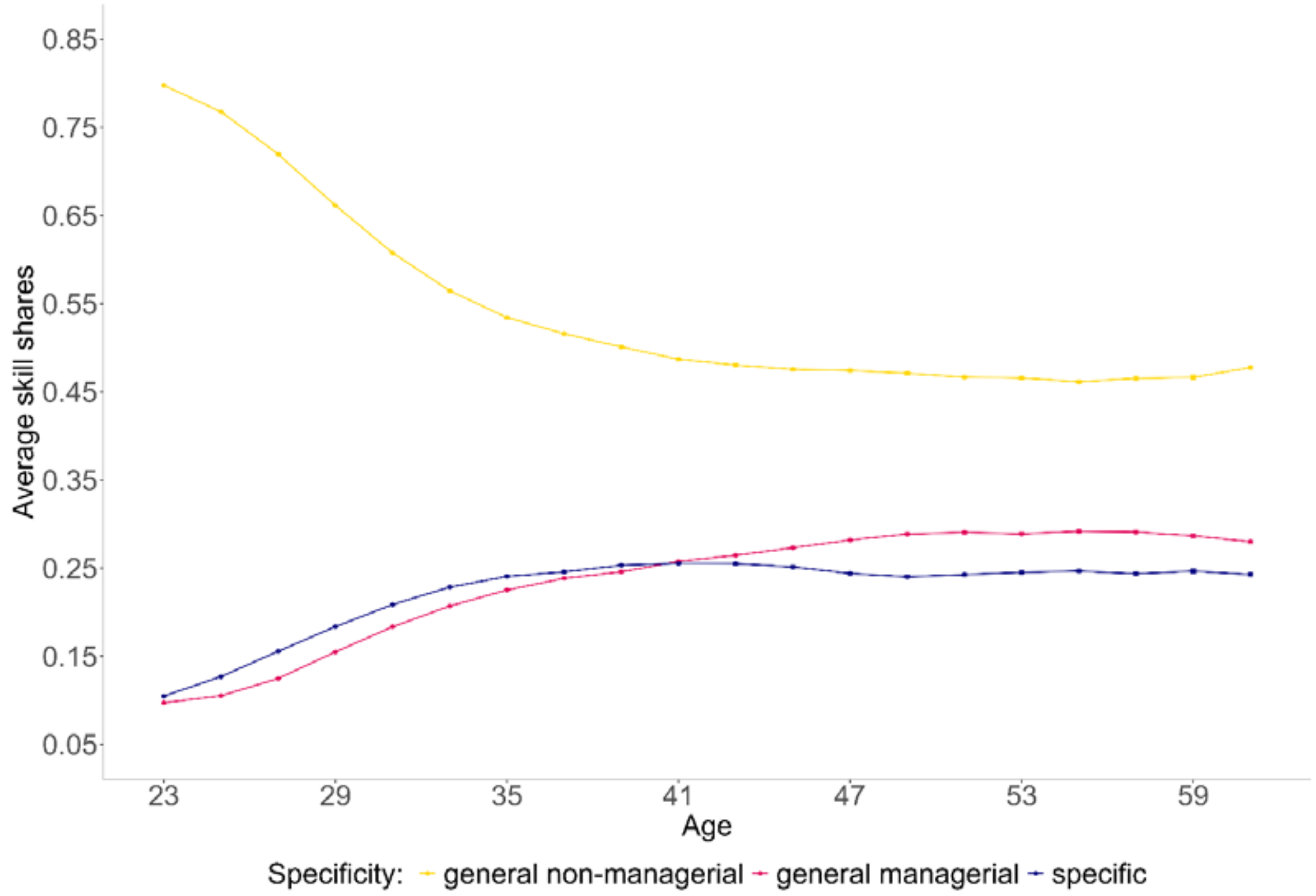

Notes: The figure shows the average fractions of general non-managerial, general managerial, and occupation-specific skills by age. Two-year age bins; last bin combines ages 61-64. Error bars depict 99 percent confidence intervals.

Patterns of skill composition by potential and actual experience and by high and low labor-market attachment again look similar to the patterns by age (Appendix Figure A4). The decline in the fraction of general non-managerial and the increase in specific and managerial skills is stronger for additional years of actual experience than for extra years of potential experience, consistent with the interpretation that skill sets evolve with on-the-job experience, rather than just with general life experience.

## 5.2 Skills by Education: Degrees, Fields of Study, and College Quality

Classical empirical implementations of human capital models such as Becker and Chiswick (1966) or Mincer (1974) proxied for the amount of human capital obtained from education with years of schooling. Subsequent work argued that human capital may depend not only on education years but also on degrees obtained, fields of study, or attending highly-ranked and prestigious universities (e.g., James et al. 1989, Altonji, Arcidiacono, and Maurel 2016).

We next conduct descriptive analyses that investigate the number of skills and the shares of general, specific, and managerial skills reported by workers who differ in terms of degree level, field of study, and college ranking. These analyses assess whether greater or different educational investments translate into corresponding patterns in terms of the number and composition of skills.

Panel A of Figure 4 indicates that the average number of reported skills is increasing with typical education program length for the three most frequent types of degrees in our data. College graduates with Associate's degrees report an average of 18.3 skills, those with Bachelor's degrees list 19.6 skills, and those with Master's degrees indicate 21.3 skills. If we assume that these graduates' total education consists of 12 years of schooling before entering college, and subsequent studies of two years for an Associate's, four years for a Bachelor's, and six years for a Master's degree, then these numbers imply a nearly linear relationship between education years and skills, with about 0.7 additional skills reported per year of education. However, this linearity is notably violated for the 9 percent of graduates in our sample who hold professional degrees or doctorates and who on average report a lower number of skills than the Master's graduates despite having equal or longer periods of education.[23]

The holders of professional degrees or PhDs, however, stand out by reporting roughly double the fraction of specific skills compared to those with other degrees. This pattern is consistent with the interpretation that many study programs for Associate's, Bachelor's, and Master's degrees emphasize general skills that are broadly applicable across a wide range of occupations, while programs for PhD studies or professional degrees such as MDs or MBAs are more targeted towards specific occupational careers and thus convey more specific skills.

[23] A potential explanation for this pattern is that holders of advanced degrees may be less inclined to report basic skills on their profiles, as discussed in section 3.3.

**Figure 4: Number of reported skills and skill composition by education**

*Panel A: Skills by highest educational degree*

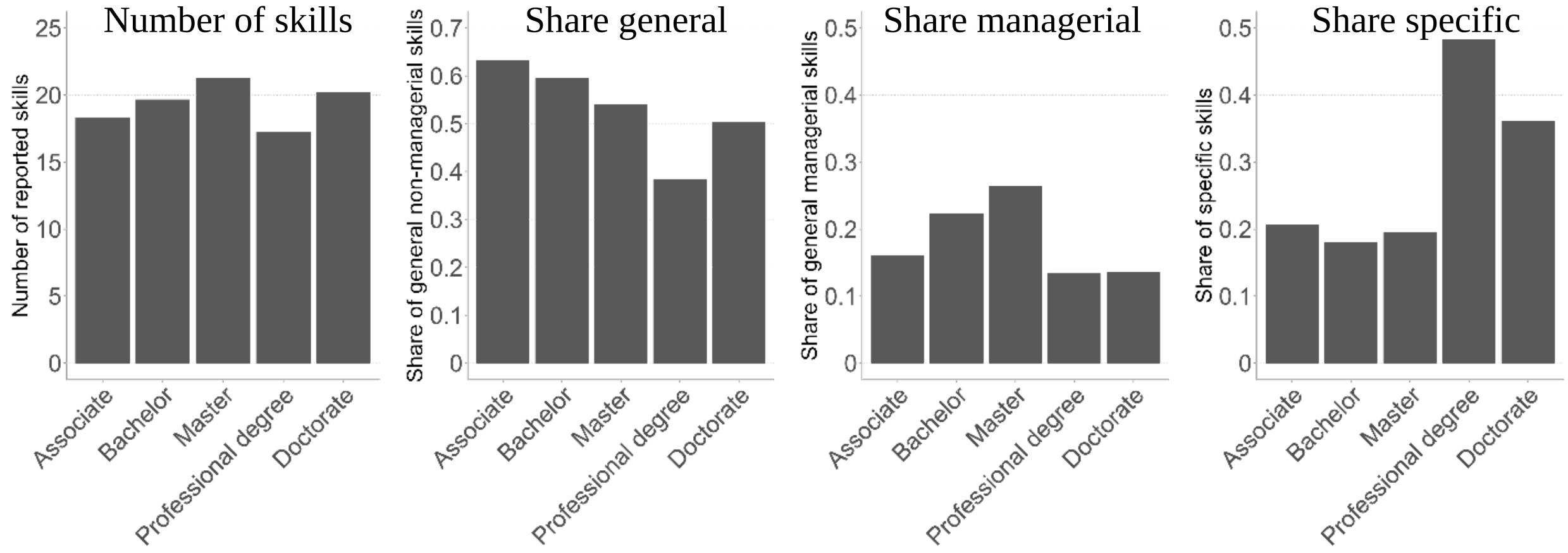

*Panel B: Skills by field of study*

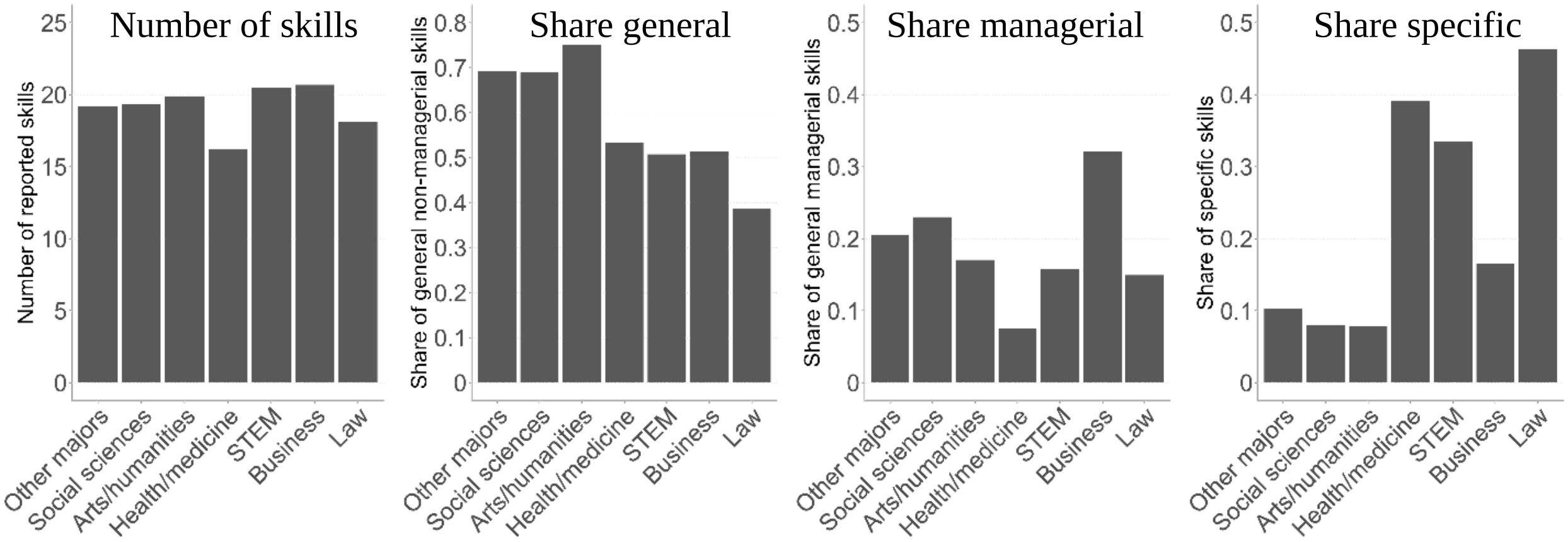

*Panel C: Skills by college ranking*

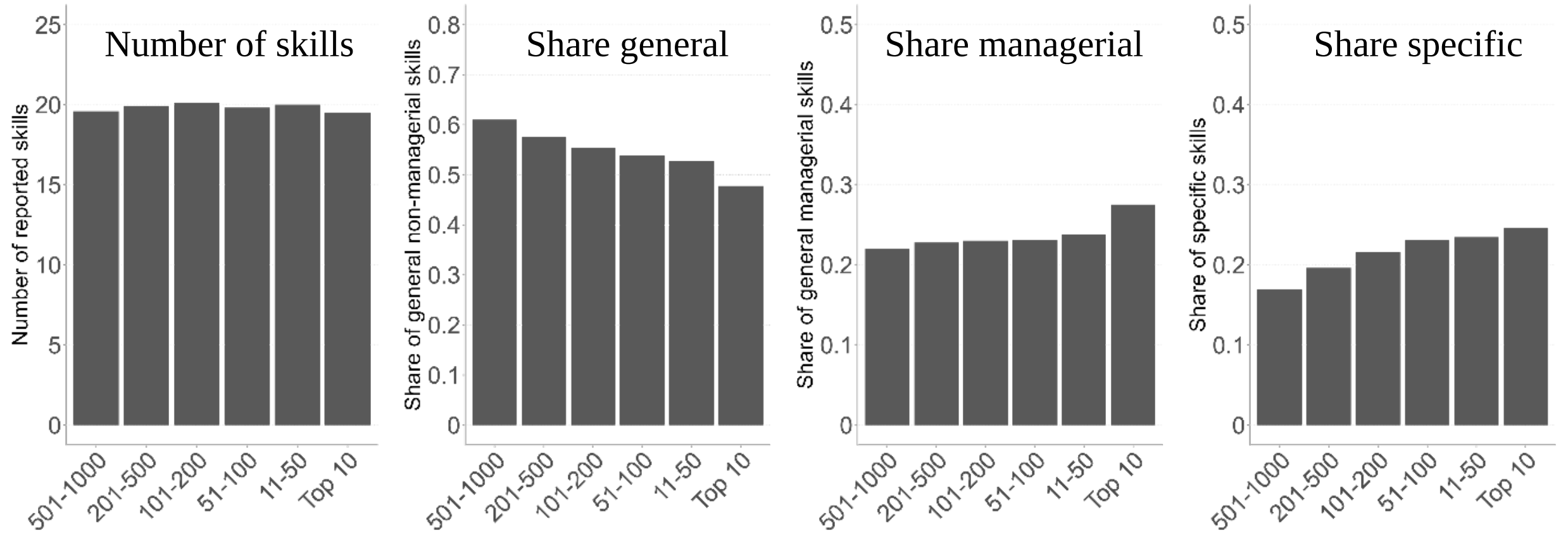

Notes: Average number of reported skills and average fractions of general non-managerial, general managerial, and occupation-specific skills by highest educational degree, field of study, and Times Higher Education (2019) U.S. college ranking, respectively. Samples: Panel A: full sample; Panel B: sample with field-of-study information (93.1 percent); Panel C: sample of graduates from U.S. colleges listed in Top 1000 of college ranking (68.5 percent).

We next explore skill patterns for graduates with different fields of study in Panel B of Figure 4. For ease of illustration, we aggregate the 37 detailed study subjects observed in the data into seven broader categories which are ordered by ascending average earnings in the figure. While there is only modest variation in terms of the number of reported skills by subject area, there are noteworthy differences in the composition of skills. Individuals who majored in disciplines whose graduates achieve higher average earnings report a higher fraction of specific skills than graduates of fields that are typically associated with lower-paid careers. Specific skills account for 30 percent or more of the skills indicated by individuals with health, STEM, or law degrees. By contrast, fewer than 10 percent of the skills are occupation-specific for graduates of arts and humanities or social sciences. Individuals with a business degree report by far the highest fraction of managerial skills.[24]

Panel C of Figure 4 reports skill statistics by college quality, which we approximate by the U.S. college ranking of Times Higher Education (2019). There is little difference in the total number of skills reported by graduates of differentially ranked institutions. However, the shares of managerial and specific skills in workers' skill sets are considerably higher for those who graduated from top U.S. universities rather than lower-ranked schools.

One challenge for the interpretation of the findings in this section is that they correlate only one dimension of education with skills at a time. It thus is not clear from the evidence of Figure 4 whether, for instance, the higher fraction of specific skills among graduates of highly ranked universities reflects a specialization of these universities in study fields or degree programs that are associated with more specific skills, or whether college quality is correlated with a higher share of specific skills also conditional on study fields and types of degrees. In Appendix Table A3, we present multivariate regressions that relate either the total number of reported skills or the shares of general non-managerial, managerial, and specific skills to indicators for degree types, study fields, college ranking, and experience. This exercise suggests that the results presented in this and in the previous section also hold as conditional correlations: While there is relatively modest variation in the number of skills reported by college graduates whose education differs in terms of degrees, field of study, or college ranking, there are notable differences in the composition of skills.

---

[24] Although our analysis in the previous section suggests that managerial skills—like specific skills—are often acquired through on-the-job experience, there is evidence that business school students obtain greater managerial skills already in college (Kang and Sharma 2012).

Holders of advanced degrees, those studying science and professional fields, and graduates of highly-ranked universities report higher shares of occupation-specific skills, and in some cases, higher shares of managerial skills. Greater actual work experience is associated with a higher number of skills and larger shares of specific and managerial skills.

## 6. Skills and Job-based Earnings

After having established that skills are systematically related to human capital investments from which these skills presumably result, we investigate the second foundational tenet of human capital theory: individuals with greater skills should command higher earnings. Section 6.1 studies whether workers with more skills, and those with higher fractions of specific and managerial skills, are employed in higher-paid jobs. We then analyze which detailed skills are most strongly associated with higher-paid jobs (section 6.2) and assess the extent to which differences in skill sets help to account for the variation in job-based earnings across workers (section 6.3).

### 6.1 Number of Skills, Skill Composition, and Earnings

The simplest metric of the skill sets reported on LinkedIn profiles is the number of reported skills. The results in Figure 1 on the average number of reported skills by age display a remarkable similarity to the well-documented concave shape of age-earnings profiles. We thus begin our analysis of the relationship between skills and earnings by investigating whether individuals who report a larger number of skills tend to have higher job-based earnings.

Results show that workers who report more skills are indeed more likely to have higher-paid jobs. Panel A of Figure 5 indicates that log earnings are consistently and, over a broad range of skill levels, nearly linearly increasing in the number of reported skills. On average, each additional skill is associated with a 0.67 log point increase in annual job-based earnings.

We established in section 5 that workers with more advanced educational degrees and those with more work experience report larger fractions of specific and managerial skills and a lower share of general non-managerial skills. Panel B of Figure 5 indicates that job-based earnings vary strongly with the skill mix across these broad skill domains. Individuals who report a larger share of occupation-specific skills tend to have substantially higher imputed earnings. Similarly, and even more strongly, earnings increase in the share of managerial skills. Consequently, the earnings gradient is negative in the share of general non-managerial skills.

**Figure 5: Job-based earnings by number of reported skills and skill composition**

*Panel A: Job-based earnings by number of reported skills*

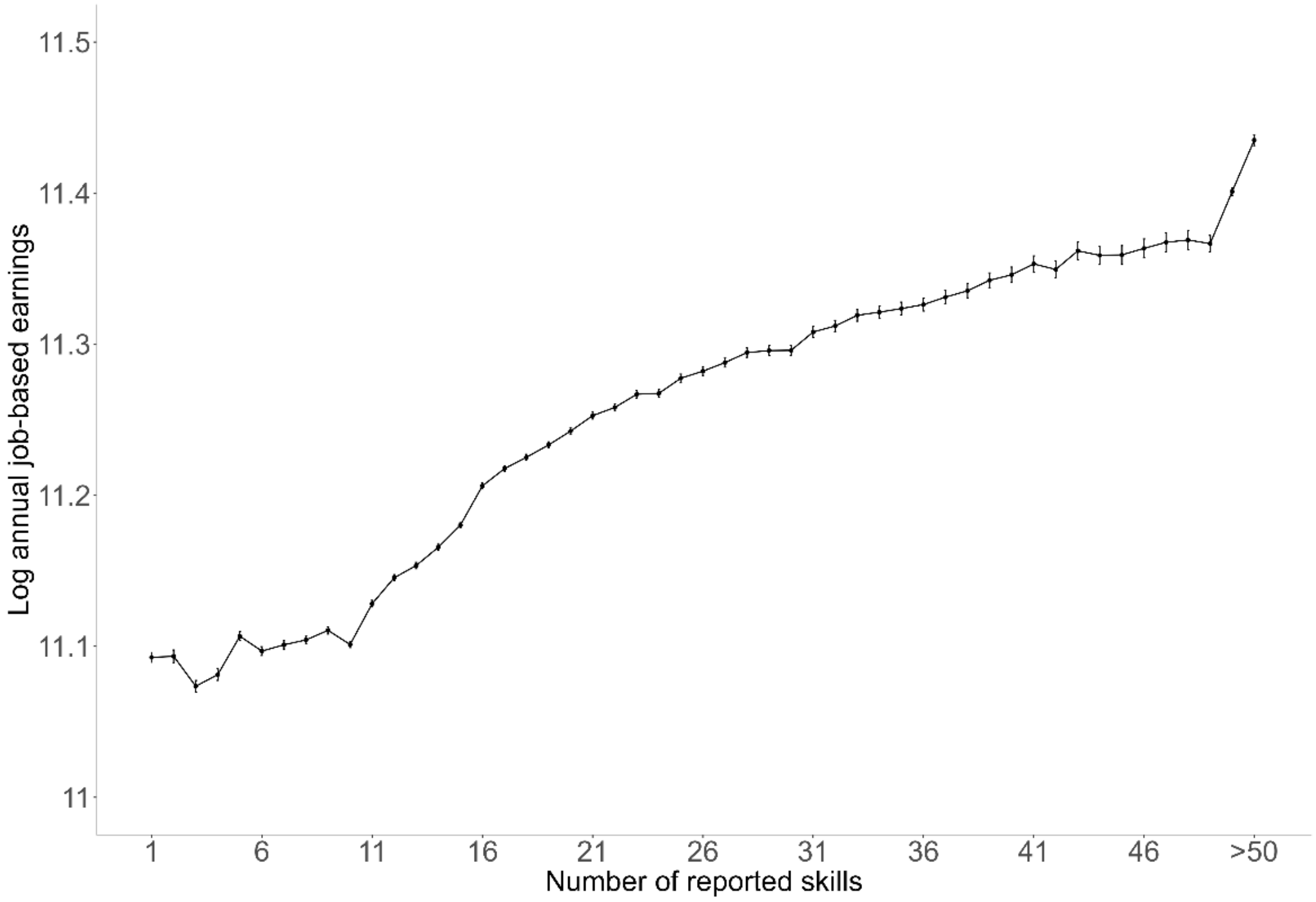

*Panel B: Job-based earnings by skill composition*

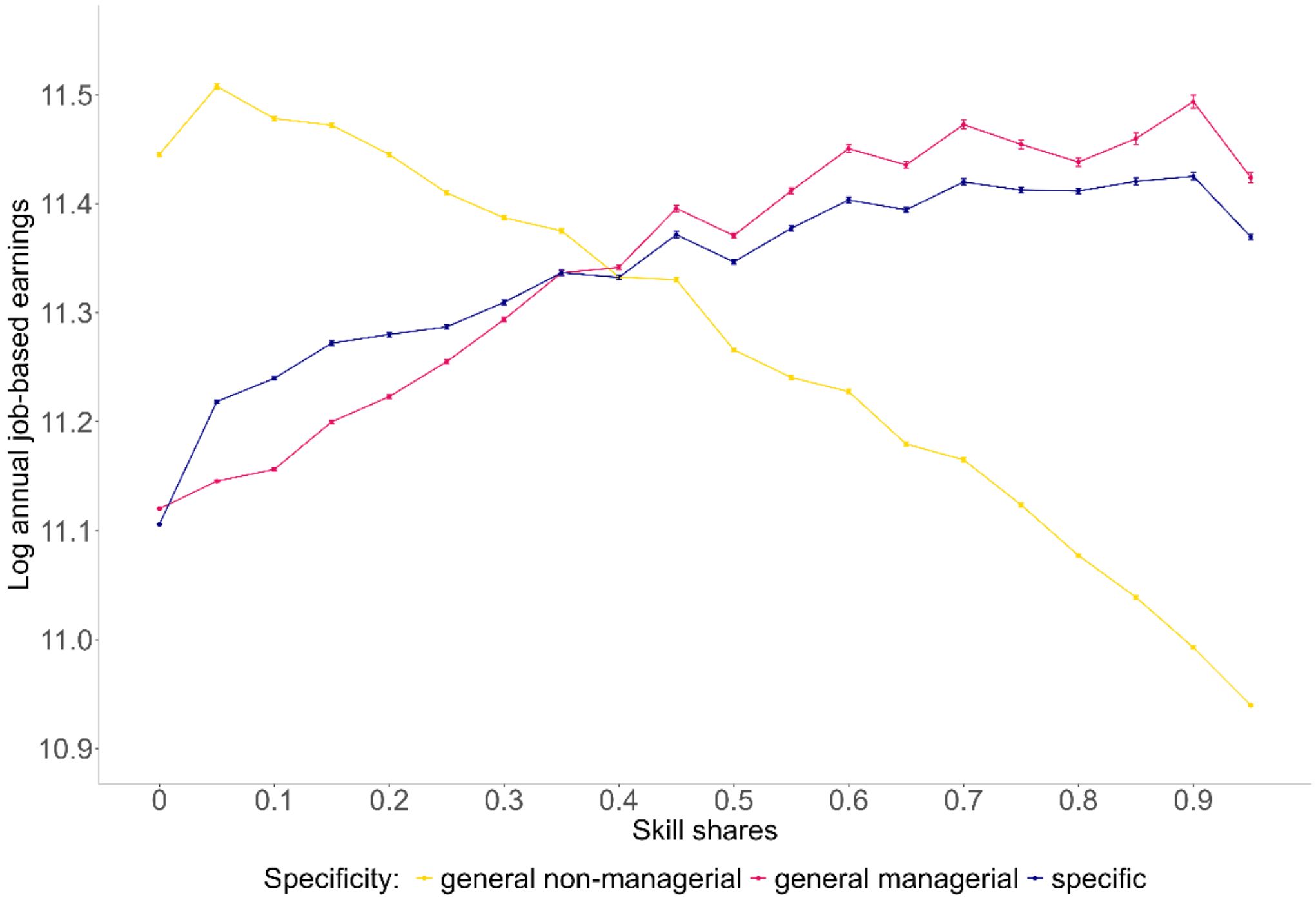

Notes: Average log annual job-based earnings are imputed by Revelio Labs' proprietary salary model. Panel A indicates average earnings for all LinkedIn profiles with the indicated number of skills, while Panel B indicates average earnings for five-percent bins of general non-managerial, general managerial, and occupation-specific skills. Error bars depict 99 percent confidence intervals.

The patterns for the association between skill categories and earnings in Panel B of Figure 5 help us to qualify the section 5.2 results on the relationship between education and skills. While we observed little variation in the number of reported skills for different fields of study, we now note that the primarily general skills reported by graduates in humanities and social sciences appear to be less valued in the labor market than the more specific skills of science, medicine, and law graduates or the managerial skills of business graduates.

### 6.2 Multidimensional Skills and Earnings

We more formally assess the association of different types of skills with job-based earnings by regressing workers' imputed log earnings on the total number of reported skills and the shares of skills reported in each of the detailed skill clusters listed in Table 1.[25] Figure 6 shows the estimated regression coefficients for each skill cluster, ordered by decreasing coefficient size within the three broad skill categories (specific, managerial, general non-managerial). The omitted reference category for the skill share variables is the cluster 'customer satisfaction/retention', which has the weakest positive association with job-based earnings.

The results in Figure 6 corroborate the previous observation that larger shares of specific and managerial skills are associated with higher job-based earnings than general skills. About four-fifths of the clusters of specific and managerial skills have above-average earnings coefficients. Workers with specialized IT skills ('mobile devices, mobile applications', 'Windows server, disaster recovery', 'SQL, software development', 'Java, Matlab'), legal skills ('legal research, legal writing'), medical skills ('clinical research, medical devices'), and various types of managerial skills ('analysis, financial analysis', 'business analysis, business process improvement', 'sales, strategic planning', 'marketing, social media marketing') obtain substantially higher earnings than other workers, with each one percentage-point increase in the share of skills in these clusters being associated with a 1.1 to 1.6 log point greater earnings premium relative to skills in the baseline skill category of 'customer satisfaction/retention'.

---

[25] We obtain qualitatively similar results from a regression that relates log earnings to the number of skills reported in each of the 48 skill clusters. The model with total number of skills and shares of skills by cluster however achieves a better goodness of fit as it is less sensitive to outlier observations from individuals who report an unusually large number of skills in a given cluster.

**Figure 6: Association between job-based earnings and skills**

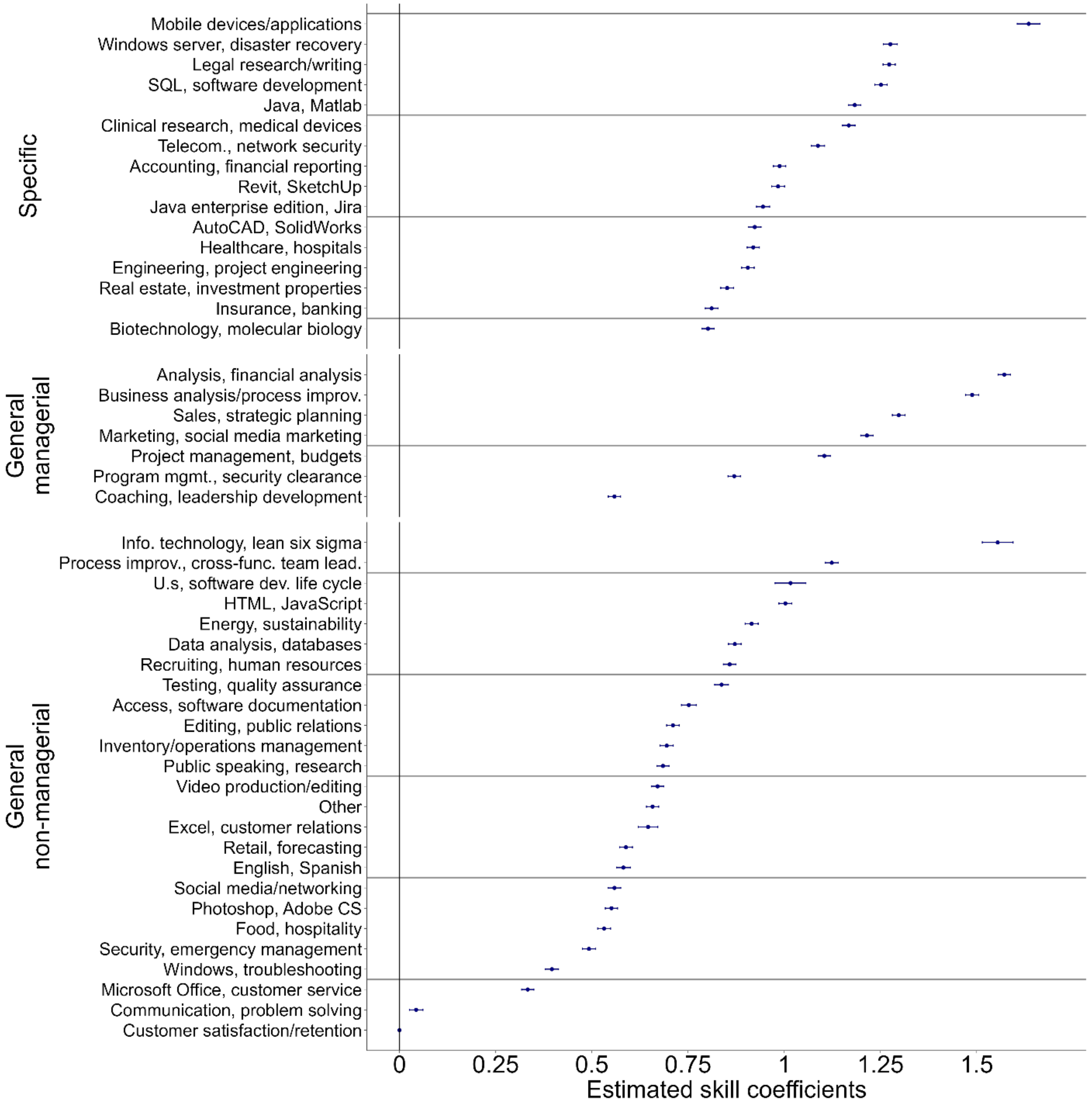

Notes: The figure plots estimated skill coefficients and 99-percent confidence intervals from an individual-level regression of log annual job-based earnings (in log points) on the total number of reported skills and shares of these skills pertaining to each of 47 skill clusters (in percentage points). The omitted skill cluster share in the regression is 'customer satisfaction/retention'. See column 4 of Appendix Table A4 for tabulated coefficients.

Conversely, four-fifths of the general skill clusters have below-average coefficients in the earnings regression. Estimated earnings coefficients are smallest for the skill clusters 'customer satisfaction, customer retention' (the reference category), 'communication, problem solving',

‘Microsoft Office, customer service’, and ‘Windows, troubleshooting’, which comprise relatively basic customer interaction, communication, and standard software skills. The few general skill clusters that have large earnings coefficients instead comprise more advanced technology and process management skills (‘information technology, lean six sigma’, ‘process improvement, cross-functional team leadership’) that are used across a broad range of occupations.

### 6.3 Do Skills Explain More Earnings Variation than Standard Human Capital Measures?

The systematic relationships between skills reported on LinkedIn profiles and job-based earnings strengthen the interpretation of the self-reported skills as a metric for human capital that is valued in the labor market. But are these skill measures better at explaining earnings variation than more conventional proxies for human capital such as experience or education? This should be the case if employers reward workers’ skills rather than their past investments into skill acquisition, and if high-quality measures of skills, experience, and education are available. However, to the extent that skills on LinkedIn profiles are reported noisily as discussed in section 3.3, it is conceivable that measures of skill investments continue to provide better proxies for the true level of worker skills and thus explain more earnings variation.

We study earnings regressions that relate job- or occupation-based earnings to either a set of education and experience variables in the spirit of Mincer (1974), or the LinkedIn skill measures, or both. These regressions ask whether workers with specific individual human capital attributes (e.g., higher levels of education, experience, or skills) are overrepresented in highly paid jobs or occupations. The regressions do not control for job-level predictors of earnings such as job titles or occupations, since these variables have been used in the construction of the earnings measures and are thus mechanically correlated with the outcome variable.[26]

We begin in Table 2 with a basic OLS regression model in the spirit of the classical Mincer equation that relates log earnings to years of education and a quadratic of years of potential experience. Since our sample consists entirely of college graduates, we replace the linear term for education years with a set of indicator variables for different types of college degrees that are typically associated with shorter or longer durations of study (Associate, Bachelor, Master, Professional, and PhD degrees). This basic Mincer model explains 14.2 percent of the overall variation in job-based earnings in our sample.[27] The second regression model in Table 2 relates log earnings to a much richer set of education and experience measures by adding 26 indicators for detailed fields of study, eight indicators for the placement of the degree-granting institution in the Times Higher Education U.S. college ranking, and a quadratic in workers' actual (in addition to potential) work experience. These measures of education and experience are unusually detailed. While we can observe them in the LinkedIn data, many of these variables would not be available in individual-level datasets that are more widely used for labor-market analyses, such as the Current Population Survey, the American Community Survey, or the Longitudinal Employer-Household Dynamics data. The augmented specification, which includes a total of 42 education and experience variables, explains 23.5 percent of the variation in job-based earnings. In the third column of Table 2, we explore whether the explanatory power can be increased further by using random forests to flexibly model the association of earnings with the large set of education and experience variables.[28] To generate comparable $R^2$ statistics for both OLS and random forests, all regression models in Table 2 are based on a randomly selected 70% of all observations for which we fit the OLS and random forest models, and we compute goodness-of-fit statistics out-of-sample by assessing the predictive power of model-predicted values in the 30% holdout sample.[29] However, we observe no additional gain in explanatory power when using random forests instead of OLS to model the relationship of earnings with education and experience variables.[30]

[26] For instance, the inclusion of occupation fixed effects or variables for experience by occupation would be mechanically correlated with the job-based and occupation-based earnings outcomes. Occupation fixed effects would even explain the entire variation in occupation-based earnings. The regressions also do not control for job tenure, which is the only person-specific variable used in the construction of the Revelio Labs earnings measure.

---

[27] To conserve space, Table 2 only indicates goodness-of-fit statistics for the different regression models, while Appendix Table A4 reports the full set of coefficient estimates for the OLS models.

[28] We use standard random forests as introduced by Breiman (2001) and estimated via the ranger package in R (Wright and Ziegler 2017) to model the association between earnings and varying sets of control variables in Table 2. Random forests are ensemble methods that combine many decision trees to improve predictive accuracy and reduce overfitting (Hastie, Tibshirani, and Friedman 2009). We use 500 trees to build each forest to trade off model performance and computational efficiency. Comparing results using different numbers of trees, we found that $R^2$ estimates stabilize from around 100 trees upward.

[29] While it is also possible to compute out-of-bag pseudo-$R^2$ values using random forests, these are not directly comparable to the adjusted $R^2$ statistics from OLS. Pseudo-$R^2$ reflects predictive performance based on internal cross-validation within trees, whereas OLS adjusted $R^2$ penalizes for model complexity. In our application, the reported out-of-sample $R^2$ statistics differ only minimally from out-of-bag pseudo-$R^2$. The $R^2$ values for the OLS models also barely differ between the split sample approach used in Table 2 and standard regressions with the full sample of observations.

[30] The random forest model has a better goodness-of-fit than OLS in the 70% training sample, but a marginally weaker fit in the out-of-sample evaluation for which the $R^2$ values are reported in Table 2.

**Table 2: Explaining earnings variation with alternative metrics of human capital**

| | Education/experience | | | Skills | | | Combined | |
|---|---|---|---|---|---|---|---|---|
| | Basic | Detailed | | Basic | Detailed | | Detailed | |
| | OLS | OLS | Random forest | OLS | OLS | Random forest | OLS | Random forest |
| | (1) | (2) | (3) | (4) | (5) | (6) | (7) | (8) |
| Potential experience (squared) | ✓ | ✓ | ✓ | | | | ✓ | ✓ |
| Actual experience (squared) | | ✓ | ✓ | | | | ✓ | ✓ |
| Highest degree | ✓ | ✓ | ✓ | | | | ✓ | ✓ |
| Field of study | | ✓ | ✓ | | | | ✓ | ✓ |
| College ranking | | ✓ | ✓ | | | | ✓ | ✓ |
| Number of skills | | | | ✓ | ✓ | ✓ | ✓ | ✓ |
| Shares of specific and managerial skills | | | | ✓ | (✓) | (✓) | (✓) | (✓) |
| Shares of skills by skill clusters | | | | | ✓ | ✓ | ✓ | ✓ |
| Job-based earnings (Revelio Labs) | | | | | | | | |
| Adj. $R^2$ | 0.142 | 0.235 | 0.234 | 0.172 | 0.253 | 0.297 | 0.325 | 0.363 |
| Occupation-based earnings (ACS) | | | | | | | | |
| Adj. $R^2$ | 0.097 | 0.192 | 0.196 | 0.113 | 0.230 | 0.267 | 0.276 | 0.319 |

Notes: N = 8,850,314 LinkedIn profiles randomly split into a 70% training and 30% holdout sample. The table reports the $R^2$ goodness-of-fit statistic (adjusted for number of regressors) for regression models whose dependent variable is either log annual earnings imputed by Revelio Labs' proprietary salary model ('job-based earnings') or the average log annual earnings at the SOC four-digit occupation level based on 2018-2019 American Community Survey (ACS) data ('occupation-based earnings'). To ascertain comparability in goodness-of-fit statistics across OLS (columns 1-2, 4-5, 7) and random forest methodologies (columns 3, 6, 8), all models are estimated on a 70% training sample, and goodness-of-fit is determined using model-predicted values in the 30% holdout sample. Regressors are defined as follows: Potential experience (squared) is the time between college graduation and scrape month (in years) and its square. Actual experience (squared) is the cumulated time of job spells in LinkedIn profile (in years) and its square. Highest degree controls for four indicator variables of educational degree levels. Field of study controls for 26 indicator variables based on the two-digit CIP taxonomy. College ranking controls for eight indicator variables based on the Times Higher Education ranking. Number of skills controls for the count of skills reported on the LinkedIn profile. The column 4 model controls for the share of specific skills and the share of managerial skills, while the models in the subsequent columns control in more detail for the share of skills in 47 skill clusters (with the share of skills in the 48th cluster being the omitted category). Regressions include fixed effects for scrape months. See Appendix Table A4 for a tabulation of coefficient estimates from OLS models for job-based earnings estimated on the full sample of LinkedIn profiles.

The regression models in the fourth to sixth columns of Table 2 replace the education and experience variables that measure inputs into human capital production with the skill variables observed on LinkedIn that measure an output of this production. Column 4 uses a parsimonious specification that relates log earnings to the number of reported skills and the shares of specific and managerial skills among total skills. This basic model with just three skill variables accounts for more variation in job-based earnings than the basic Mincer model (17.2 vs. 14.2 percent). The more detailed skill specification in column 5 considers the share of skills that a worker reports across the 48 skill clusters (corresponding to the OLS regression underlying Figure 6), while column 6 more flexibly models the association between earnings and detailed skill variables using random forests. The corresponding estimates indicate that the models with skill variables outperform those with detailed education and experience variables in terms of explanatory power. With random forest specifications, the fraction of job-based earnings variation explained by skills is one-fourth larger than the fraction explained by education and experience (29.7 vs. 23.4 percent).

The final two columns of Table 2 relate earnings to the full set of education, experience, and skill variables using either OLS or random forest estimation. Relative to the random forest model with detailed education and experience variables in columns 4, the addition of skill variables in column 8 raises the explanatory power by more than one-half from 23.4 percent to 36.3 percent. This sizable improvement suggests that the skill variables provide substantial additional information about workers' human capital that cannot already be inferred from detailed metrics of education and experience.

By contrast, the gain from adding education and experience variables to a skills-only model is considerably smaller (6.6 percentage points for the random forest models of columns 6 and 8, or 7.2 percentage points for the OLS models of columns 5 and 7). This observation implies that the impact of education and experience on earnings as measured in columns 2 and 3 is to a substantial extent mediated through skills. A closer inspection of OLS coefficient estimates, which are tabulated in Appendix Table A4, indicates that notably the field-of-study variables lose explanatory power in models that also account for skills. In the column 2 model that considers only education and experience, the three study fields with largest earnings effects (computer science, engineering, and business studies) are on average associated with 36.4 log points higher earnings than the three lowest-paying fields (education, family and consumer sciences, and public administration). This difference shrinks by two-thirds to just 12.6 log points in the model that additionally controls for worker skills.[31] Conversely, the coefficient difference between the three skill clusters with highest and lowest earnings associations (1.46 log points in the skills-only model of column 5) declines by less than a quarter in the model that also accounts for education and experience (1.13 log points in column 7).[32]

The lower panel of Table 2 repeats the same set of earnings regressions for the alternative earnings outcome measure that is based on average occupational earnings in the Census Bureau's 2018-2019 ACS data, instead of Revelio Labs' earnings model of job-based earnings. The pattern of the goodness-of-fit statistics across the column 1 to 8 models remains qualitatively similar to the results with the job-based earnings metric in the upper panel of the table. The main quantitative difference is that human capital variables explain a lower fraction of variation in occupation-based earnings than in job-based earnings. This finding implies that the education, experience, and skill variables considered help to explain earnings variation within occupations, for instance across jobs with or without supervisory functions, that is captured by the earnings model of Revelio Labs but not by the simple measure of average occupational earnings.

Our overall conclusion from the analyses in sections 5 and 6 is that skills observed on LinkedIn profiles are related in plausible ways to human capital investments such as education and experience, but contain additional information on workers' human capital that cannot be directly inferred from the education and experience variables alone. As such, worker skills provide a useful complement to education and experience when measuring human capital, and may provide a superior metric if only one type of variables were available.

## 7. The Gender Skill Gap

We conclude our analysis by studying skill patterns by gender. Extant evidence indicates that women and men differ only modestly in terms of traditional human capital proxies such as years of education and experience, while there is an important gender gap in occupational choice that explains a substantial part of the gender earnings gap (Blau and Kahn 2017, Olivetti, Pan, and Petrongolo 2024). Our results of the previous sections suggest that the LinkedIn skill data can capture additional human capital differences that are not captured by education and experience variables, and that these differences in skills help to explain the employment of workers in higher- versus lower-paid jobs and occupations. We bring these insights to bear on the analysis of gender-specific patterns in the labor market by first documenting and discussing gender differences in reported skills (section 7.1) and then analyzing to what extent these skill differences contribute to a gender gap in job- and occupation-based earnings (section 7.2).

[31] In the model without skill variables, the average earnings coefficient of the three high-paying fields is 28.0 log points whereas the average coefficient for the three low-paying fields is -8.3 log points (relative to the reference category of agricultural sciences). In the combined model that includes skill variables, the average earnings coefficients are 8.2 log points for the former and -4.4 log points for the latter fields.

[32] As indicated in Figure 6, the three skill clusters with highest earnings coefficients are 'mobile devices/applications', 'analysis, financial analysis', and 'information technology, lean six sigma'. The average coefficient estimate for these skills is 1.59 in the skills-only model and 1.24 in the combined model. The three clusters with lowest earnings coefficients are 'customer service/retention', 'communication, problem solving', and 'Microsoft Office, customer service', with an average coefficient of 0.12 in the skills-only model and 0.11 in the combined model.

### 7.1 Skills by Gender

We showed in Figure 1 that the cross-sectional relationship between age and the average number of reported skills has a remarkable similarity to the well-documented concave shape of age-earnings profiles. Panel A of Figure 7 repeats this analysis separately for women and men. It documents a striking new pattern of gender differences in the labor market: Women report fewer skills than men, especially at higher ages. The slopes of the gender-specific age-skill profiles are quite similar for both genders in the mid-twenties when these profiles slope steeply upwards, and from the mid-forties, when the profiles flatten and ultimately decline slightly. There is a remarkable difference, however, between the late twenties and early forties where the age-skill profile increases much less for women than for men. While the gender difference in the average number of reported skills is 0.5 in the age range 23-26, it rises steadily to 3.1 at age 43-44 and remains in the range 3.1 to 3.3 until age 64. These gender-specific age-skill profiles are similar to the shapes of gender-specific age-earnings profiles (see Appendix Figure A5).

**Figure 7: Number of reported skills and skill composition by age and gender**

*Panel A: Number of reported skills by age and gender*

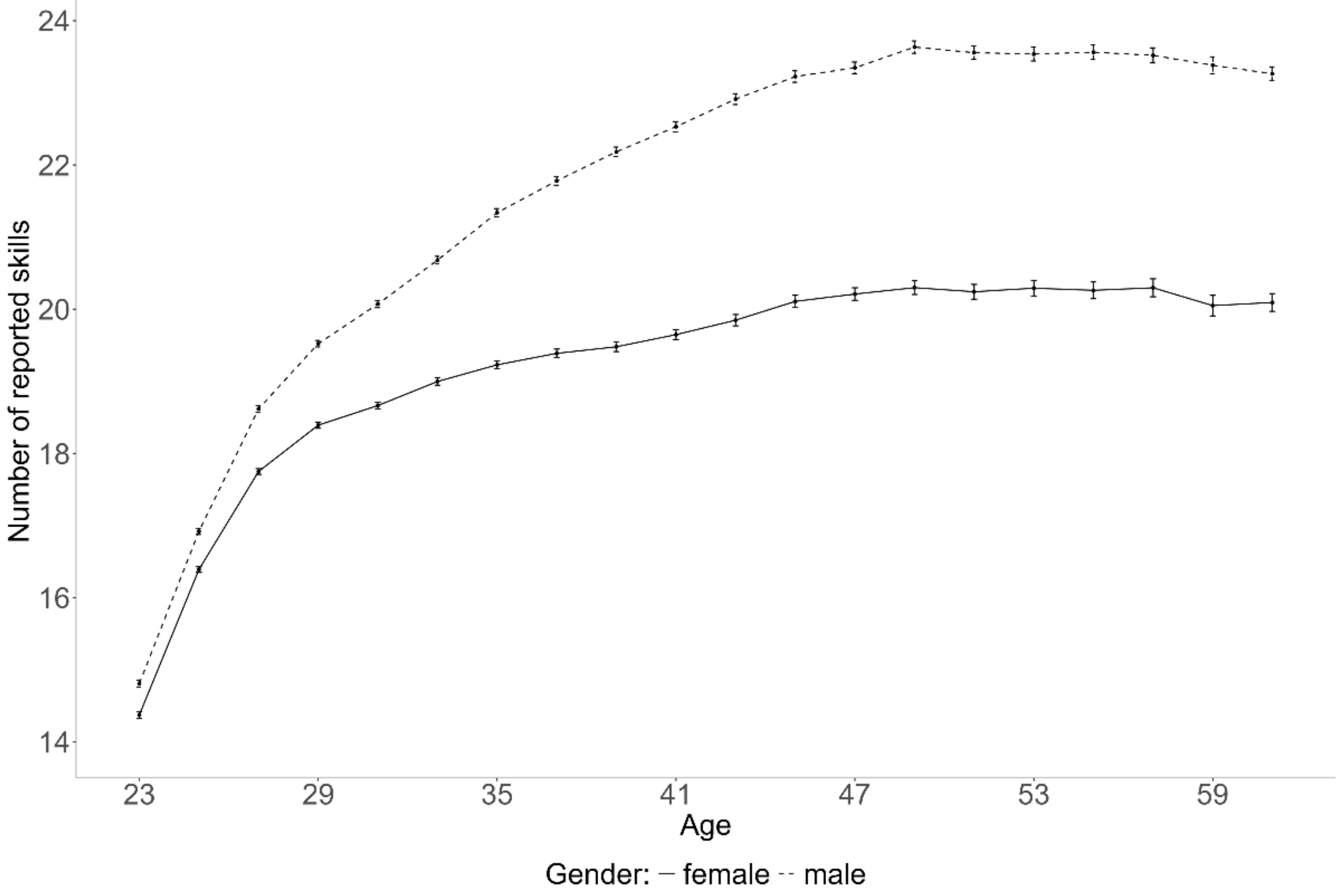

*Panel B: Share of general non-managerial, general managerial, and specific skills by age and gender*

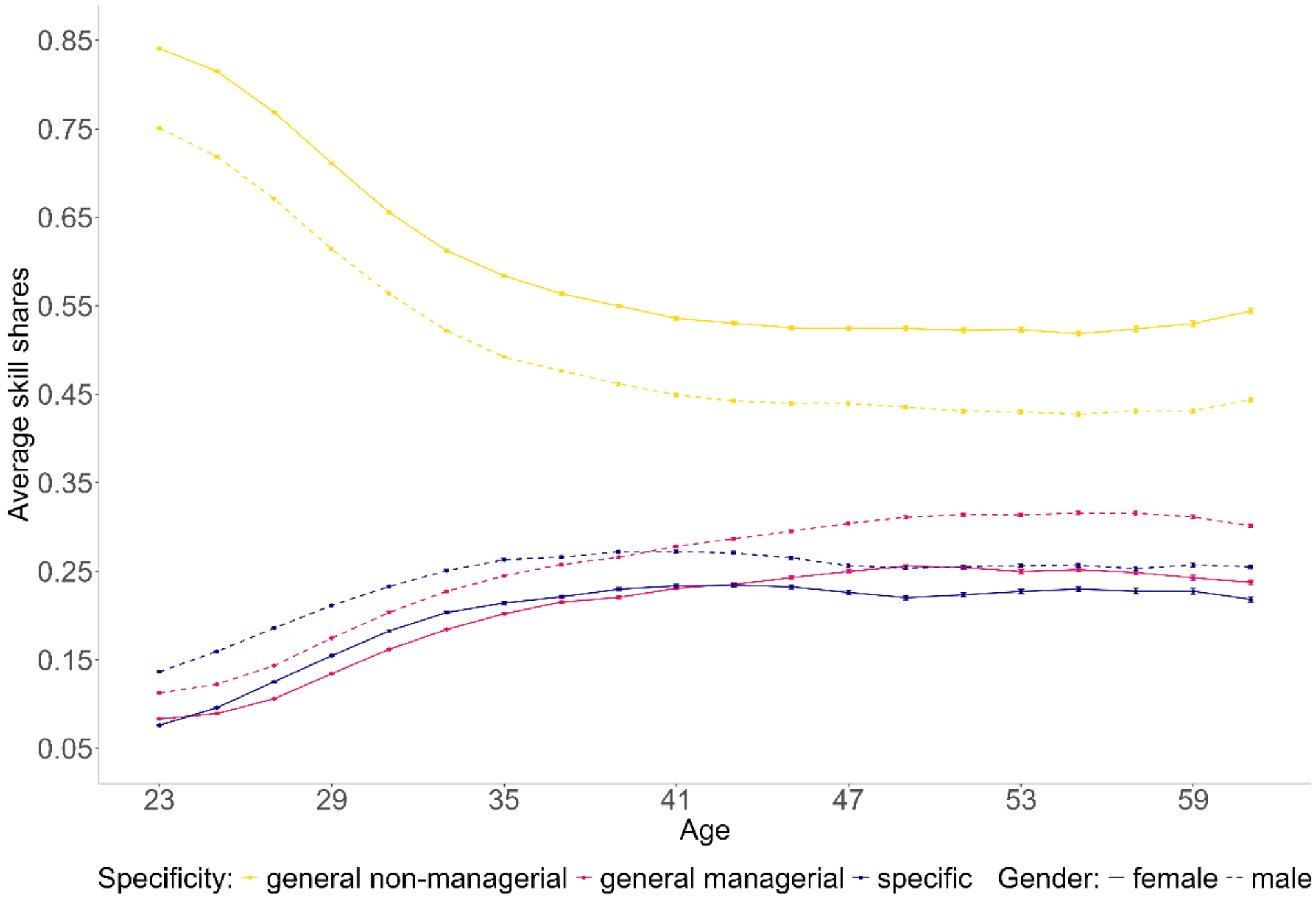

Notes: The figures show the average number of reported skills and the average fractions of general non-managerial, general managerial, and occupation-specific skills, respectively, by age and gender. Two-year age bins; last bin combines ages 61-64. Error bars depict 99 percent confidence intervals.

Panel B of Figure 7 indicates the shares of general, managerial, and specific skills reported by women and men of different ages. Women consistently report a higher fraction of general skills and lower shares of managerial and specific skills than men of the same age. This pattern suggests that women enter the labor market with a considerably different skill mix than men, which may in part be explained by gendered patterns of field-of-study choice while in college. As shown in Figure 4, individuals who majored in the relatively female-dominated fields of social sciences, arts, and humanities report higher shares of general skills than those who majored in STEM, business, health, or law. The difference in the share of specific skills narrows slightly with age, whereas the difference in the share of managerial skills slightly widens.

We next discuss four potential explanations for the observed gender skill gap: male overconfidence, lack of profile updating among women, occupation-specific skill accumulation, and lower skill growth due to reduced labor supply induced by motherhood.

***Male Overreporting***. Experimental evidence indicates that men are overconfident in their abilities (Niederle and Vesterlund 2007) and more likely to self-promote (Exley and Kessler 2022). There is ongoing debate on where and how this phenomenon is quantitatively important, as gender differences in overconfidence may be sensitive to the subject domain (Bordalo et al. 2019) and close to zero on average (Bandiera et al 2022). Since there is no objective benchmark for the self-reported multidimensional skills that we analyze, we cannot directly quantify the contribution of overconfidence and self-promotion, if any, to the gender gaps in self-reported skills of Figure 7. However, two results speak against a predominant role of male overreporting. First, if gender differences in the number of reported skills were driven primarily by male overreporting, it would have to be the case that such overconfidence is much more prevalent among older individuals but nearly absent among younger ones. We are not aware of any literature that establishes such a strong age or cohort gradient in gender-specific overconfidence.

Second, we test whether skills reported by men are as strongly linked to higher-paying jobs as those reported by women. If men were widely overreporting, employers—who gather more accurate skill information during recruitment and on the job—would discount men's self-reported skills. This would result in smaller earnings gains per reported skill for men compared to women. However, gender-specific earnings regressions (using the specification of column 4 in Appendix Table A4) show similar associations between reported skills and pay, with a slightly higher coefficient for men (0.402) than for women (0.374). Thus, employers do not seem to discount the skills reported by men.

***Gender Differences in Profile Updating***. A different type of reporting bias emerges when women update their LinkedIn profiles less frequently than men. In this case, the average man's profile may indicate the current skill set while the average woman's profile shows the skills that she possessed a few years ago, thus failing to account for the skill accumulation that occurred since. If one assumes that both men and women are especially likely to update their profiles when they are looking for a new job, then the profiles of both genders should be equally up-to-date at young ages where all college graduates had to look for a job, while males' profiles may be more current at later ages due to greater subsequent male job mobility (Loprest 1992, Keith and McWilliams 1999, Manning and Swaffield 2008) or stronger male career aspirations (Azmat, Cunat, and Henry 2020).

In Panel A of Figure 8, we analyze a subsample of workers who report an employer change in the last two years, and for whom we thus know that their profiles have been updated relatively recently. The recent firm switchers have steeper age-skill profiles, which could be consistent both with a more thorough updating of the skill information on their LinkedIn profiles, or a positive selection effect where workers who rapidly acquire new skills are more likely to seek better jobs at new employers as job mobility is an important source of earnings growth (Topel and Ward 1992, Guvenen et al. 2021). The focus on the subsample of recent firm switchers however does not narrow the gender skill gap which remains similar in timing and magnitude as in the full sample of LinkedIn profiles. We therefore conjecture that gender-specific patterns of profile updating are not the main driver of the gender skill gap seen in Panel A of Figure 7.[33]

[33] Appendix B provides additional statistics and results for the subsample of recent firm switchers.

**Figure 8: Alternative age-skill profiles by gender**

*Panel A: Full sample versus subsample of recent firm switchers*

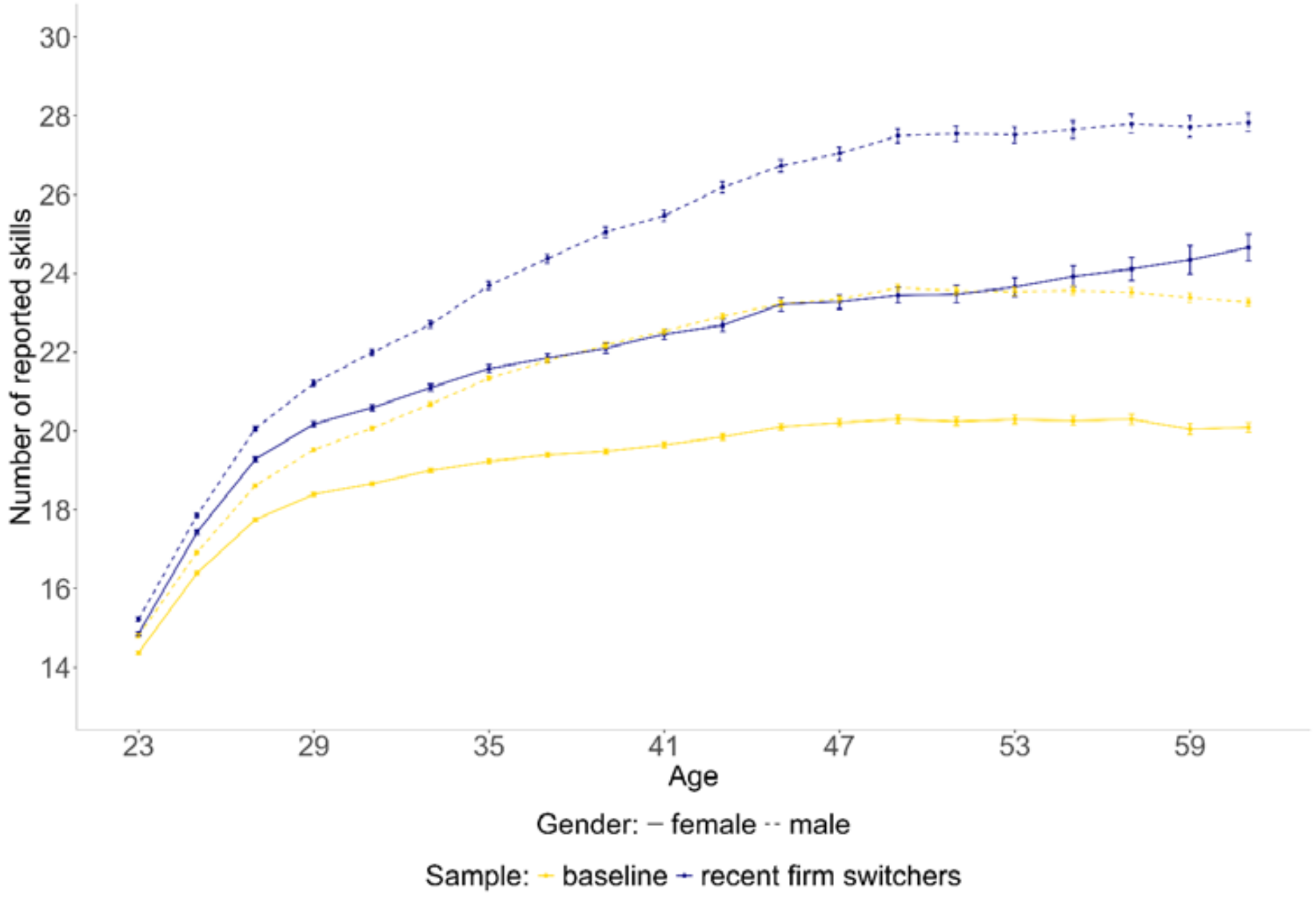

*Panel B: Correcting for gender differences in occupational composition*

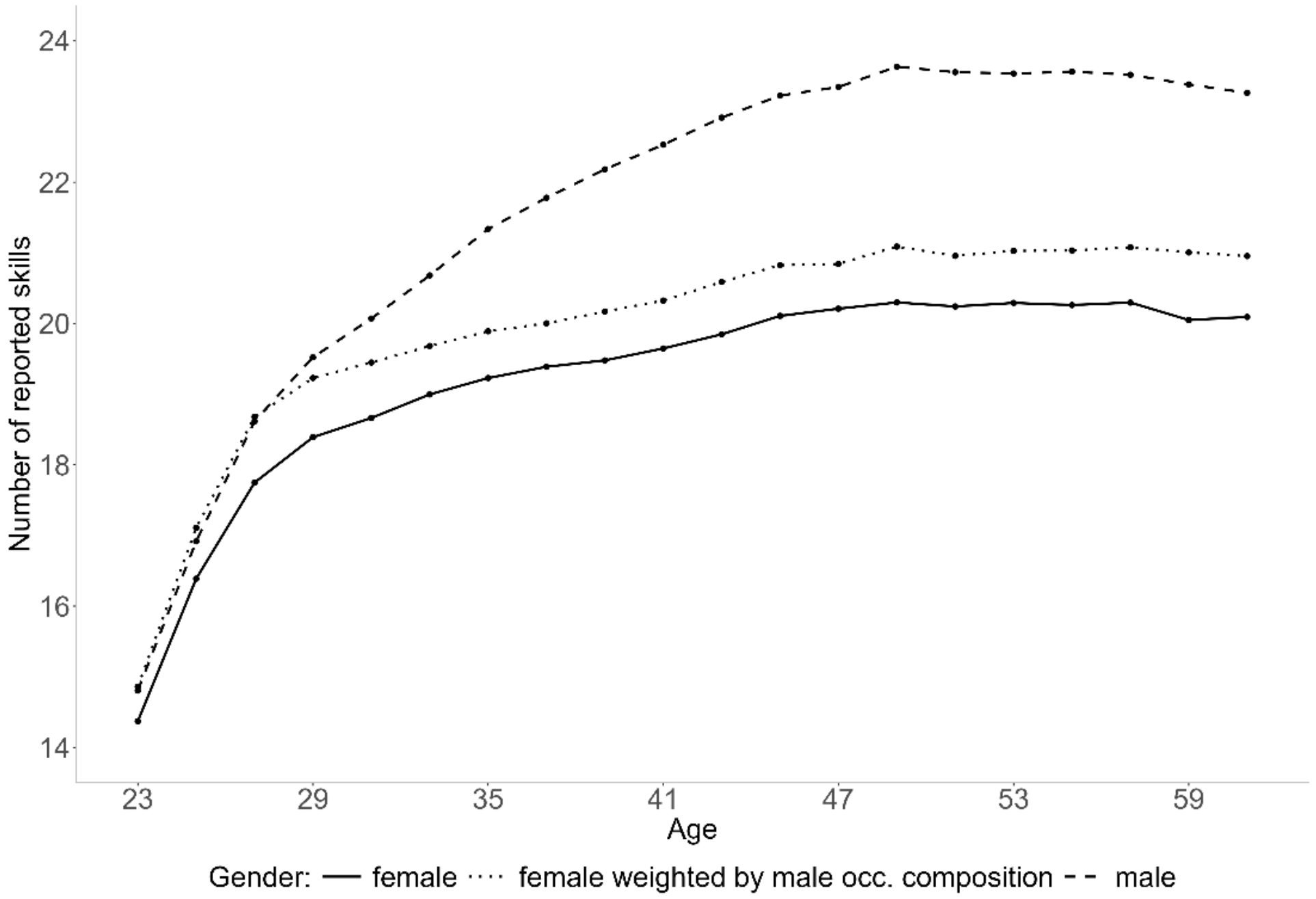

Notes: The figures show the average number of reported skills by age and gender. In Panel A, the data series 'recent firm switchers' retains only the LinkedIn profiles of individuals who report a change of employer within the last two years. In Panel B, the data series 'female weighted by male occ. composition' is based on the average number of skills for women of a given age group and detailed occupation, weighted by the employment share of each occupation among the men of the same age group. The difference between this data series and the male series indicates within-occupation differences in skills across gender, while the difference with regard to the female series indicates between-occupation differences in skills across gender. Two-year age bins; last bin combines ages 61-64.

***Between-Occupation Differences in Skill Accumulation***. Occupations may vary in the extent that they provide workers with a potential for on-the-job skill accumulation. The flatter slope of women's age-skill profile may thus result from a concentration of women in occupations that offer less learning opportunities and thus lower returns to experience over the life cycle (Adda, Dustmann, and Stevens 2017).[34]

To assess this hypothesis, Panel B of Figure 8 indicates average skills by age group for women and men, as in Panel A of Figure 7, but additionally shows a data series that weights women's average reported number of skills by occupation and age with the occupational employment shares of men in the corresponding age group. If the gender skill gap resulted entirely from women and men's differential distribution across occupations, the counterfactual data series ('female weighted by male occupational composition') should match the data series for males. The results show that up to age 28, the small gender gap in reported skills is indeed entirely a between-occupation gap, while women and men in the same occupation report the same average number of skills. However, after age 28, a sizable within-occupation gender gap in skills begins to accumulate until about age 40, and then stays constant. Consequently, gender differences in occupational choice can account for only a small fraction of the overall gender gap in reported skills at older ages.

Appendix Figure A6 shows that a large part (though not all) of the gender gaps in shares of general, managerial, and specific skills at young ages are also between occupations, while within-occupation differences become more important at higher ages. We thus conclude that whereas female and male workers in the same occupation report very similar skill sets at young ages, women subsequently appear to experience less within-occupation growth in overall, managerial, and specific skills.

***Gender Differences in Labor Supply and Effects of Parenthood***. An important gender difference in career trajectories is that women tend to have lower work hours and more career interruptions than men, especially after becoming parents (Bertrand, Goldin, and Katz 2010, Cortes and Pan 2023). Following childbirth, mothers' earnings decline substantially relative to fathers' earnings, and this 'motherhood penalty' accounts for a substantial part of the overall gender earnings gap (Kleven, Landais, and Sogaard 2019). Goldin (2014) shows that shorter weekly work hours are associated not only with reduced total earnings proportional to the lower labor supply, but also with lower hourly wage rates in a range of highly paid occupations that reward long hours. We hypothesize that in addition to these static effects of lower labor supply on earnings, there may be a dynamic effect where shorter work hours of women slow their on-the-job accumulation of skills and thus reduce their career progression to more highly paid jobs.[35]

[34] Adda, Dustmann, and Stevens (2017) show that women are more likely to choose occupations with lower returns to experience at labor market entry in anticipation of later motherhood. There is also evidence that women move to more family-friendly occupations later in their career after childbirth (Kleven, Landais, and Sogaard 2019), which intensifies occupational sorting by gender and perhaps concentrates women in occupations whose workers tend to report lower numbers of skills.

The LinkedIn data unfortunately provide no information on motherhood and do not indicate individuals' work hours in current and past jobs.[36] To assess the relationships between skill growth by gender, work hours, and motherhood, we therefore combine information from the LinkedIn profiles with American Community Survey (ACS) data at the level of cells delineated by age, education, and geographic region. Each cell $c$ is defined by a two-year age range, two levels of educational degrees (undergraduate/postgraduate), and nine geographic census divisions. For each such cell, we use the ACS to compute the ratio of average annual work hours of women versus men in the cell.[37] We also use the same source to compute the fraction of women in the cell who live in a household that comprises a child under the age of 18, which provides a proxy for motherhood.[38] Finally, we draw on the LinkedIn data to approximate the two-year skill growth for either women or men in the cell by subtracting the average number of skills reported by that group from the average skills of individuals of the same gender, education, and geographic location who are two years older.

Panel A of Figure 9 shows the cell-level relationship between skill growth among the females in a cell $c$, $ds_c^f$, and skill growth among the men in the cell, $ds_c^m$. While there is a strong positive correlation between these variables, female skill growth lags male skill growth in most cells, and the slope of the regression line indicates that women on average acquire only 0.82 skills per skill obtained by men. Panel B of Figure 9 explores the relationship between this gender gap in skill accumulation and gender differences in labor supply. It fits separate regression lines for worker cells in which the ratio of average female versus average male work hours is relatively high (female/male hours ratio exceeding the median value of 0.82), and cells in which women provide considerably fewer worker hours than men (else). In the cells with relatively high female labor supply, women acquire nearly 0.9 skills per every skill gained by men, whereas in cells with low female labor supply, women obtain only 0.4 skills per added skill of men.

[35] Cook et al. (2021) observe that female ride-share drivers accumulate fewer hours of work experience per calendar month and therefore exhibit slower growth in productivity than their male counterparts.

[36] The LinkedIn data only allow to derive the number of months worked since college graduation as a relatively coarse measure of labor supply. Panel A of Appendix Figure A7 indicates that for a given level of potential experience, women have accumulated less actual experience than men on average. Panel B in the figure additionally reports that women are considerably more likely than men to have a low labor-market attachment in the sense that past career breaks lead to a low ratio of actual to potential work experience.

[37] We compute average female and male work hours for individuals in the cell inclusive of individuals who provide no work hours in a given year, since both reduced hours and spells of non-employment contribute to lower labor-market experience.

[38] The ACS reports the relationship of every household member to the household head. If a woman is not the household head, it is not always clear whether she is the mother of a child that is present in the household, though motherhood is likely in most cases. We do not simultaneously consider the share of men with children in the household since parenthood is associated with little labor supply change among men (Kleven, Landais, and Sogaard 2019).

We explore more formally how the relationship between female skill growth $ds_c^f$ and male skill growth $ds_c^m$ depends on the ratio of female vs. male annual work hours in the cell ($h_c^f/h_c^m$) by estimating the cell-level regression

$$ds_c^f = \alpha + \beta ds_c^m + \gamma \frac{h_c^f}{h_c^m} ds_c^m + \delta \frac{h_c^f}{h_c^m} + \varepsilon_c \qquad (1)$$

In this regression, a coefficient estimate of $\hat{\gamma} = 1$ would imply that in a cell where women provide 10 percent fewer work hours than men, women's skill growth is 10 percent smaller than that of men. Appendix Table A5 provides estimates. It indicates a coefficient estimate $\hat{\gamma} = 2.28$ (s.e. 0.40) that is positive and significantly larger than unity. This suggests that lower work hours of women relative to men may be associated with a disproportional penalty in terms of skill accumulation. Such a penalty would, for instance, result if firms are more likely to provide training or job tasks that offer learning opportunities to employees who work longer hours, in addition to paying higher hourly wage rates (Goldin 2014).[39]

[39] There is some evidence for a lower incidence of formal on-the-job training for women relative to men (Barron, Black, and Loewenstein 1993, Knoke and Ishio 1998) and for mothers in particular (Pan, Seward, and Dhuey 2021).

**Figure 9: Skill accumulation by gender, relative female/male work hours, and motherhood**

*Panel A: Female vs. male skill growth*

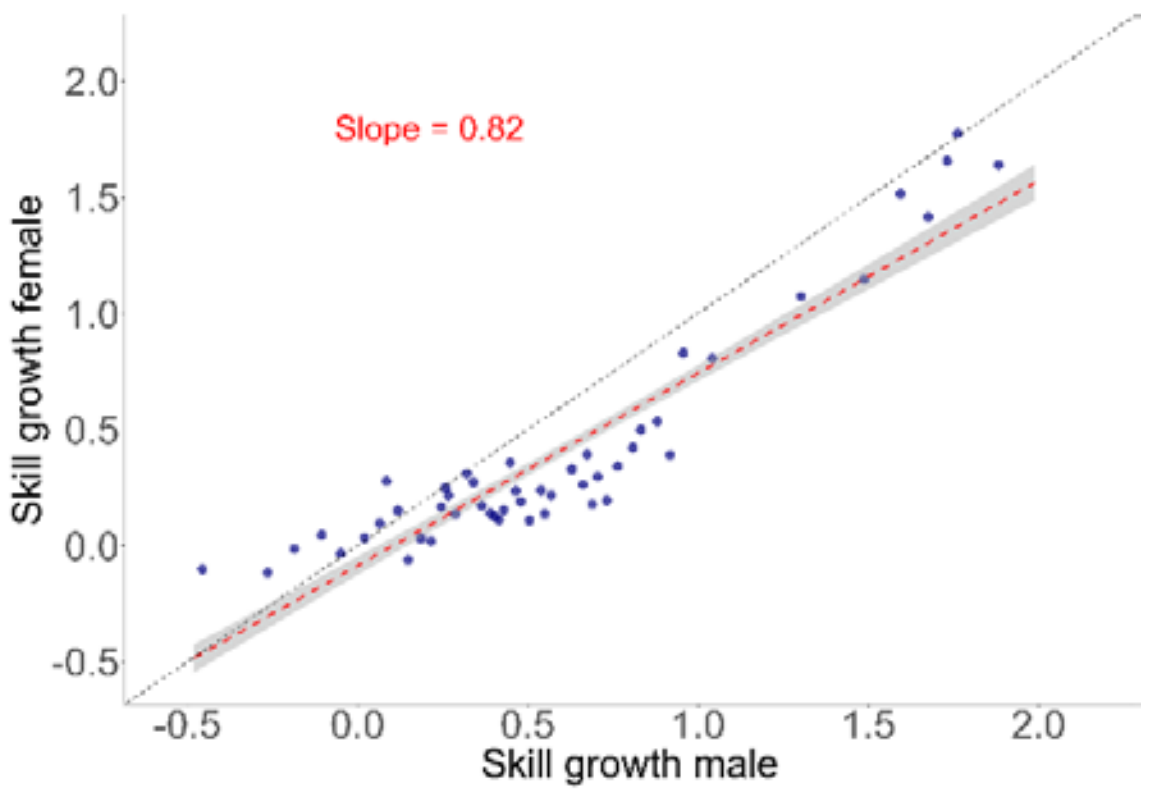

*Panel B: Skill growth by F/M hours ratio*

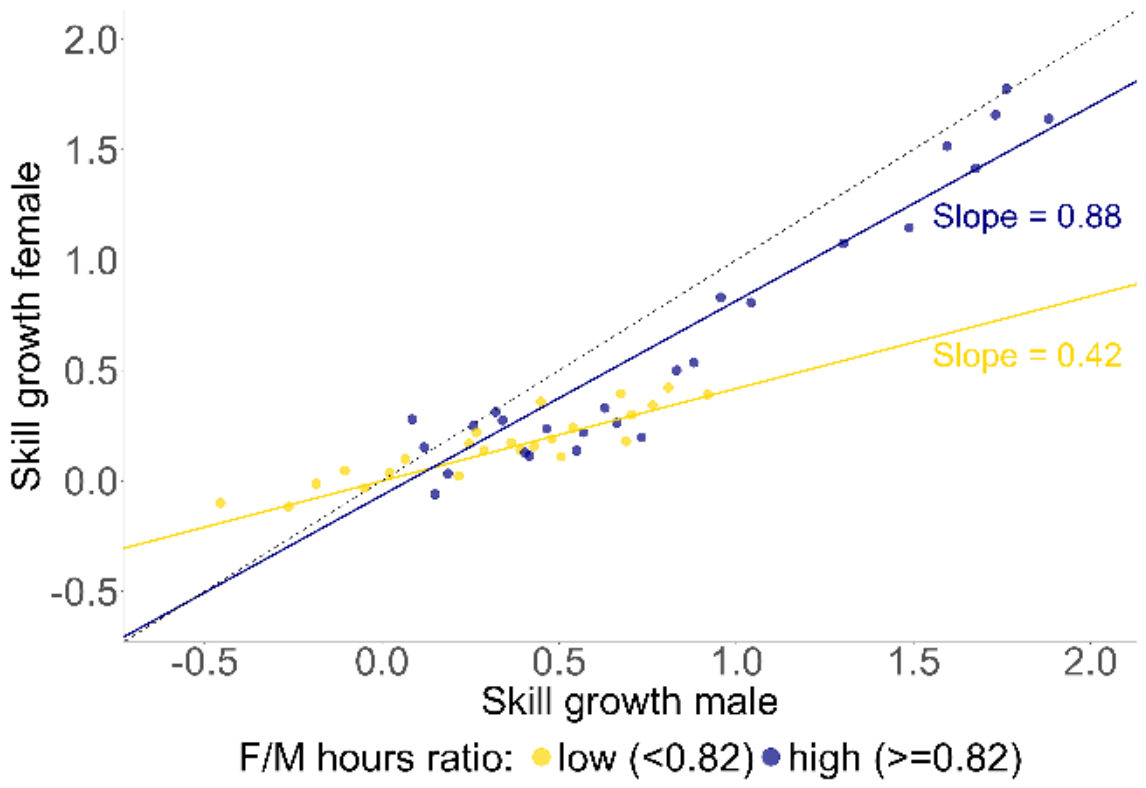

*Panel C: F/M hours ratio vs. motherhood*

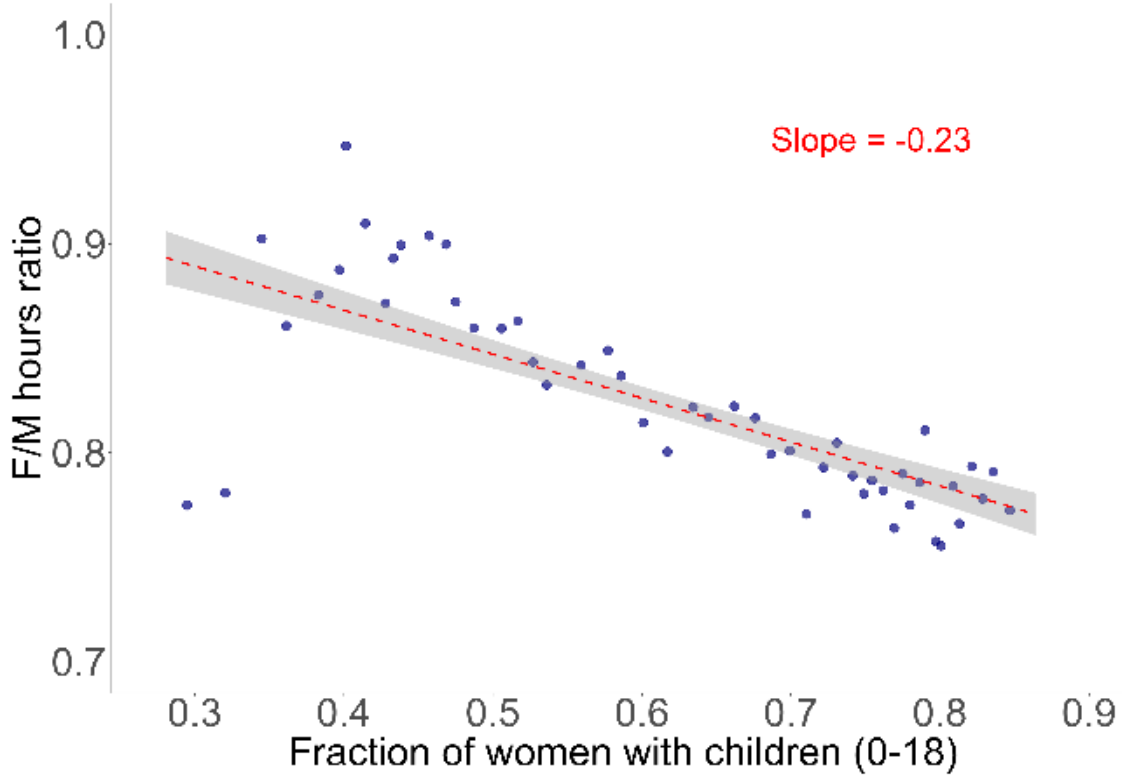

*Panel D: Counterfactual female skill growth*

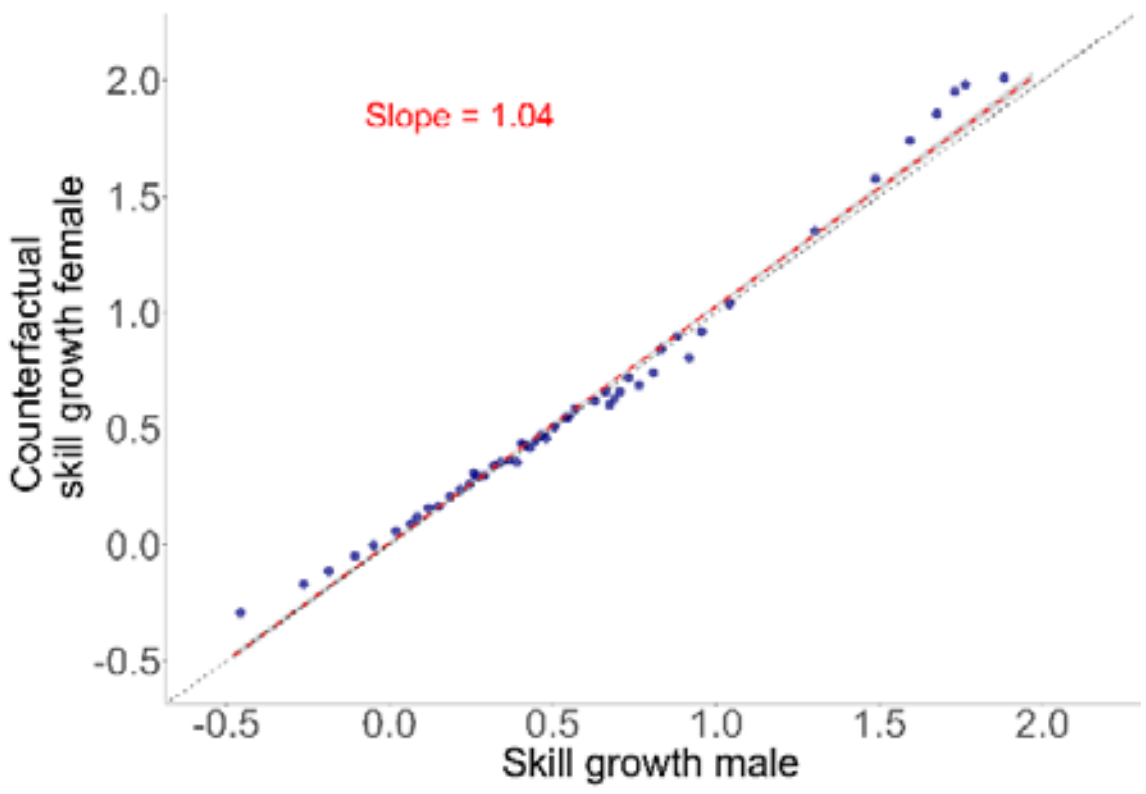

*Panel E: Counterfactual female age-skill profile*

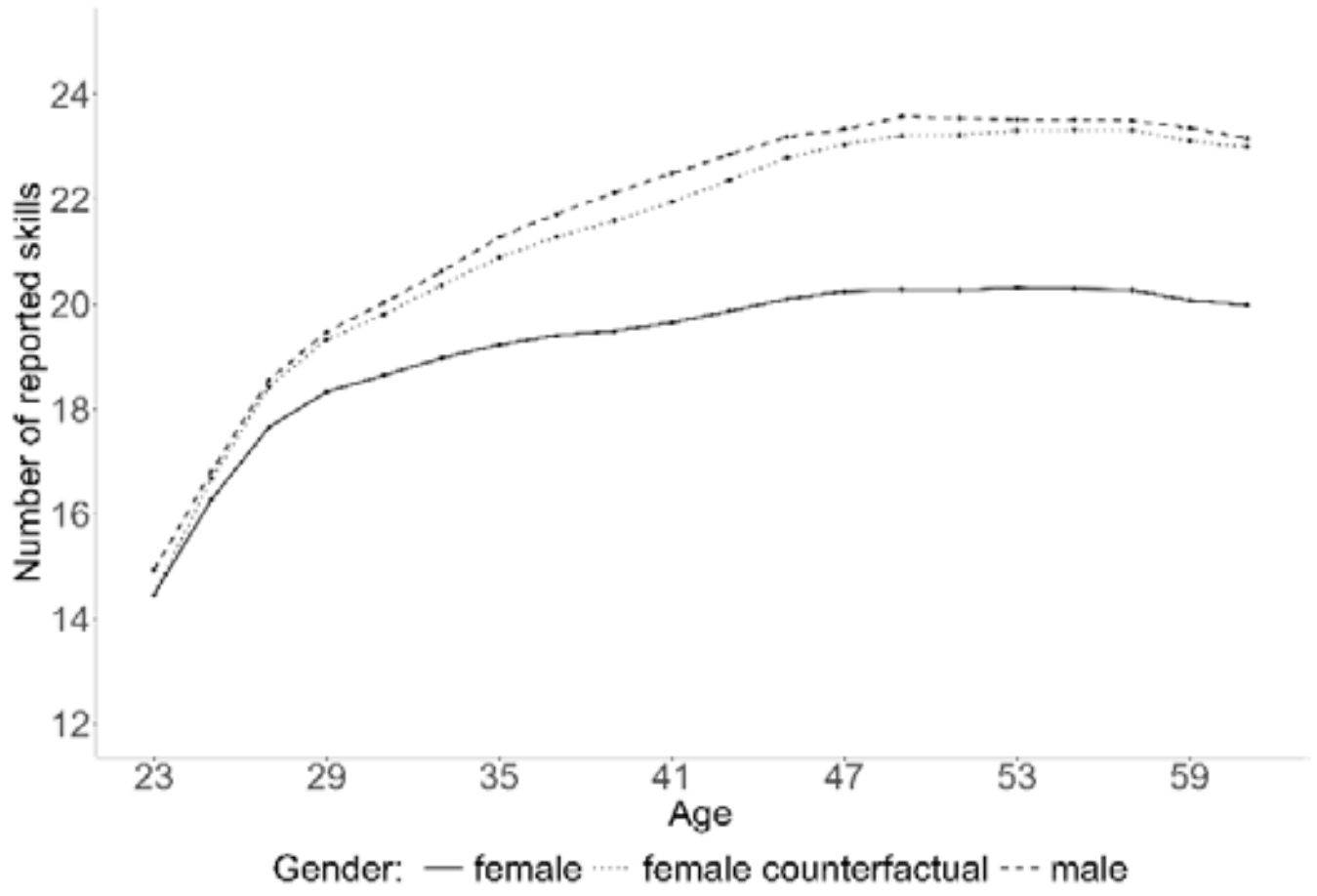

Notes: Panels A-D are binned scatterplots of 342 cells defined by two-year age bins, two educational degrees, and nine census divisions. Panel A shows the relationship of the growth in the number of LinkedIn skills between women and men. Panel B fits separate regression lines for cells with above and below median female/male (F/M) annual work hours ratio from the American Community Survey (ACS). Panel C shows the relationship between the F/M hours ratio and the fraction of women with children aged 0-18 in ACS data. Panel D indicates the relationship between male skill growth and a counterfactual female skill growth based on an adjusted F/M hours ratio that corrects for the impact of motherhood on work hours suggested by Panel C. Panel E shows actual age-skill profiles for males and females, and a counterfactual profile for females consistent with the counterfactual skill growth of Panel D.

We next relate the female-to-male work hours ratio to the fraction of women in the cell who live in a household with an underage child. Panel C of Figure 9 shows a highly significant negative association between the proxy for motherhood and the relative labor supply of women. For every 10 percentage-point increase in the share of women with children in the cell, the female-to-male ratio of work hours is 2.3 percentage points lower. If one extrapolates this cell-level relationship to an individual person who becomes mother (i.e., for whom the child variable changes from 0 to 100 percent), then having a child is associated with a 23 percent decline in work hours relative to men. While this relationship indicates only a correlation, its magnitude is consistent with event study evidence on parenthood. Kleven, Landais, and Leite-Mariante (2024) estimate that parenthood reduces women's labor supply by 25 percent relative to men in North America.

We finally assess how much of the gender skill gap would disappear absent the motherhood-related reduction in female work hours. To construct a counterfactual, motherhood-adjusted skill growth for women, we proceed as follows: First, we compute for every cell a motherhood-adjusted counterfactual female-male hours ratio by adding to the actual hours ratio another 0.23 percentage points for every percentage point of the share of women with children in the cell. Next, we use the estimates for equation (1) from Appendix Table A5 to compute the female skill growth for the cell that is consistent with the motherhood-adjusted hours ratio. Panel D of Figure 9 shows that absent the motherhood-related reduction in female work hours, female skill accumulation would look strikingly like the skill accumulation of males in the same cell. Finally, we construct a counterfactual age-skill profile for women as the sum of average skills that women reported in the youngest age cell (age 23-24), plus the cumulative counterfactual cell-level female skill growth for every subsequent higher age. Panel E of Figure 9 shows that this counterfactual, which is based on higher female work hours absent motherhood, yields an age-skill profile for women that closely traces men's skill profile at every age. The gender skill gap of Panel A of Figure 7, which is small at career start but starts to widen when workers reach their late twenties and thus typical ages of first motherhood for college-educated women in the United States (Nitsche and Brückner 2021), may thus largely be a result of motherhood, which reduces work hours and skill accumulation among women.

## 7.2 Skills and the Gender Earnings Gap

There is a long tradition in labor economics to investigate the contribution of gender differences in human capital to earnings differences between women and men. As women's educational attainment and labor force attachment increased, gender differences in human capital measured by education and experience variables declined to an extent that they can no longer explain a meaningful portion of the remaining gender earnings gap (Goldin 2014, Blau and Kahn 2017, Olivetti, Pan, and Petrongolo 2024).[40] Given our new finding of a substantial gender skill gap, we next investigate whether our skill measures of human capital account for a larger portion of the earnings differential between women and men.

Since our job- and occupation-based earnings variables provide a metric of the types of jobs that people hold, rather than their actual individual earnings, it is useful to first consider to what extent these job- and occupation-based earnings measures capture a relevant part of the overall gender earnings gap. In the sample of all employed college graduates in the ACS 2019, which we use as comparison set to the LinkedIn data, the raw gender gap in average log annual earnings of full-time, full-year workers is 30.3 log points.[41] The primarily job-based earnings measure of Revelio Labs, which uses information about job title, firm, location, and tenure but *not* gender, shows a 21.7 log point raw gender earnings gap in the LinkedIn sample. This suggests that slightly less than three quarters of the overall gender earnings gap (21.7/30.3 = 71.7%) are due to gender differences in these job-based earnings. Our alternative earnings measure, which assigns to each worker the average ACS-based earnings of college graduates in the worker's occupation, yields a smaller but still sizable raw gender earnings gap of 13.7 log points. These statistics confirm prior research which found that gender differences in occupation (Sloane, Hurst, and Black 2021) and additional job-related characteristics such as industry (Blau and Kahn 2017, Olivetti, Pan, and Petrongolo 2024) or firm wage premia (Goldin et al. 2017) account for a sizable part of the overall gender earnings gap. However, the literature often found it difficult to interpret whether the differential sorting of women and men across jobs reflects job-based discrimination of equally qualified workers, or underlying differences in skills that are not captured by education and experience variables (Blau and Kahn 2017).

---

[40] In the words of Goldin (2014, p. 1093), "For a long time the gender gap in wages has been viewed as summarizing human capital differences between men's and women's productivity as well as differential treatment of men and women in the labor market. As the grand gender convergence has proceeded, underlying differences between the human capital capabilities of women and men have been vastly reduced and in many cases eliminated."

[41] This statistic is based on the wage and salary income of all workers in the ACS 2019 who report having worked at least 51.5 weeks last year and having at least 35 usual weekly work hours.

Our subsequent regression analysis estimates to what extent different sets of human capital variables can statistically explain the gender differentials in job- and occupation-based earnings, which in turn account for a large portion of the overall gender earnings gap. The upper panel of Table 3 regresses workers' log job-based earnings on a female indicator and the different vectors of education, experience, and skill variables that we explored in Table 2 above.

The first column of Table 3, which includes only the female indicator, shows the indicated raw gender gap of 21.7 log points in full-time equivalent job-based earnings for our sample of college graduates. The basic Mincer-style regression in column 2 that controls for five levels of educational degrees and a quartic in potential experience decreases the coefficient estimate of the female indicator only slightly, which confirms the conclusion of prior research that basic human capital proxies explain only a small part (1-(-0.196/-0.217) = 9.7%) of the gender earnings gap. The augmented OLS regression model in column 3, which adds a large vector of education and experience variables including detailed indicators for field of study, can account for over a third (35.5%) of the raw earnings gap, and that proportion rises further (43.8%) when using a more flexible random forest specification in column 4.[42]

These proportions, however, fall well short of the explanatory power of the skill variables. The basic skill model in column 5 which controls for just three skill variables (number of reported skills and shares of specific and managerial skills) already accounts for nearly the same fraction (33.6%) of the gender gap in job-based earnings as the highly detailed regression in column 3 that comprises 42 education and experience variables. Models that consider the distribution of skills across detailed skill clusters explain nearly half (48.8%) of the raw gender gap in the OLS estimation of column 6, and even three-fifths (59.9%) of that gap in the random forest estimation of column 7.

[42] We use the random forest method introduced in Wager and Athey (2018) and extended in Athey, Tibshirani, and Wager (2019), using 750 trees, to estimate the covariate-adjusted difference in earnings between females and males in Table 3. This method estimates treatment effects by orthogonalizing the treatment and outcome models, providing point estimates of the treatment parameter while flexibly accounting for nonlinear effects and high-order interactions of the covariates.

**Table 3: Skills and the gender earnings gap**

| | Gender wage gap | Education/experience | | | Skills | | | Combined | |
|---|---|---|---|---|---|---|---|---|---|
| | | Basic | Detailed | | Basic | Detailed | | Detailed | |
| | OLS | OLS | OLS | Random forest | OLS | OLS | Random forest | OLS | Random forest |
| | (1) | (2) | (3) | (4) | (5) | (6) | (7) | (8) | (9) |
| Job-based earnings (Revelio Labs) | | | | | | | | | |
| Female | -0.217 | -0.196 | -0.140 | -0.122 | -0.144 | -0.111 | -0.087 | -0.094 | -0.072 |
| | *(0.0003)* | *(0.0003)* | *(0.0003)* | *(0.0004)* | *(0.0003)* | *(0.0003)* | *(0.0004)* | *(0.0003)* | *(0.0003)* |
| Reduction vs. raw gap (in percent) | | 9.7 | 35.5 | 43.8 | 33.6 | 48.8 | 59.9 | 56.7 | 66.8 |
| Occupation-based earnings (ACS) | | | | | | | | | |
| Female | -0.137 | -0.129 | -0.082 | -0.072 | -0.098 | -0.061 | -0.046 | -0.048 | -0.034 |
| | *(0.0002)* | *(0.0002)* | *(0.0002)* | *(0.0003)* | *(0.0002)* | *(0.0002)* | *(0.0003)* | *(0.0002)* | *(0.0002)* |
| Reduction vs. raw gap (in percent) | | 5.8 | 40.1 | 47.4 | 28.5 | 55.5 | 66.4 | 65.0 | 75.2 |
| Potential experience (squared) | | ✓ | ✓ | ✓ | | | | ✓ | ✓ |
| Actual experience (squared) | | | ✓ | ✓ | | | | ✓ | ✓ |
| Highest degree | | ✓ | ✓ | ✓ | | | | ✓ | ✓ |
| Field of study | | | ✓ | ✓ | | | | ✓ | ✓ |
| College ranking | | | ✓ | ✓ | | | | ✓ | ✓ |
| Number of skills | | | | | ✓ | ✓ | ✓ | ✓ | ✓ |
| Shares of specific, managerial skills | | | | | ✓ | (✓) | (✓) | (✓) | (✓) |
| Shares of skills by skill clusters | | | | | | ✓ | ✓ | ✓ | ✓ |

Notes: Columns 1-3, 5, 6, and 8: OLS regressions. Columns 4, 7, and 9: random forests. N = 8,850,314. The dependent variable in the upper panel is log annual earnings imputed by Revelio Labs' proprietary salary model ('job-based earnings') and in the lower panel is log annual earnings at the SOC four-digit occupation level based on 2018-2019 ACS ('occupation-based earnings'). All regressions include an indicator for females whose coefficient estimate is reported in the table. The 'reduction vs. raw gap' measures the percentage point reduction in the female coefficient relative to the raw gender gap reported in column 1. Additional regressors are defined as in Table 2. Robust standard errors in parentheses.

Finally, the combination of the full set of skill, education, and experience variables explains more than half (56.7%) of the raw gender earnings gap in the OLS model of column 8, and fully two thirds (66.4%) of this differential in the more flexible random forest model of column 9.[43] These combined models can thus account for only a modestly larger proportion of the raw gender earnings gap than the skills-only specifications in column 6 and 7.

The lower panel of Table 3 repeats the analysis of the gender earnings gap for the alternative earnings metric that is based on average occupational earnings. The results are qualitatively similar to those obtained with the more detailed earnings measure of Revelio Labs. The regression model with detailed skill variables explains over half (55.5%) of the occupation-based gender earnings differential in the OLS estimation of column 6, and two thirds (66.4%) of that gap in the random forest model of column 7. The combination with detailed education and experience variable raises these proportions further to nearly two thirds (65.0%) in the OLS specification and three quarters (75.2%) in the random forest model.

Overall, these results suggest that human capital differences in terms of skills account for a sizable portion of the gender earnings gap among college graduates. Gender differences in human capital thus appear to be a more important contributor to the gender earnings gap than previously suggested by studies that measured human capital with basic education and experience variables. It is however important to note that our findings do not need to imply that gender discrimination in the labor market is less severe than previously thought. Our analysis does not rule out that the observed gender skill gap could arise from discriminatory barriers that limit women's access to training and skill development. The correlational nature of our earnings results also does not preclude the possibility that some men have better access to high-paying jobs than equally qualified women, which may in turn enable them to accumulate more valuable skills through on-the-job experience. However, the fact that recruiters frequently rely on LinkedIn skill listings when searching for candidates suggests that possessing relevant skills is often a prerequisite for accessing such jobs in the first place. What our gender earnings analysis does show is that, while women on average hold considerably lower-paid jobs than men, the extent of their ‘underplacement’ is substantially reduced when one conditions on their self-reported skills.

---

[43] Since gender differences in job-based earnings account for about 72 percent of the overall gender earnings gap among college graduates, the two thirds of the gender gap in job-based earnings explained by the human capital variables in column 9 of Table 3 correspond to about half of the overall gender earnings gap (66.8%*71.7%=47.9%). This simple quantification may provide only a lower bound for the contribution of human capital variables to the gender earnings gap as it disregards any additional contribution of gender differences in detailed skill, education, and experience variables to earnings differences between women and men within job types.

## 8. Conclusions

Modern theories of the labor market purport that workers require a multitude of skills to perform a broad range of job tasks, and evidence from job advertisements confirms that employers search for a wide variety of worker skills. Empirical research, however, mostly reverts to proxying for individual human capital by educational attainment and work experience variables that capture inputs into human capital production rather than the skills that result from that production.

Rich data from online professional profiles on the LinkedIn platform provide us with a new opportunity to study the role of human capital in the labor market. Users of the platform self-report CV information including skills to signal their qualifications to potential employers. We gather this information for 8.85 million college graduates in the U.S. labor market for whom we also observe detailed information on education and experience, and can infer job-based earnings from information such as job titles and occupations. We aggregate the thousands of reported skill strings into 48 skill clusters using a word association algorithm and use the ensuing database to perform three sets of analyses.

First, we show that self-reported skills correlate sensibly with experience and education variables that capture inputs into skill production. Older and more experienced workers report a larger number of skills. Their skill sets also contain a larger share of skills that are occupation-specific rather than widespread across many occupations, as well as a larger share of managerial skills. These patterns are consistent with the notion that young workers primarily obtain general skills through education while older workers additionally acquire specific and managerial skills through work experience and post-university training. Skill patterns also vary across workers with different types of education, where studies in STEM, health, and law fields and more advanced degrees provide a higher fraction of occupation-specific skills.

Second, there is a clear tendency for workers with more reported skills to be employed in higher-paying jobs. Many specific and managerial skills are particularly strongly associated with higher job-based earnings. While one may be concerned that self-reported skills provide only a noisy measure of human capital, we show that skills explain a much higher fraction of earnings variation than basic education and experience variables, and even have higher explanatory power than an unusually detailed vector of education and experience variables that accounts for field of study, college quality, and actual post-college work experience. Indeed, a sizable part of the well-known relationship between education, experience, and earnings appears to be mediated by the skills observed on LinkedIn profiles.

Third, we document a substantial gender gap in self-reported skills between female and male college graduates, particularly at older ages. At labor market entry, women report slightly fewer skills than men and are less likely to list specific or managerial skills. However, much of this initial gap in skill quantity and composition vanishes when comparing men and women within the same occupations. Among workers in their late twenties to early forties, a notable divergence emerges: the number of reported skills grows significantly more for men than for women, even within the same occupational categories. We show that this pattern aligns with the effects of reduced labor supply following motherhood, which limits women's opportunities for on-the-job skill accumulation. Importantly, these gender differences in skill profiles explain a substantial portion of the gender earnings gap—far more than traditional variables such as educational degrees or years of experience, which now show minimal gender disparities.

Overall, our analysis underscores that conventional measures of human capital based on education and work experience may not sufficiently reflect the breadth and complexity of workers' skills. Our findings suggest that LinkedIn skill data offer a valuable complement to traditional metrics, enabling a richer understanding of human capital differences across diverse groups in the labor market.

# Appendix A: Additional Tables and Figures

## Table A1: Most frequent skill strings in each skill cluster

| Skill cluster | Most frequent skill strings |
|---|---|
| **Specific skills** | |
| Accounting, financial reporting | Accounting, financial reporting, auditing, financial accounting, accounts payable |
| AutoCAD, SolidWorks | AutoCad, SolidWorks, CAD, mechanical engineering, finite element analysis |
| Biotechnology, molecular biology | Biotechnology, molecular biology, chemistry, cell culture, PCR |
| Clinical research, medical devices | Clinical research, medical devices, pharmaceutical industry, clinical trials, oncology |
| Engineering, project engineering | Engineering, project engineering, petroleum, gas, electrical engineering |
| Healthcare, hospitals | Healthcare, hospitals, healthcare management, nursing, EMR |
| Insurance, banking | Insurance, banking, credit, loans, property |
| Java enterprise edition, Jira | Java Enterprise Edition, Jira, JSP, test automation, Hibernate |
| Java, Matlab | Java, Matlab, C++, Python, C |
| Legal research/writing | Legal research, legal writing, litigation, civil litigation, courts |
| Mobile devices/applications | Mobile devices, mobile applications, Objective-C, iOS, iOS development |
| Real estate, investment properties | Real estate, investment properties, real estate transactions, residential homes, commercial real estate |
| Revit, SketchUp | Revit, SketchUp, product design, architecture, interior design |
| SQL, software development | SQL, software development, Linux, agile methodologies, Microsoft SQL Server |
| Telecom., network security | Telecommunications, network security, wireless, Cisco technologies, VoIP |
| Windows server, disaster recovery | Windows server, disaster recovery, servers, data center, VMware |
| **General managerial skills** | |
| Analysis, financial analysis | Analysis, financial analysis, finance, risk management, financial modeling |
| Business analysis/process improv. | Business analysis, business process improvement, integration, change management, SDLC |
| Coaching, leadership development | Coaching, leadership development, curriculum development, staff development, curriculum design |
| Marketing, social media marketing | Marketing, social media marketing, marketing strategy, advertising, marketing communications |
| Program mgmt., security clearance | Program management, security clearance, military, DoD, proposal writing |
| Project management, budgets | Project management, budgets, project planning, policy, government |
| Sales, strategic planning | Sales, strategic planning, team leadership, account management, strategy |

(continued on next page)

**Table A1 (continued)**

| Skill cluster | Most frequent skill strings |
|---|---|
| **General non-managerial skills** | |
| Access, software documentation | Access, software documentation, Visio, SAP, ERP |
| Communication, problem solving | Communication, problem solving, organization, organization skills, organizational leadership |
| Customer satisfaction/retention | Customer satisfaction, customer retention, automotive, call centers, customer experience |
| Data analysis, databases | Data analysis, databases, statistics, SPSS, qualitative research |
| Editing, public relations | Editing, public relations, writing, blogging, creative writing |
| Energy, sustainability | Energy, sustainability, environmental awareness, environmental science, renewable energy |
| English, Spanish | English, Spanish, education, history, French |
| Excel, customer relations | Excel, customer relations, Word, planning, IIS |
| Food, hospitality | Food, hospitality, beverage, restaurants, hospitality management |
| HTML, JavaScript | HTML, JavaScript, CSS, web design, WordPress |
| Info. technology, lean six sigma | Information technology, lean six sigma, people management, people development, IT solutions |
| Inventory/operations management | Inventory management, operations management, logistics, supply chain management, purchasing |
| Microsoft Office, customer service | Microsoft Office, customer service, leadership, Microsoft Excel, management |
| Photoshop, Adobe CS | Photoshop, Adobe Creative Suite, graphic design, photography, InDesign |
| Process improv., cross-func. team lead. | Process improvement, cross-functional team leadership, manufacturing, continuous improvement, Six Sigma |
| Public speaking, research | Public speaking, research, teaching, community outreach, nonprofits |
| Recruiting, human resources | Recruiting, human resources, employee relations, interviews, employee benefits |
| Retail, forecasting | Retail, forecasting, merchandising, visual merchandising, pricing |
| Security, emergency management | Security, emergency management, first aid, criminal justice, homeland security |
| Social media/networking | Social media, social networking, Facebook, sports, wellness |
| Testing, quality assurance | Testing, quality assurance, SharePoint, technical writing, FDA |
| U.s, software dev. life cycle | U.s, software development life cycle, relationship building, waterfall methodologies, communications |
| Video production/editing | Video production, video editing, music, Final Cut Pro, video |
| Windows, troubleshooting | Windows, troubleshooting, networking, technical support, system administration |

Notes: The table shows the five most commonly reported raw skills for the 48 skill clusters.

## Table A2: Sample selection

| Filtering step | Observations | Share: Cumulative | Share: Of previous step | Share: Just this step | Age | Female | College degree |
|---|---|---|---|---|---|---|---|
| | (1) | (2) | (3) | (4) | (5) | (6) | (7) |
| All LinkedIn profiles, U.S. workers, 2019 | 61,850,913 | 1.00 | | 1.00 | 37.75 | 0.48 | 0.44 |
| **Non-missing baseline data** | | | | | | | |
| Non-missing gender | 57,835,032 | 0.94 | 0.94 | 0.94 | 37.86 | 0.48 | 0.43 |
| Non-missing occupation | 57,834,793 | 0.94 | 1.00 | 1.00 | 37.86 | 0.48 | 0.43 |
| Non-missing state | 55,418,826 | 0.90 | 0.96 | 0.96 | 37.88 | 0.48 | 0.44 |
| **Non-missing education data** | | | | | | | |
| Any education stage | 43,450,459 | 0.70 | 0.78 | 0.78 | 37.88 | 0.48 | 0.56 |
| Mapped highest degree | 33,673,272 | 0.54 | 0.78 | 0.60 | 37.88 | 0.49 | 0.73 |
| **Valid age and experience indicators** | | | | | | | |
| Highest degree with valid date | 28,343,736 | 0.46 | 0.84 | 0.51 | 37.88 | 0.48 | 0.86 |
| Graduation date before scrape date | 25,922,593 | 0.42 | 0.92 | 0.46 | 37.88 | 0.47 | 0.94 |
| Imputed age at first job at least 14 | 24,320,432 | 0.39 | 0.94 | 0.43 | 38.39 | 0.48 | 0.94 |
| First job at most 5 years after graduation | 20,658,665 | 0.33 | 0.85 | 0.40 | 35.92 | 0.48 | 0.95 |
| **Age range and college degree** | | | | | | | |
| Age 23-64 | 18,755,466 | 0.30 | 0.91 | 0.41 | 36.08 | 0.48 | 0.97 |
| At least Associate’s degree | 18,136,202 | 0.29 | 0.97 | 0.44 | 36.17 | 0.48 | 1.00 |
| **Non-missing skill data** | | | | | | | |
| At least one reported skill | 8,950,793 | 0.14 | 0.49 | 0.38 | 37.46 | 0.46 | 1.00 |
| No non-English skill clusters | 8,850,314 | 0.14 | 0.99 | 1.00 | 37.47 | 0.46 | 1.00 |

Notes: Profile characteristics of the U.S. workforce on LinkedIn in 2019 after different sample selection steps. Column 2 reports the share of observations that is retained at each selection step relative to the baseline sample, column 3 indicates the share of observations retained relative to the previous selection step, and column 4 provides the share of observations that would be retained if a given selection step were applied to the baseline sample. Columns 5 to 7 report the mean age, fraction of females, and fraction of college graduates at each step of the sample selection.

## Table A3: Skill count and composition by experience and education: Regressions

| | | Share of skill domain in total skills | | |
|---|---|---|---|---|
| Dependent variable: | Number of reported skills | General non-managerial | General managerial | Specific |
| | (1) | (2) | (3) | (4) |
| **Experience** | | | | |
| Potential experience | -2.135 | -0.176 | 0.088 | 0.088 |
| | *(0.037)* | *(0.0009)* | *(0.0007)* | *(0.0008)* |
| Potential experience$^2$ | 1.762 | 0.390 | -0.175 | -0.214 |
| | *(0.081)* | *(0.002)* | *(0.002)* | *(0.002)* |
| Actual experience | 7.755 | -0.107 | 0.078 | 0.029 |
| | *(0.040)* | *(0.001)* | *(0.0007)* | *(0.0009)* |
| Actual experience$^2$ | -12.146 | 0.118 | -0.095 | -0.023 |
| | *(0.097)* | *(0.002)* | *(0.002)* | *(0.002)* |
| **Highest degree (Ref: associate)** | | | | |
| Bachelor | 1.794 | -0.050 | 0.040 | 0.010 |
| | *(0.018)* | *(0.0004)* | *(0.0003)* | *(0.0004)* |
| Master | 2.847 | -0.055 | 0.045 | 0.010 |
| | *(0.018)* | *(0.0004)* | *(0.0003)* | *(0.0004)* |
| Professional degree | -0.653 | -0.187 | -0.044 | 0.231 |
| | *(0.027)* | *(0.0007)* | *(0.0005)* | *(0.0008)* |
| Doctorate | 1.992 | -0.070 | -0.054 | 0.124 |
| | *(0.027)* | *(0.0006)* | *(0.0004)* | *(0.0006)* |
| **Field of study (Ref: arts/humanities)** | | | | |
| Business | 0.150 | -0.226 | 0.133 | 0.093 |
| | *(0.015)* | *(0.0003)* | *(0.0003)* | *(0.0003)* |
| Health/medicine | -3.099 | -0.223 | -0.073 | 0.296 |
| | *(0.020)* | *(0.0005)* | *(0.0003)* | *(0.0005)* |
| Law | -0.218 | -0.234 | 0.011 | 0.223 |
| | *(0.032)* | *(0.0008)* | *(0.0005)* | *(0.0009)* |
| Missing | 0.486 | -0.173 | 0.039 | 0.134 |
| | *(0.021)* | *(0.0005)* | *(0.0004)* | *(0.0004)* |
| Other majors | -0.485 | -0.069 | 0.037 | 0.031 |
| | *(0.024)* | *(0.0005)* | *(0.0004)* | *(0.0004)* |
| Social sciences | -0.750 | -0.059 | 0.060 | -0.0004 |
| | *(0.016)* | *(0.0004)* | *(0.0003)* | *(0.0002)* |
| STEM | 0.349 | -0.240 | -0.009 | 0.249 |
| | *(0.016)* | *(0.0003)* | *(0.0003)* | *(0.0003)* |
| **College ranking (Ref: Top 10)** | | | | |
| 11-50 | 0.679 | 0.041 | -0.041 | 0.0001 |
| | *(0.036)* | *(0.0009)* | *(0.0008)* | *(0.0008)* |
| 51-100 | 0.482 | 0.048 | -0.051 | 0.002 |
| | *(0.035)* | *(0.0008)* | *(0.0007)* | *(0.0008)* |
| 101-200 | 0.857 | 0.069 | -0.060 | -0.008 |
| | *(0.035)* | *(0.0008)* | *(0.0007)* | *(0.0008)* |
| 201-500 | 0.686 | 0.082 | -0.068 | -0.015 |
| | *(0.033)* | *(0.0008)* | *(0.0007)* | *(0.0007)* |
| 501-1000 | 0.413 | 0.106 | -0.083 | -0.022 |
| | *(0.033)* | *(0.0008)* | *(0.0007)* | *(0.0007)* |
| No US rank | 1.299 | 0.088 | -0.070 | -0.019 |
| | *(0.034)* | *(0.0008)* | *(0.0007)* | *(0.0008)* |
| Non-US college | 1.888 | 0.007 | -0.049 | 0.043 |
| | *(0.039)* | *(0.0009)* | *(0.0008)* | *(0.0009)* |
| Missing college | 1.238 | 0.094 | -0.081 | -0.013 |
| | *(0.035)* | *(0.0008)* | *(0.0007)* | *(0.0008)* |
| Observations | 8,850,314 | 8,850,314 | 8,850,314 | 8,850,314 |
| $R^2$ | 0.052 | 0.228 | 0.180 | 0.221 |

Notes: OLS regressions. Dependent variable indicated in column header. Regressions include fixed effects for scrape months. Robust standard errors in parentheses.

## Table A4: Alternative metrics of human capital and job-based earnings

| | Education/experience | | Skills | | Combined |
|---|---|---|---|---|---|
| | Basic | Detailed | Basic | Detailed | Detailed |
| | (1) | (2) | (3) | (4) | (5) |
| **Experience** | | | | | |
| Potential experience | 2.939 | 1.164 | | | 0.288 |
| | *(0.006)* | *(0.014)* | | | *(0.013)* |
| Potential experience$^2$ / 100 | -4.743 | -3.590 | | | -1.370 |
| | *(0.015)* | *(0.030)* | | | *(0.029)* |
| Actual experience | | 2.110 | | | 1.129 |
| | | *(0.015)* | | | *(0.014)* |
| Actual experience$^2$ / 100 | | -1.665 | | | -0.477 |
| | | *(0.035)* | | | *(0.034)* |
| **Highest degree** | | | | | |
| Bachelor | 26.236 | 20.933 | | | 16.576 |
| | *(0.058)* | *(0.064)* | | | *(0.060)* |
| Master | 32.330 | 29.099 | | | 24.318 |
| | *(0.061)* | *(0.067)* | | | *(0.064)* |
| Professional degree | 56.310 | 54.889 | | | 49.973 |
| | *(0.086)* | *(0.105)* | | | *(0.110)* |
| Doctorate | 41.277 | 40.419 | | | 40.097 |
| | *(0.089)* | *(0.098)* | | | *(0.099)* |
| **Field of study** | | | | | |
| Architecture and related services | | 17.643 | | | 8.793 |
| | | *(0.173)* | | | *(0.178)* |
| Biological and biomedical sciences | | 5.157 | | | 3.164 |
| | | *(0.176)* | | | *(0.171)* |
| Business, management, marketing | | 22.549 | | | 10.157 |
| | | *(0.161)* | | | *(0.157)* |
| Communication, journalism, related programs | | 10.566 | | | 3.601 |
| | | *(0.169)* | | | *(0.166)* |
| Computer+information sciences, support services | | 33.734 | | | 11.085 |
| | | *(0.167)* | | | *(0.167)* |
| Education | | -12.305 | | | -5.425 |
| | | *(0.171)* | | | *(0.169)* |
| Engineering | | 27.933 | | | 14.530 |
| | | *(0.164)* | | | *(0.164)* |
| Engineering/engin.-rel. technologies/technicians | | 12.737 | | | 3.219 |
| | | *(0.198)* | | | *(0.191)* |
| Family and consumer sciences/human sciences | | -8.005 | | | -5.165 |
| | | *(0.290)* | | | *(0.280)* |
| Health professions and related programs | | 9.470 | | | 4.896 |
| | | *(0.168)* | | | *(0.167)* |
| History | | 3.985 | | | 2.053 |
| | | *(0.216)* | | | *(0.207)* |
| Homeland security, law enforcement, firefighting | | -1.473 | | | -2.208 |
| | | *(0.205)* | | | *(0.201)* |
| Legal professions and studies | | 13.395 | | | 4.703 |
| | | *(0.195)* | | | *(0.195)* |
| Liberal arts+sciences, general studies, humanities | | 2.064 | | | -0.091 |
| | | *(0.186)* | | | *(0.180)* |
| Mathematics, statistics, and physical sciences | | 15.819 | | | 7.230 |
| | | *(0.185)* | | | *(0.178)* |
| Missing field of study | | 17.359 | | | 8.158 |
| | | *(0.169)* | | | *(0.164)* |

(continued on next page)

**Table A4 (continued)**

| | Education/experience | | Skills | | Combined |
|---|---|---|---|---|---|
| | Basic | Detailed | Basic | Detailed | Detailed |
| | (1) | (2) | (3) | (4) | (5) |
| Multi/interdisciplinary studies | | 10.539 | | | 6.110 |
| | | *(0.207)* | | | *(0.199)* |
| Natural resources and conservation | | 5.199 | | | 1.683 |
| | | *(0.251)* | | | *(0.245)* |
| Other field of study | | 7.530 | | | 4.747 |
| | | *(0.191)* | | | *(0.184)* |
| Parks, recreation, leisure, fitness, kinesiology | | -3.461 | | | -2.906 |
| | | *(0.238)* | | | *(0.233)* |
| Philosophy and religious studies | | 7.112 | | | 3.409 |
| | | *(0.259)* | | | *(0.247)* |
| Psychology | | -2.165 | | | -1.500 |
| | | *(0.180)* | | | *(0.175)* |
| Public administration, social service professions | | -4.728 | | | -2.576 |
| | | *(0.197)* | | | *(0.192)* |
| Social sciences | | 14.347 | | | 7.128 |
| | | *(0.180)* | | | *(0.174)* |
| Theology and religious vocations | | -3.699 | | | -0.452 |
| | | *(0.269)* | | | *(0.261)* |
| Visual and performing arts | | 5.770 | | | 2.672 |
| | | *(0.179)* | | | *(0.177)* |
| **College ranking** | | | | | |
| 11-50 | | -11.539 | | | -8.395 |
| | | *(0.135)* | | | *(0.127)* |
| 51-100 | | -18.455 | | | -13.495 |
| | | *(0.132)* | | | *(0.124)* |
| 101-200 | | -21.144 | | | -15.541 |
| | | *(0.132)* | | | *(0.123)* |
| 201-500 | | -27.229 | | | -20.059 |
| | | *(0.127)* | | | *(0.119)* |
| 501-1000 | | -33.450 | | | -24.643 |
| | | *(0.129)* | | | *(0.121)* |
| No US rank | | -31.329 | | | -23.609 |
| | | *(0.129)* | | | *(0.121)* |
| Non-US college | | -17.793 | | | -16.524 |
| | | *(0.144)* | | | *(0.135)* |
| Missing college | | -33.937 | | | -25.798 |
| | | *(0.130)* | | | *(0.122)* |
| **Number of skills** | | | 0.588 | 0.408 | 0.319 |
| | | | *(0.001)* | *(0.001)* | *(0.001)* |
| **Shares of specific and managerial skills** | | | | | |
| Specific skills | | | 0.536 | | |
| | | | *(0.0005)* | | |
| Managerial skills | | | 0.614 | | |
| | | | *(0.0007)* | | |
| **Shares of skills by skill clusters** | | | | | |
| ***Specific skills*** | | | | | |
| Mobile devices/applications | | | | 1.635 | 1.349 |
| | | | | *(0.012)* | *(0.011)* |
| SQL, software development | | | | 1.251 | 0.960 |
| | | | | *(0.006)* | *(0.006)* |
| Windows server, disaster recovery | | | | 1.276 | 0.948 |
| | | | | *(0.007)* | *(0.006)* |

(continued on next page)

**Table A4 (continued)**

| | Education/experience | | Skills | | Combined |
|---|---|---|---|---|---|
| | Basic | Detailed | Basic | Detailed | Detailed |
| | (1) | (2) | (3) | (4) | (5) |
| Telecom., network security | | | | 1.088 | 0.800 |
| | | | | *(0.007)* | *(0.006)* |
| Java enterprise edition, Jira | | | | 0.945 | 0.717 |
| | | | | *(0.007)* | *(0.006)* |
| Java, Matlab | | | | 1.183 | 0.709 |
| | | | | *(0.006)* | *(0.006)* |
| Clinical research, medical devices | | | | 1.168 | 0.706 |
| | | | | *(0.006)* | *(0.006)* |
| Accounting, financial reporting | | | | 0.988 | 0.695 |
| | | | | *(0.006)* | *(0.006)* |
| Revit, SketchUp | | | | 0.984 | 0.690 |
| | | | | *(0.006)* | *(0.006)* |
| Healthcare, hospitals | | | | 0.919 | 0.663 |
| | | | | *(0.006)* | *(0.006)* |
| Legal research/writing | | | | 1.273 | 0.655 |
| | | | | *(0.006)* | *(0.006)* |
| Engineering, project engineering | | | | 0.906 | 0.648 |
| | | | | *(0.006)* | *(0.006)* |
| AutoCAD, SolidWorks | | | | 0.923 | 0.632 |
| | | | | *(0.006)* | *(0.006)* |
| Real estate, investment properties | | | | 0.852 | 0.581 |
| | | | | *(0.006)* | *(0.006)* |
| Insurance, banking | | | | 0.811 | 0.570 |
| | | | | *(0.006)* | *(0.006)* |
| Biotechnology, molecular biology | | | | 0.802 | 0.424 |
| | | | | *(0.006)* | *(0.006)* |
| ***General managerial skills*** | | | | | |
| Analysis, financial analysis | | | | 1.572 | 1.154 |
| | | | | *(0.006)* | *(0.006)* |
| Business analysis/process improv. | | | | 1.488 | 1.033 |
| | | | | *(0.006)* | *(0.006)* |
| Sales, strategic planning | | | | 1.298 | 0.966 |
| | | | | *(0.006)* | *(0.006)* |
| Marketing, social media marketing | | | | 1.215 | 0.931 |
| | | | | *(0.006)* | *(0.006)* |
| Project management, budgets | | | | 1.104 | 0.769 |
| | | | | *(0.006)* | *(0.006)* |
| Program mgmt., security clearance | | | | 0.870 | 0.563 |
| | | | | *(0.006)* | *(0.006)* |
| Coaching, leadership development | | | | 0.559 | 0.327 |
| | | | | *(0.006)* | *(0.006)* |
| ***General non-managerial skills*** | | | | | |
| Info. technology, lean six sigma | | | | 1.555 | 1.230 |
| | | | | *(0.016)* | *(0.014)* |
| HTML, JavaScript | | | | 1.003 | 0.860 |
| | | | | *(0.006)* | *(0.006)* |
| U.s, software dev. life cycle | | | | 1.016 | 0.851 |
| | | | | *(0.016)* | *(0.014)* |
| Process improv., cross-func. team lead. | | | | 1.124 | 0.730 |
| | | | | *(0.006)* | *(0.006)* |
| Testing, quality assurance | | | | 0.837 | 0.640 |
| | | | | *(0.007)* | *(0.006)* |

(continued on next page)

**Table A4 (continued)**

| | Education/experience | | Skills | | Combined |
|---|---|---|---|---|---|
| | Basic | Detailed | Basic | Detailed | Detailed |
| | (1) | (2) | (3) | (4) | (5) |
| Recruiting, human resources | | | | 0.858 | 0.584 |
| | | | | *(0.006)* | *(0.006)* |
| Energy, sustainability | | | | 0.915 | 0.562 |
| | | | | *(0.007)* | *(0.006)* |
| Access, software documentation | | | | 0.752 | 0.521 |
| | | | | *(0.007)* | *(0.007)* |
| Data analysis, databases | | | | 0.872 | 0.508 |
| | | | | *(0.006)* | *(0.006)* |
| Video production/editing | | | | 0.671 | 0.500 |
| | | | | *(0.006)* | *(0.006)* |
| Inventory/operations management | | | | 0.695 | 0.486 |
| | | | | *(0.007)* | *(0.006)* |
| Editing, public relations | | | | 0.711 | 0.461 |
| | | | | *(0.006)* | *(0.006)* |
| Social media/networking | | | | 0.559 | 0.412 |
| | | | | *(0.006)* | *(0.006)* |
| Other Skills | | | | 0.658 | 0.412 |
| | | | | *(0.006)* | *(0.006)* |
| Photoshop, Adobe CS | | | | 0.551 | 0.408 |
| | | | | *(0.006)* | *(0.006)* |
| Public speaking, research | | | | 0.685 | 0.407 |
| | | | | *(0.006)* | *(0.005)* |
| Retail, forecasting | | | | 0.589 | 0.397 |
| | | | | *(0.007)* | *(0.006)* |
| Food, hospitality | | | | 0.532 | 0.385 |
| | | | | *(0.007)* | *(0.006)* |
| Windows, troubleshooting | | | | 0.396 | 0.384 |
| | | | | *(0.007)* | *(0.006)* |
| Security, emergency management | | | | 0.492 | 0.369 |
| | | | | *(0.007)* | *(0.006)* |
| English, Spanish | | | | 0.582 | 0.357 |
| | | | | *(0.007)* | *(0.006)* |
| Excel, customer relations | | | | 0.646 | 0.335 |
| | | | | *(0.010)* | *(0.009)* |
| Microsoft Office, customer service | | | | 0.333 | 0.268 |
| | | | | *(0.006)* | *(0.006)* |
| Communication, problem solving | | | | 0.043 | 0.071 |
| | | | | *(0.007)* | *(0.006)* |
| Adj. $R^2$ | 0.142 | 0.235 | 0.172 | 0.253 | 0.325 |

Notes: N = 8,850,314 LinkedIn profiles. The dependent variable is the log of job-based annual earnings imputed by Revelio Labs' proprietary salary model. All regressions are estimated using OLS. The model in column 4 provides the coefficient estimates shown in Figure 6. The regressions in columns 1 to 5 of this table use the same regressors as the models in columns 1, 2, 4, 5, and 7 of Table 2, but are based on the full sample of LinkedIn profiles instead of a random 70% subsample. Potential experience is the time between college graduation and scrape month in years, and actual experience is the cumulative period of job spells since college graduation in years. Number of skills is the count of skill strings reported on a LinkedIn profile. Shares of specific and managerial skills and shares of skills by skill clusters refer to the share of skill strings falling in the respective skill domain and cluster, respectively. Omitted categories are Associate for highest degree, agricultural/animal/plant science for field of study, Top 10 for college ranking, general skills for skill domain shares, and customer satisfaction/retention for skill cluster shares. Regressions include fixed effects for scrape months. Robust standard errors in parentheses.

**Table A5: Female skill growth as a function of male skill growth and the female/male work hours ratio**

| | (1) | (2) |
|---|---|---|
| Male skill growth | 0.820 | -1.303 |
| | *(0.036)* | *(0.351)* |
| Female/male work hours ratio * Male skill growth | | 2.278 |
| | | *(0.403)* |
| Female/male work hours ratio | | 0.276 |
| | | *(0.360)* |
| Observations | 342 | 342 |
| $R^2$ | 0.750 | 0.821 |

Notes: Regression of female skill growth on male skill growth, the female/male annual work hours ratio, and their interaction. Cell-level regressions with cells defined by two-year age bins, two educational degrees, and nine census divisions. Female and male skill growth is approximated by subtracting the average number of LinkedIn skills reported by a gender-age-education-geography group from the average skills of individuals of the same gender, education, and geographic location who are two years older. The ratio of average annual work hours of women versus men is computed from the pooled 2018-2019 American Community Survey (ACS) data and based on both individuals with zero and positive work hours. Robust standard errors in parentheses.

**Figure A1: Job-based (Revelio Labs) and occupation-based (ACS) earnings measures**

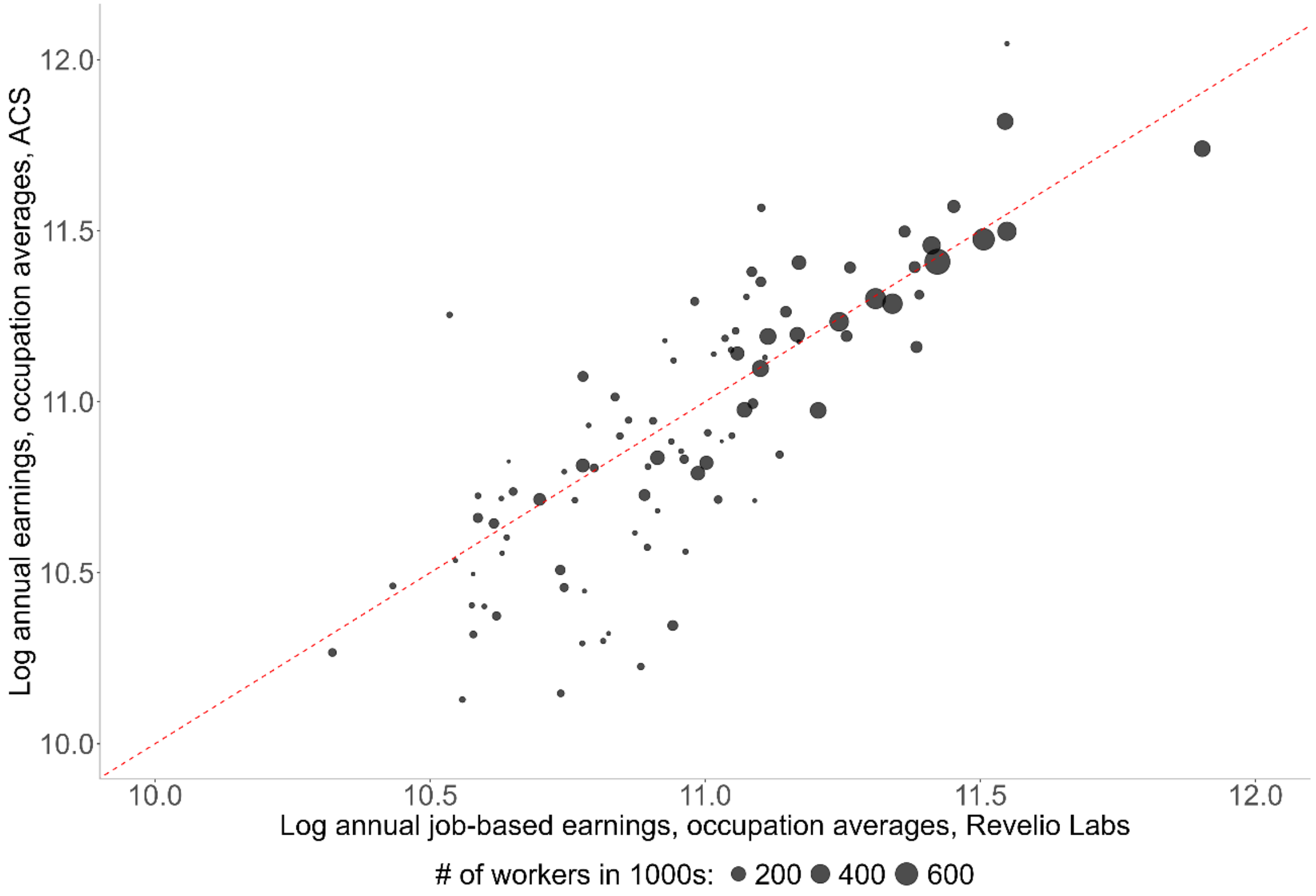

Notes: The x-axis indicates the average log annual job-based earnings that Revelio Labs computes for all employees of the 2019 LinkedIn sample, aggregated to 95 occupations. The y-axis indicates the average log annual earnings of the corresponding occupation in the 2018-2019 American Community Survey (ACS).

## Figure A2: Comparison of LinkedIn sample to ACS

*Panel A: By highest educational degree*

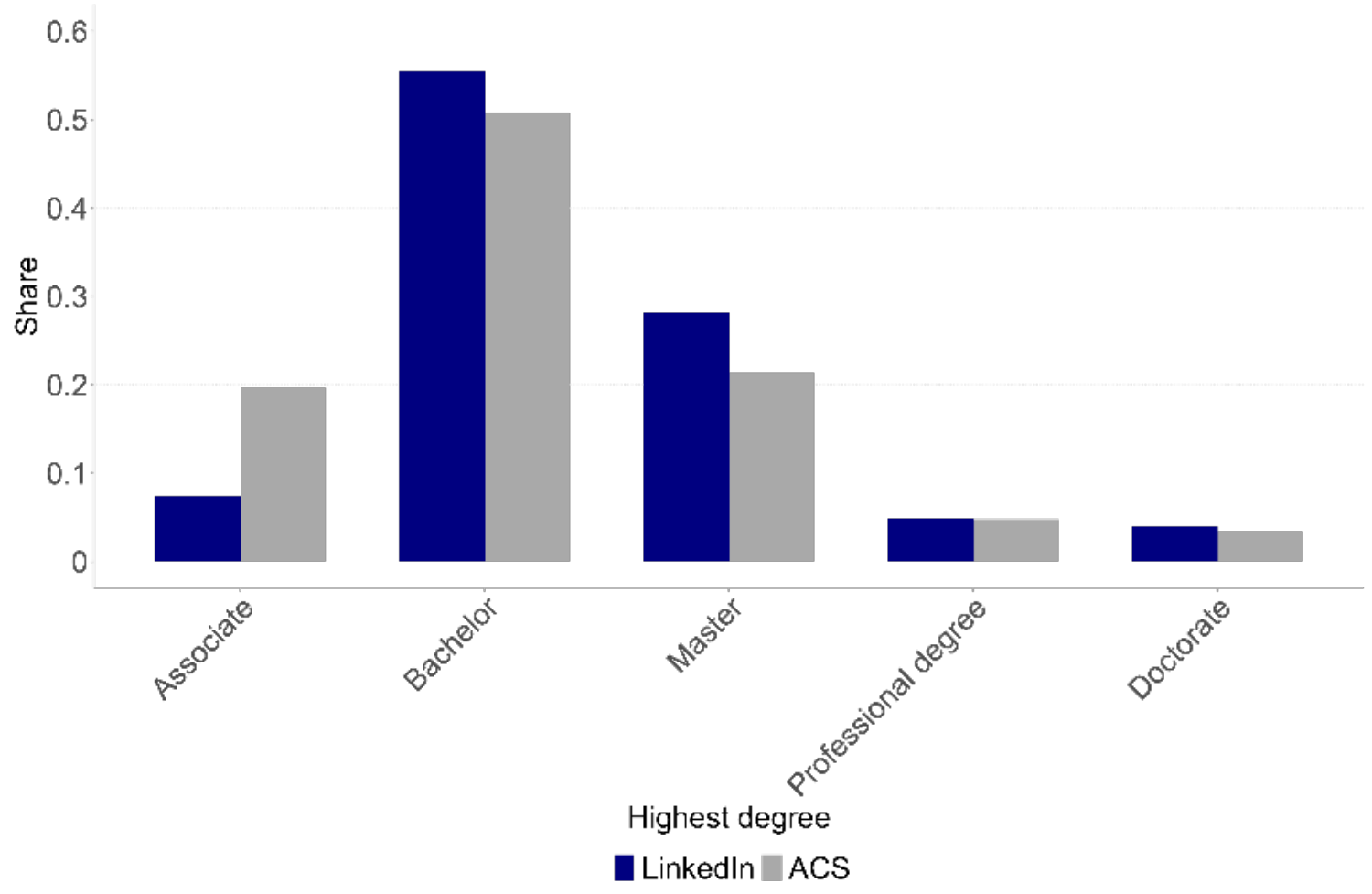

*Panel B: By occupational groups*

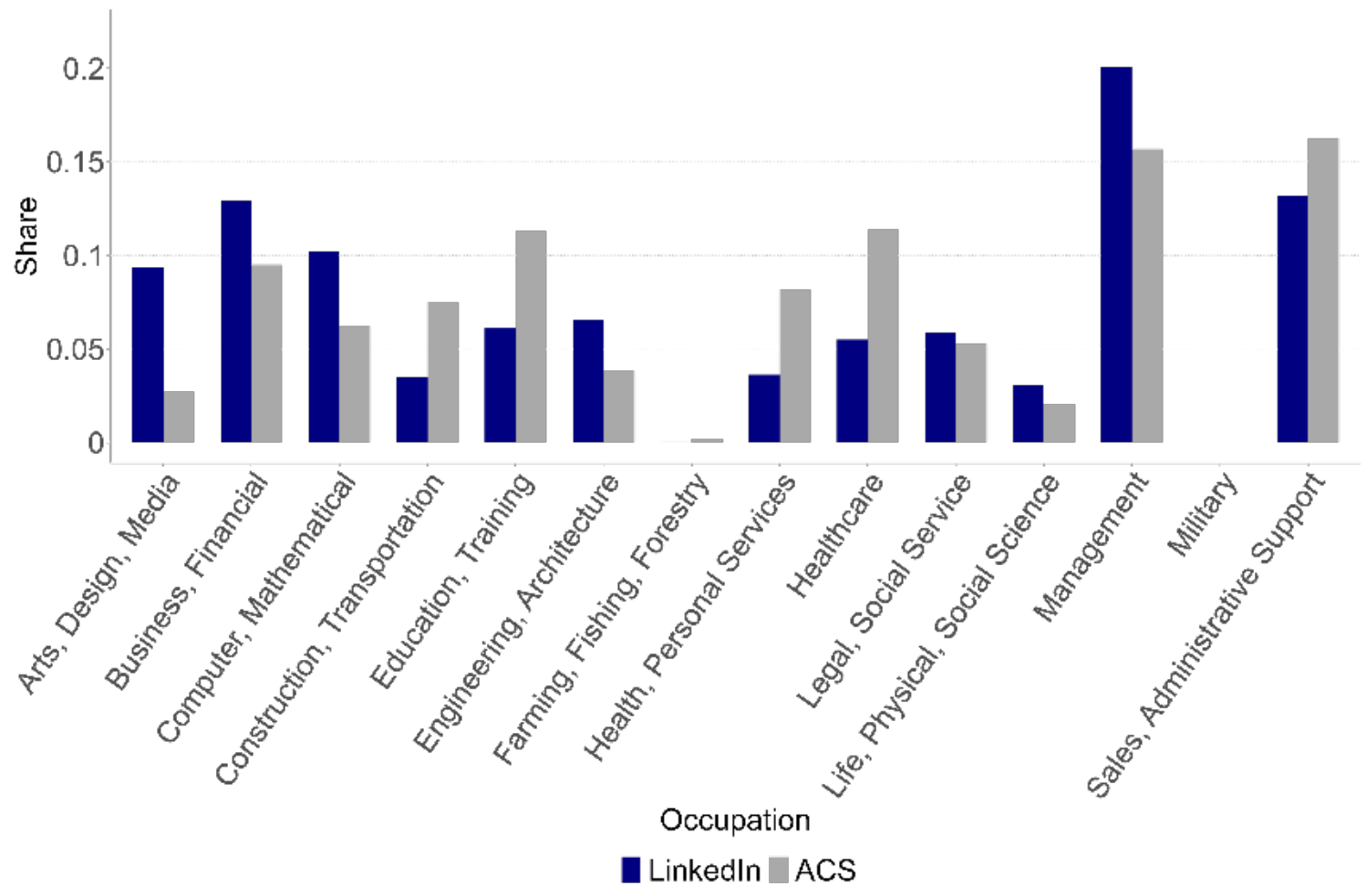

*Panel C: By bachelor’s field of study*

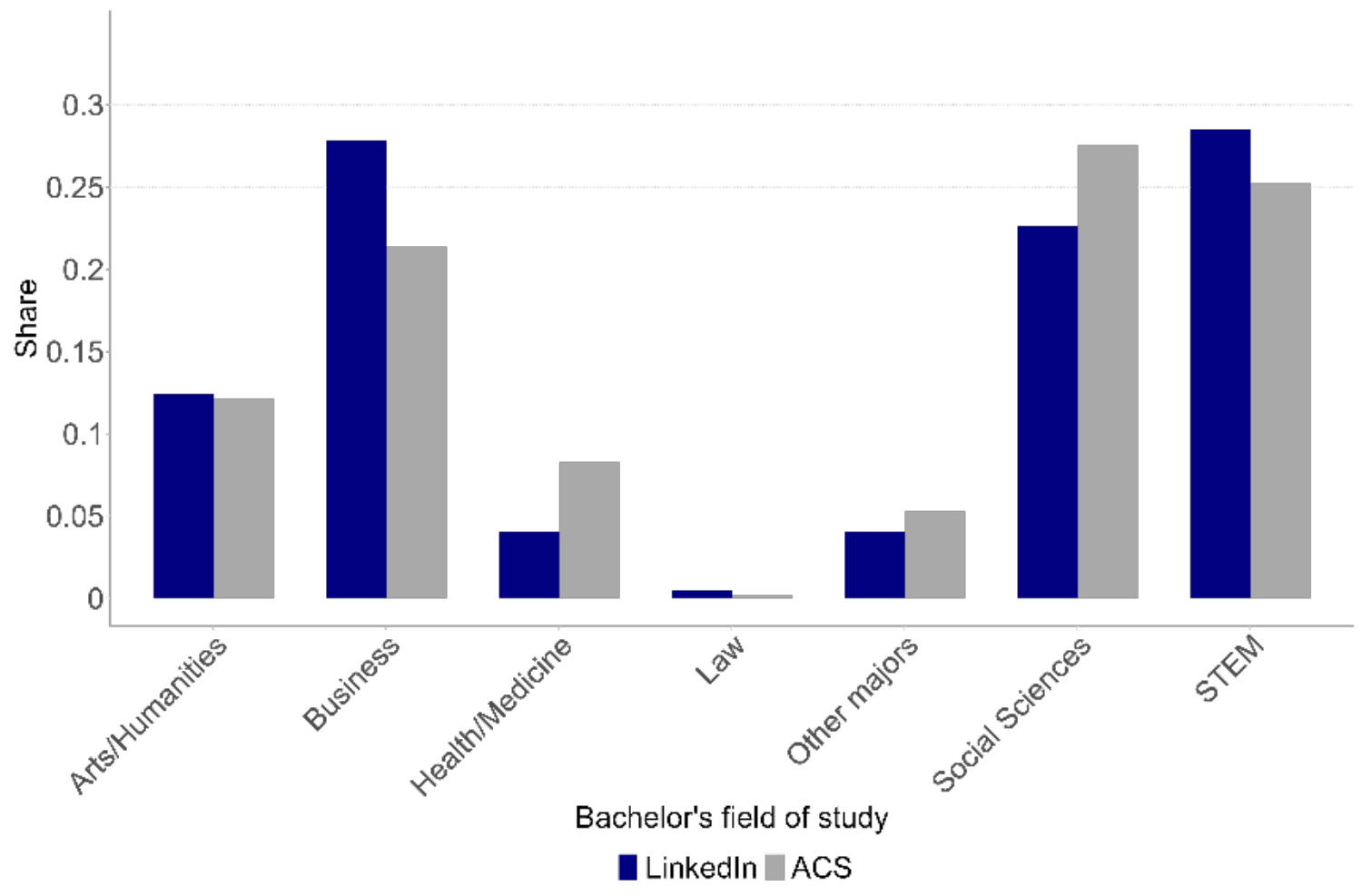

*Panel D: By U.S. census division*

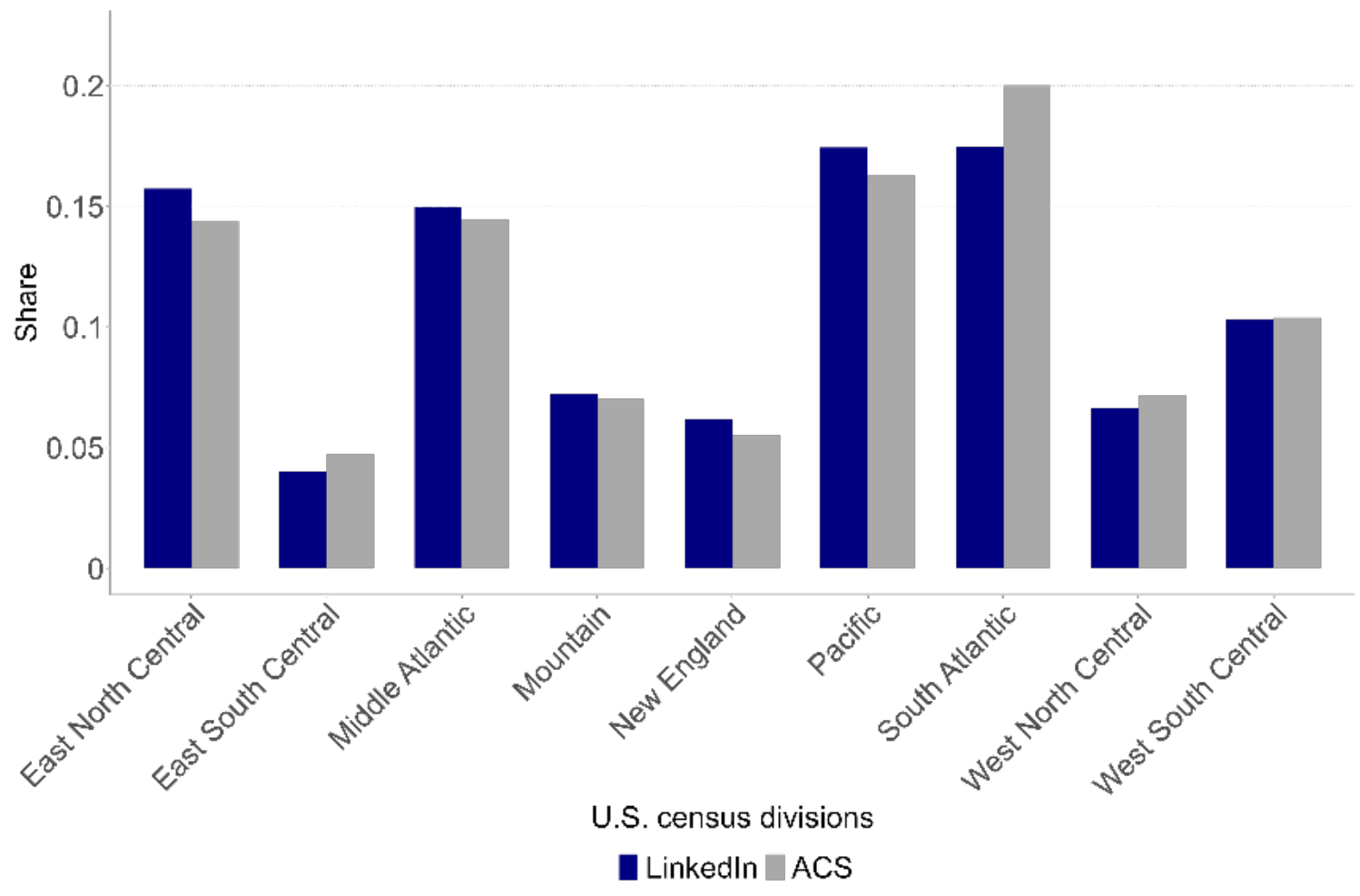

Notes: The figures compare the LinkedIn estimation sample (blue) to the employed college graduates aged 23-64 in the American Community Survey (ACS) 2019 (grey). The ACS 1-percent sample contains 6.5m employed college graduates. Panel B classifies occupation groups as aggregates of SOC two-digit codes as in Deming and Noray (2020). The occupation groups ‘military’ and ‘farming, fishing, forestry’ represent only trivial fractions of the estimation sample and total US employment. The statistics in Panel C are based only on individuals with at least a Bachelor’s degree and omit the 7 percent of individuals in our sample whose highest qualification is an Associate’s degree. Field of study is classified at the CIP two-digit level and aggregated into seven categories.

**Figure A3: Age-earnings profile for college graduates**

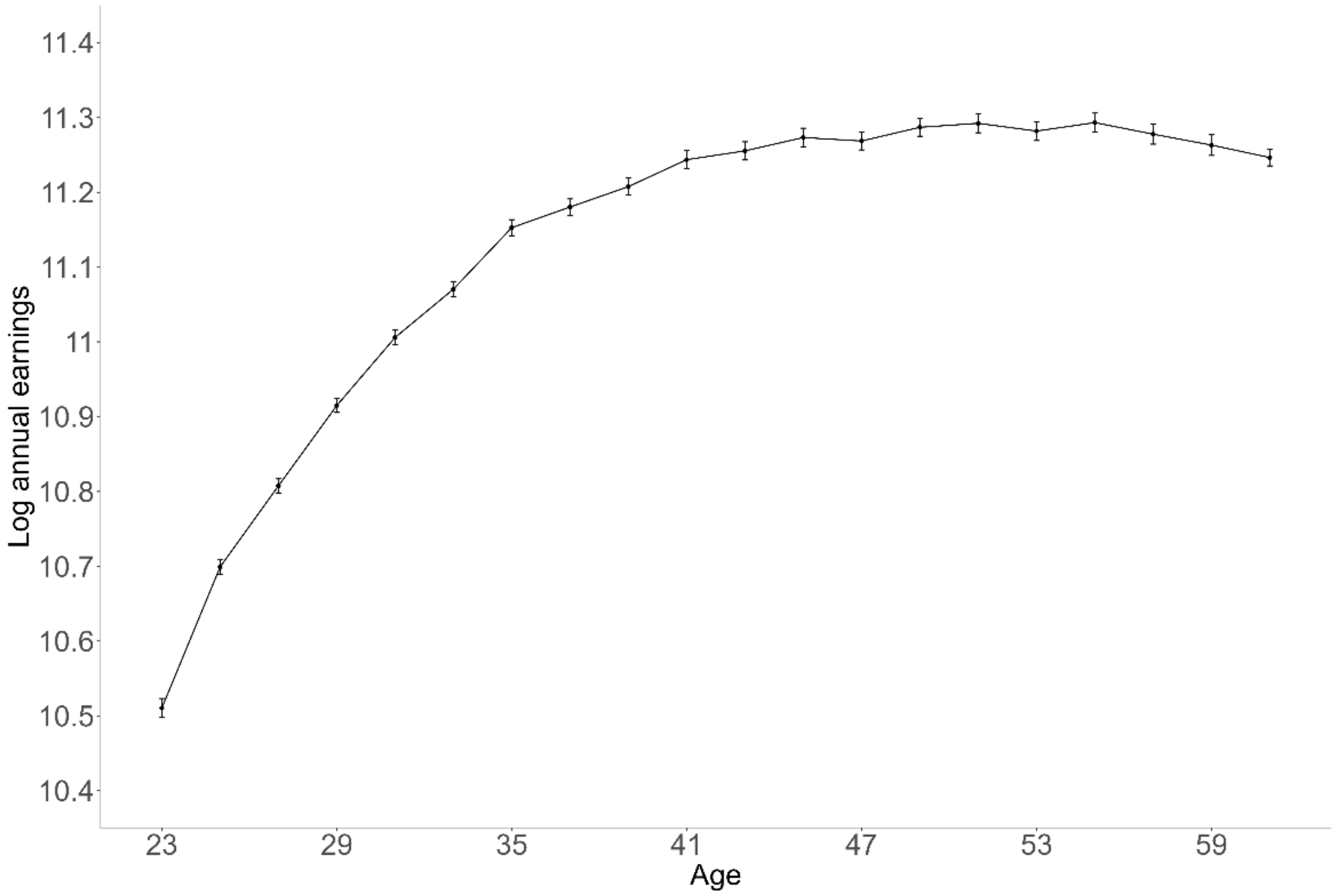

Notes: Average log annual earnings of employed college graduates aged 23-64 in the American Community Survey (ACS) 2019 by age (two-year age bins). Error bars depict 99 percent confidence intervals.

**Figure A4: Skill composition by experience and labor-market attachment**

*Panel A: Skill composition by actual and potential work experience*

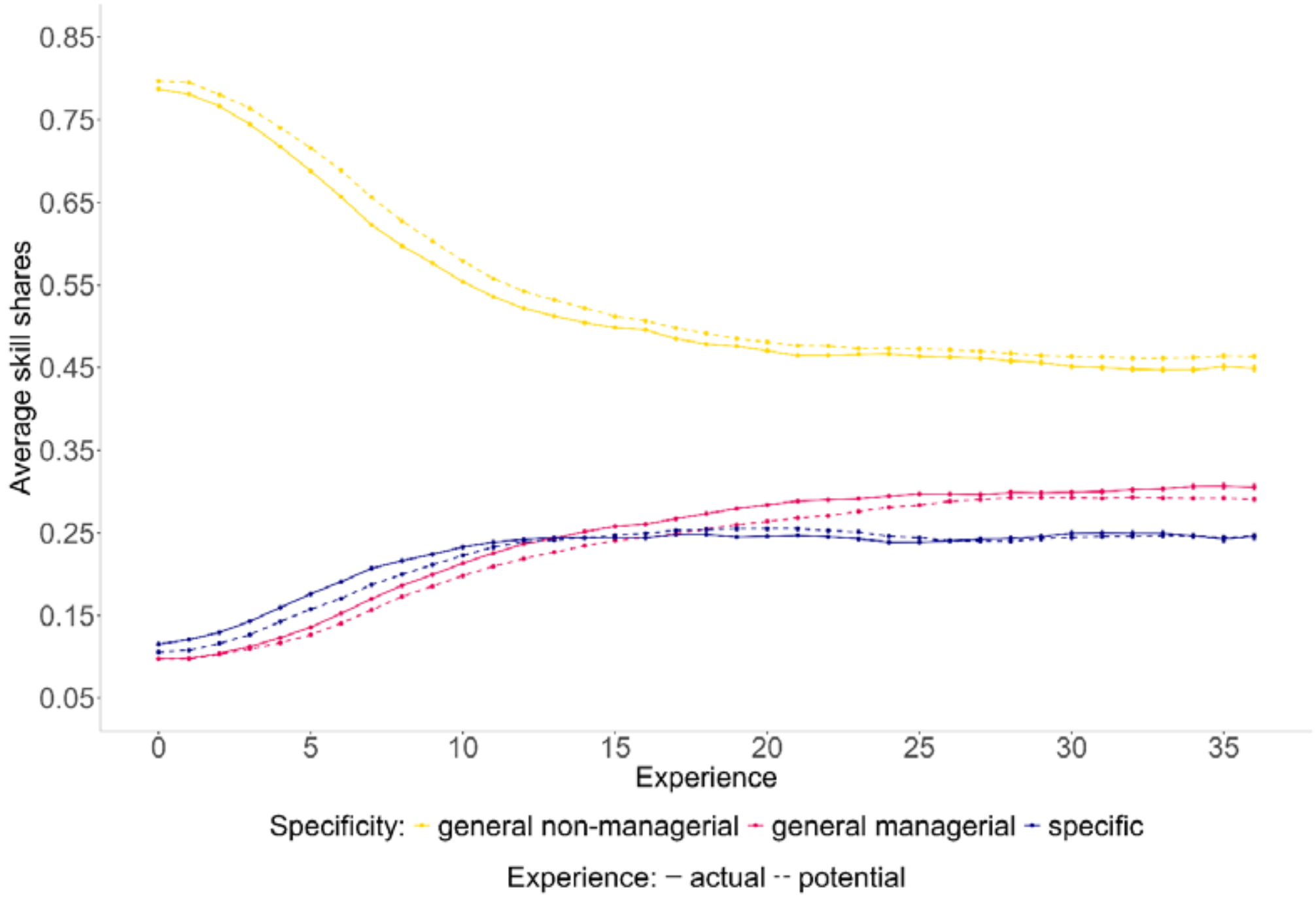

*Panel B: Skill composition by high and low labor-market attachment*

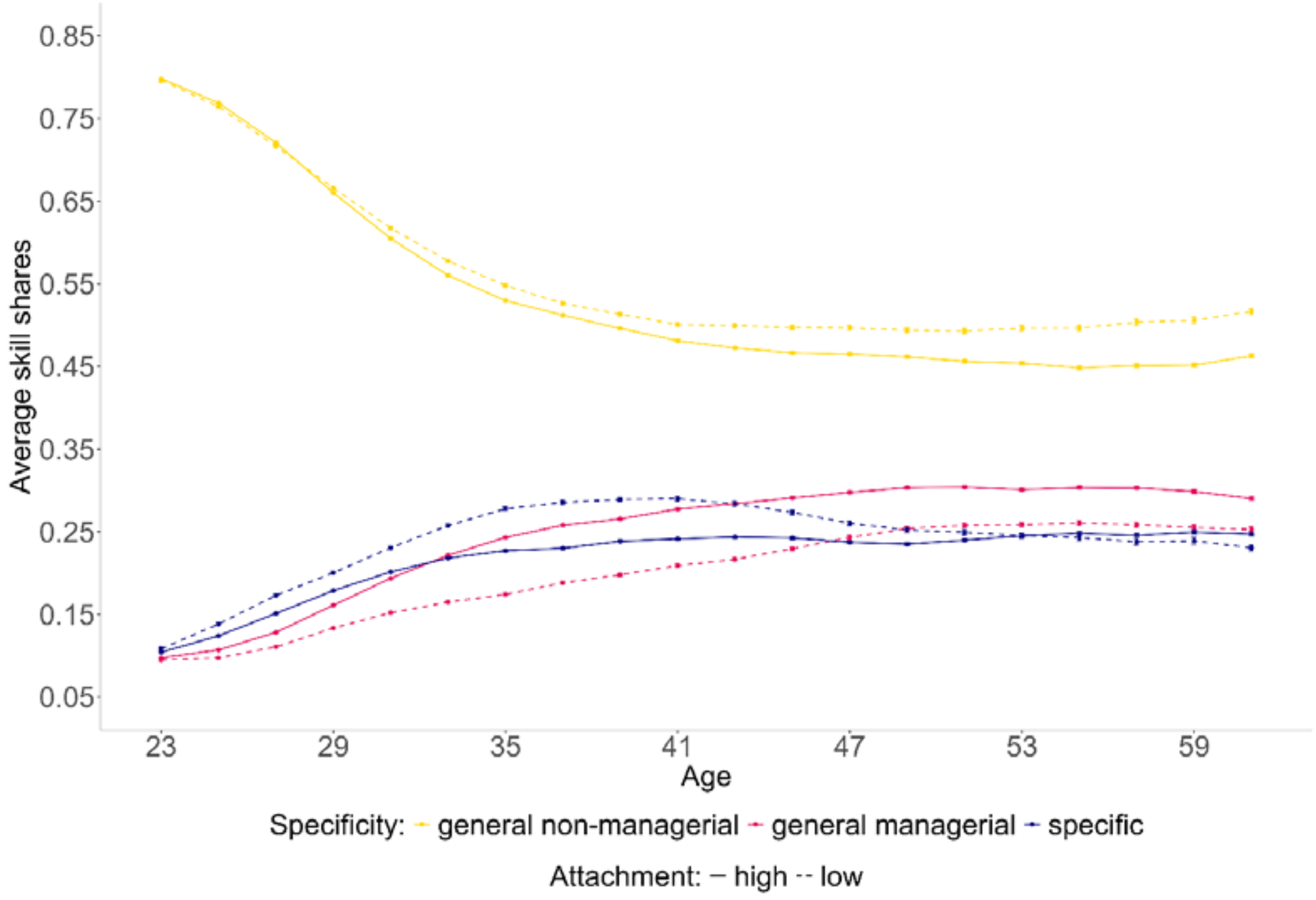

Notes: Panel A reports the average fractions of general non-managerial, general managerial, and occupation-specific skills by potential and actual work experience. Panel B reports the average fractions of general non-managerial, general managerial, and occupation-specific skills by for workers with low labor force attachment (defined as a ratio of actual to potential experience below 0.8 and cumulative non-employment spells of at least half a year) and for workers with high attachment (everyone else). Error bars depict 99 percent confidence intervals.

## Figure A5: Age-earnings profiles for college graduates by gender

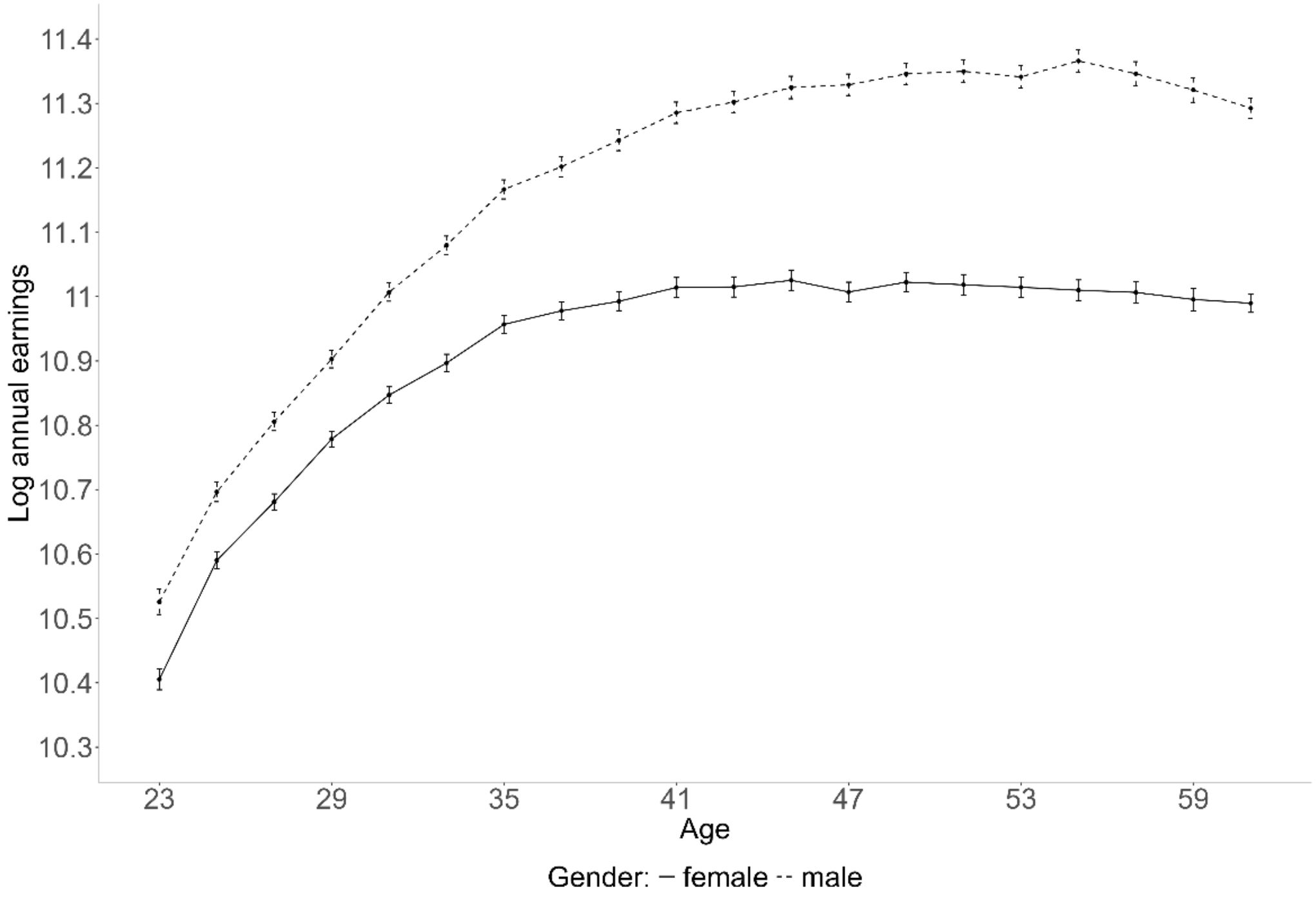

Notes: Average log annual earnings of employed college graduates aged 23-64 in the American Community Survey (ACS) 2019 by age (two-year age bins) and gender. Error bars depict 99 percent confidence intervals.

**Figure A6: Skill composition by age and gender conditional on occupation**

*Panel A: Share of general non-managerial skills by age and gender*

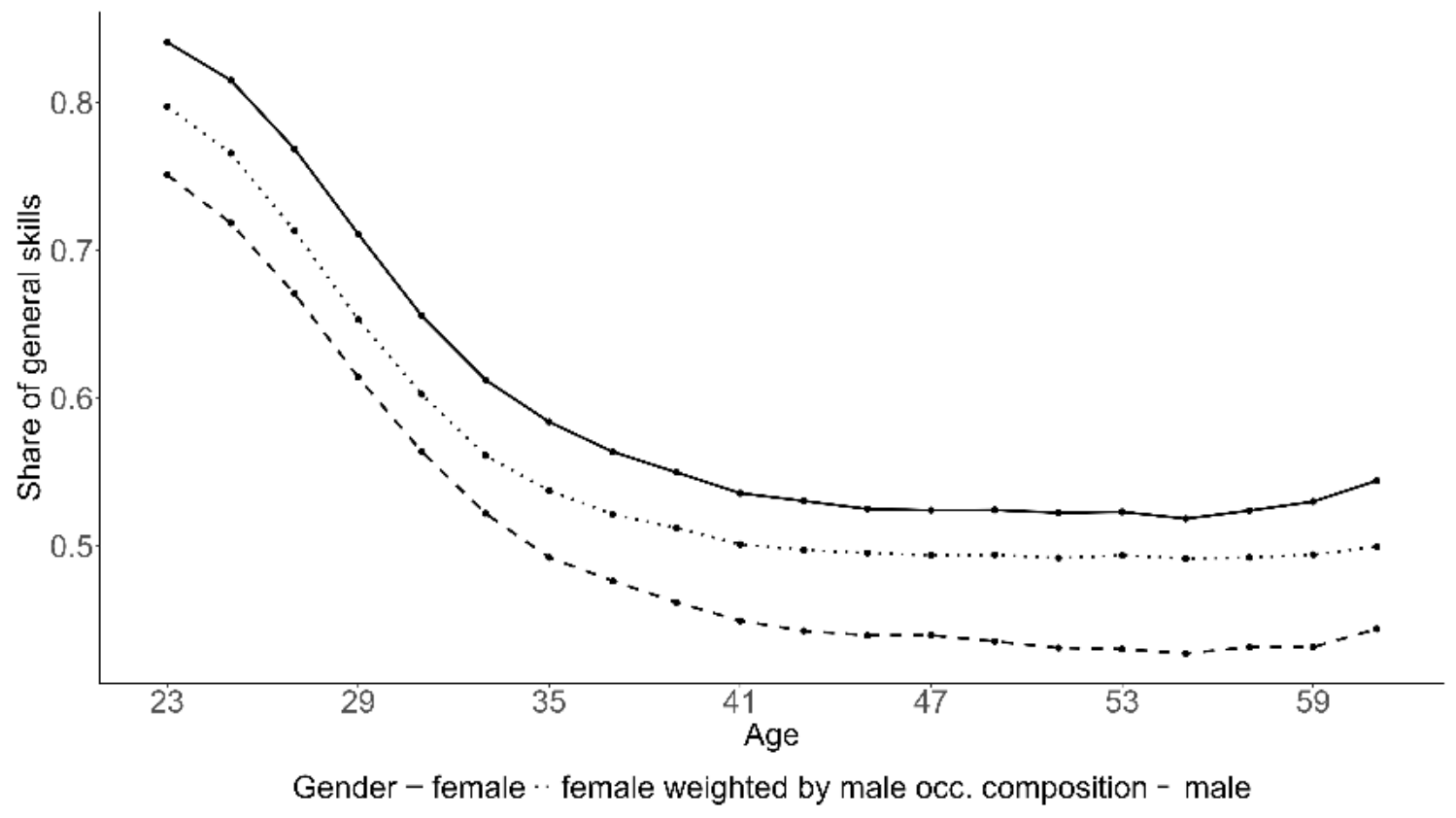

*Panel B: Share of general managerial skills by age and gender*

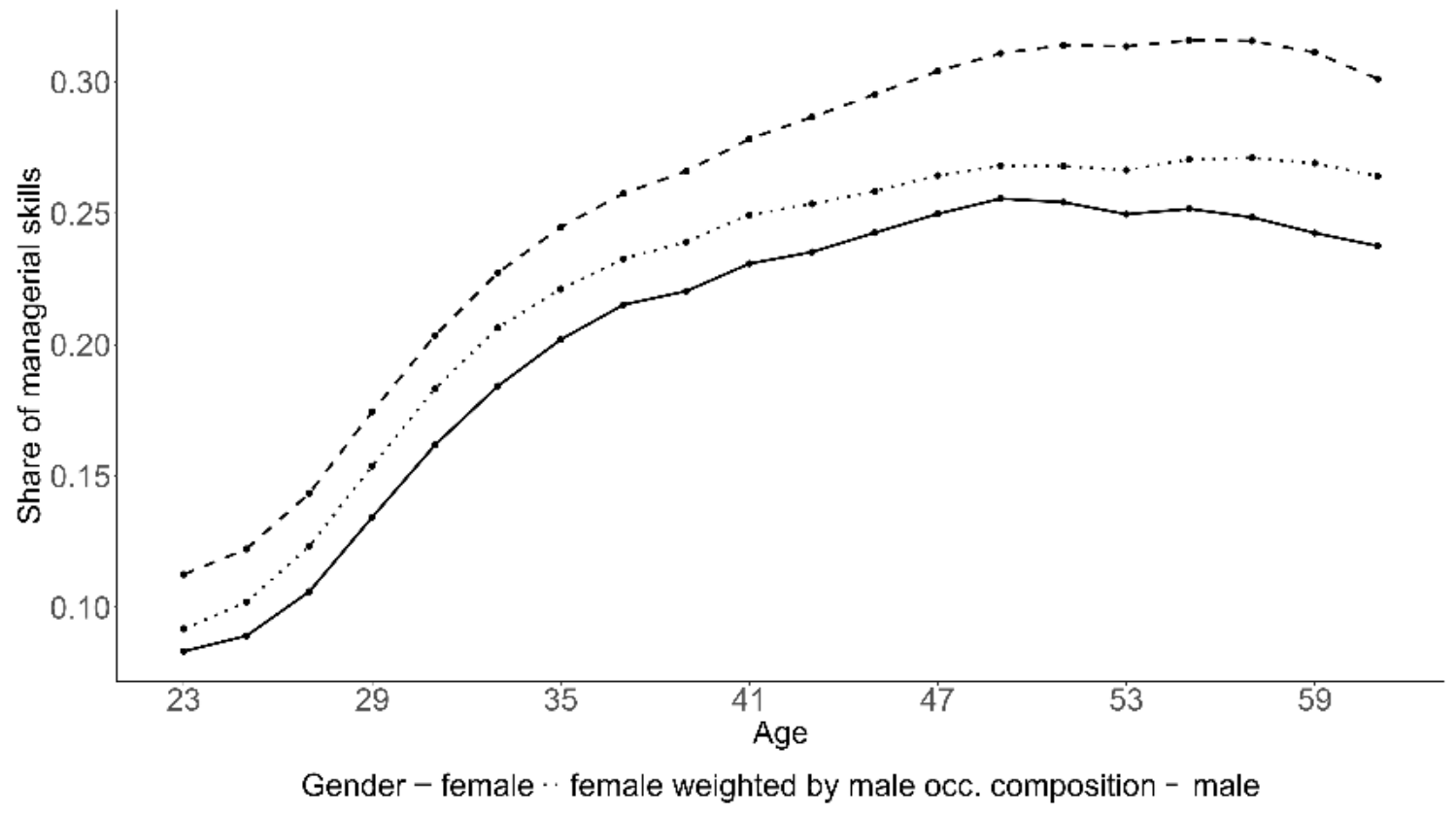

*Panel C: Share of specific skills by age and gender*

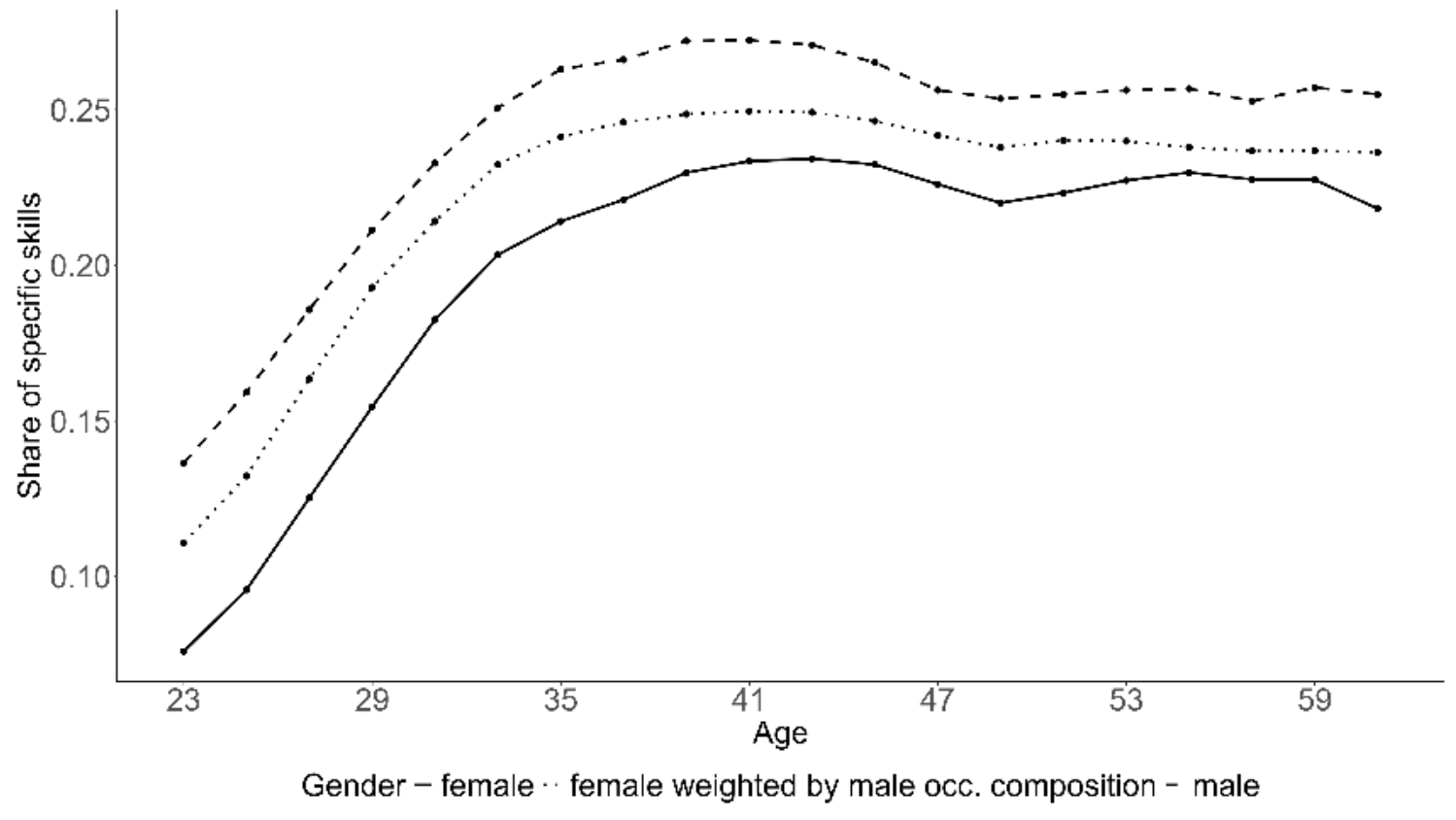

Notes: The figure shows the average fractions of general non-managerial, general managerial, and occupation-specific skills, respectively, by age and gender. The data series 'female weighted by male occ. composition' is based on the average number of skills (in Panel A) or average shares of general non-managerial, general managerial, and specific skills for women of a given age group and detailed occupation, weighted by the employment share of each occupation among the men of the same age group. The difference between this data series and the male series indicates within-occupation differences in skills across gender, while the difference with regard to the female series indicates between-occupation differences in skills across gender. Two-year age bins; last bin combines ages 61-64.

**Figure A7: Potential experience, actual experience, and labor force attachment by gender**

*Panel A: Potential and actual experience by gender*

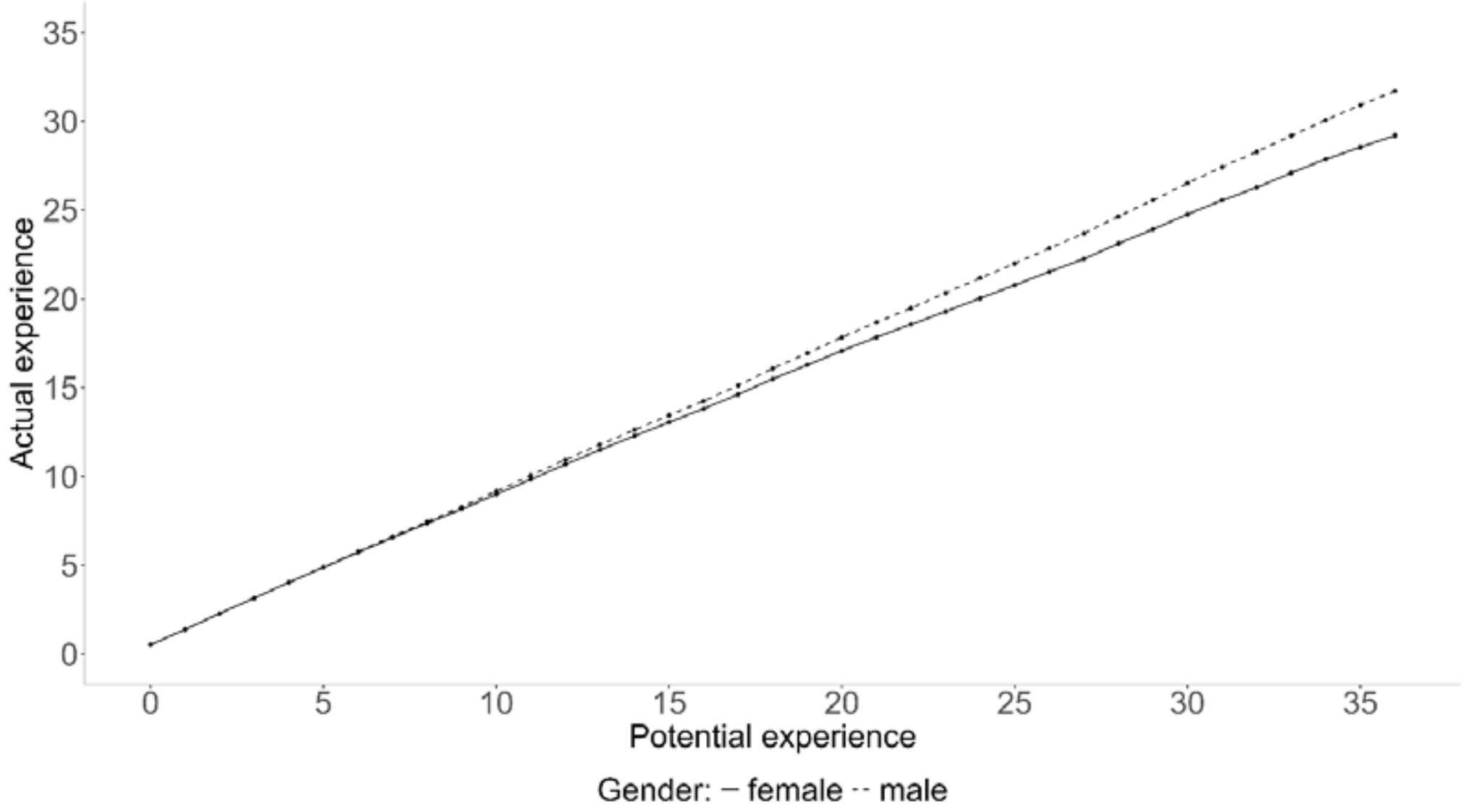

*Panel B: Fraction of workers with low labor force attachment by gender*

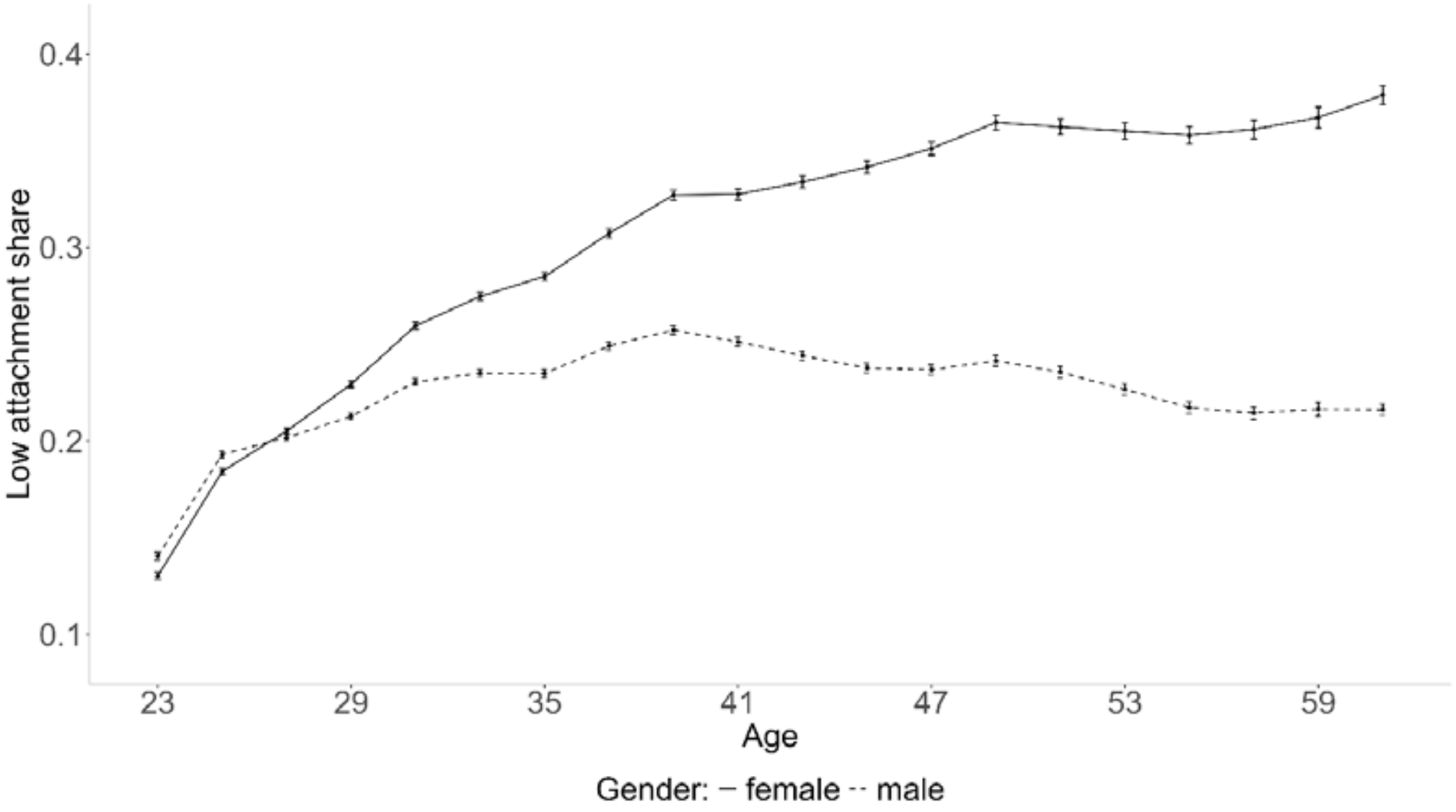

Notes: Low attachment share refers to the share of workers with low labor force attachment, defined as a ratio of actual to potential experience below 0.8 and cumulative non-employment spells of at least half a year. Error bars depict 99 percent confidence intervals.

## Appendix B: Results for LinkedIn Users with Recently Updated Profiles

The LinkedIn data may be inaccurate for users who do not regularly update their profiles. While we do not observe the date when a user last modified any profile information, we are able to observe whether a user recently changed to a job at a new firm, in which case the user may also have updated other relevant profile information such as skills prior to applying for the new job.

In this appendix, we study the robustness of key findings of our analysis to constraining the LinkedIn sample to users who changed firms within the last two years (in 2017 to 2019).[44] The resulting subset of profiles will likely be more up to date, but also less representative of the college-educated workforce at large. This is the case because a sample of recent firm switchers will oversample individuals who have an elevated job-to-job mobility, such as early-career workers. Nevertheless, we consider this robustness analysis useful for instance for the comparison of the skill profiles of women and men, where the relatively flatter age-skill profile for women observed in Panel A of Figure 7 may potentially result from less profile updating among women.

Roughly one third of the workers in our baseline sample (3,238,071 of 8,850,314 individuals) report starting a job at a new employer since 2017. Compared to the baseline sample, the movers are considerably younger on average (34.4 vs. 37.5 years), while the fraction of males is quite similar in both samples (53.2 vs. 54.5 percent).

Panel A of Figure B1 shows the age-skill profile from the baseline sample (in yellow), which is replicated from Figure 1, and the corresponding profile for the sample of recent firm switchers (in blue). The age-skill profile of recent firm switchers is notably steeper than for the baseline sample, consistent with switchers being either positively selected on skills, or caring more to update their skill profiles. Still, the qualitative pattern of a concave age-skill profile is just as evident in the sample of recent firm switchers, indicating that this shape is unlikely to be driven by underreporting by inactive users at higher ages.

[44] Results are similar in the smaller sample of those who changed firms within the last year.

**Figure B1: Number of reported skills and skill composition by age: Baseline sample vs. recent firm switchers**

*Panel A: Number of reported skills by age*

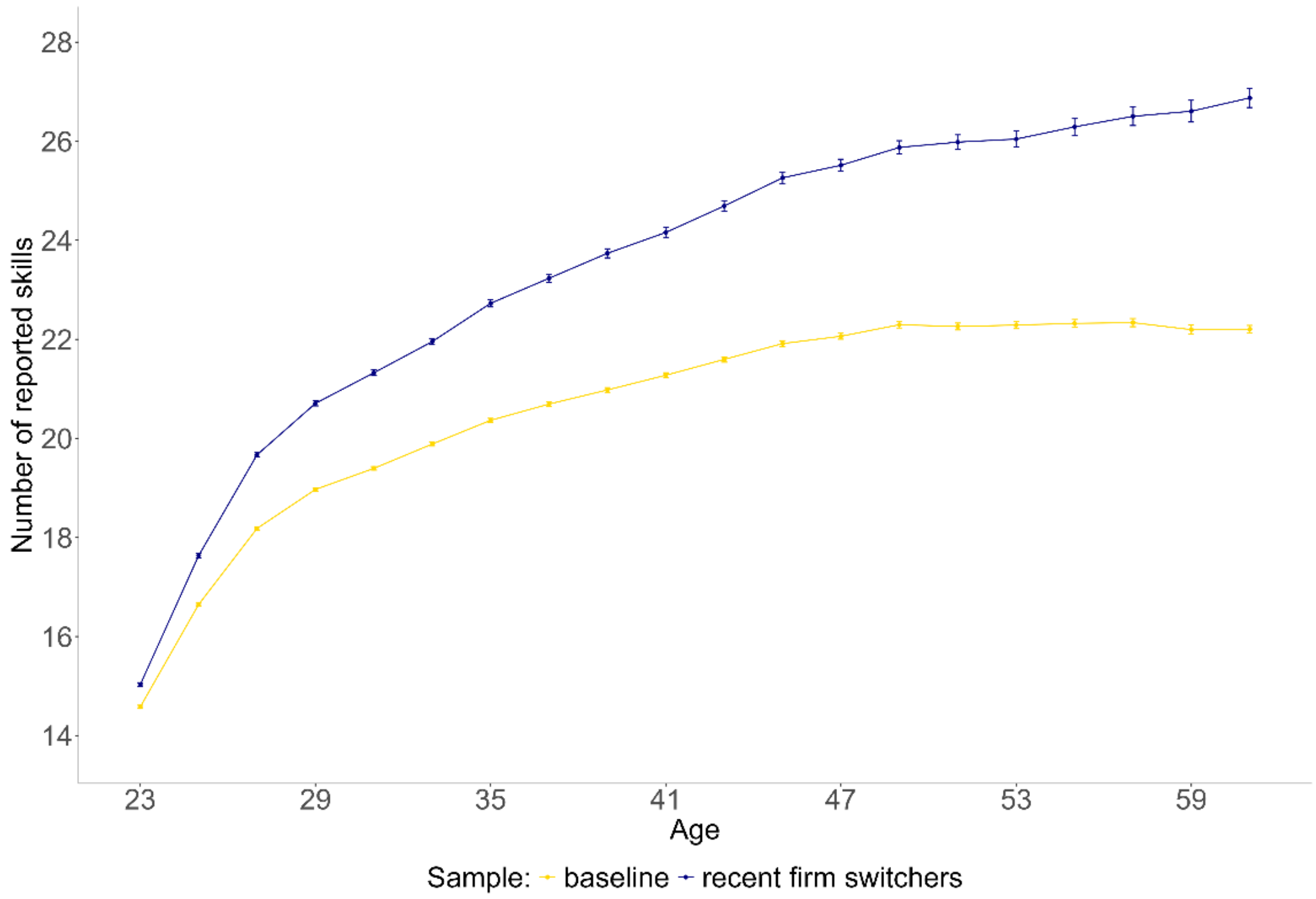

*Panel B: Share of general, managerial, and specific skills by age*

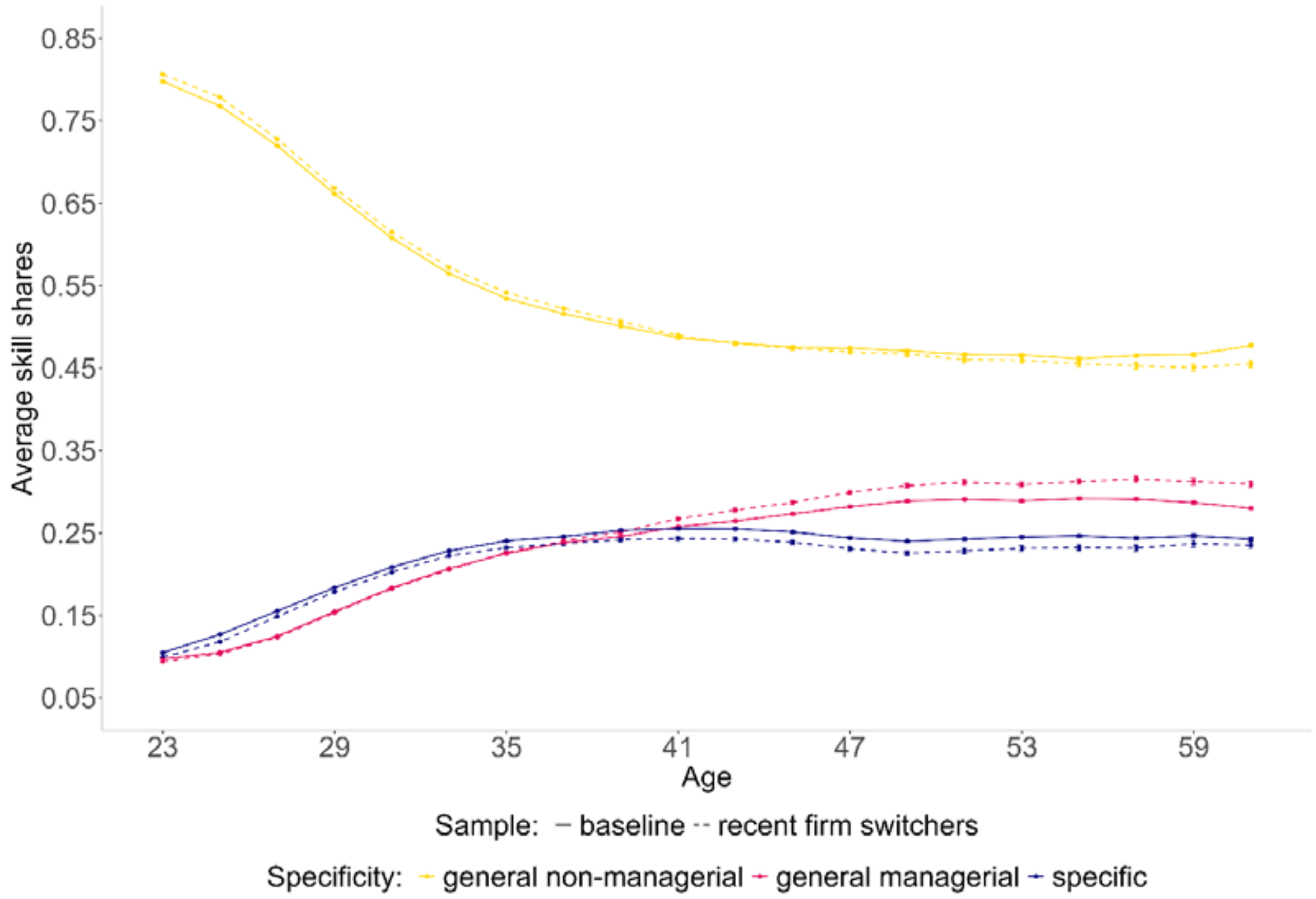

Notes: Average number of reported skills and average fractions of general non-managerial, general managerial, and occupation-specific skills, respectively, by age. Two-year age bins; last bin combines ages 61-64. Error bars depict 99 percent confidence intervals.

Panel B of Figure B1 shows the share of general, managerial, and specific skills by age, as in Figure 3, for the baseline sample and for recent firm switchers. The skill composition by age is remarkably similar between the two samples. Likewise, Figure B2 shows that the associations of job-based earnings with the number of reported skills (Panel A) and with the skill composition (Panel B) among workers who have recently changed their employer are very similar to those shown for the baseline sample in Figure 5.

Figure B3 reproduces Panel A of Figure 8, which reports separate age-skill profiles by gender in both samples. As in the baseline sample, there is only a very small gender gap in the number of reported skills among recent firm switchers in the 20s. But during the 30s and 40s, the gender skill gap grows also among recent job switchers and attains a magnitude that is comparable to the baseline sample. The increasing gender differences in the number of reported skills with age in the baseline analysis are thus unlikely to be driven by gender differences in job mobility.

Overall, these key results of our analysis are qualitatively similar for the sample of recent firm switchers, suggesting that the main characteristics of the baseline sample are unlikely to be driven by incomplete updating of inactive users.

**Figure B2: Job-based earnings by number of reported skills and skill composition: Baseline sample vs. recent firm switchers**

*Panel A: Job-based earnings by number of reported skills*

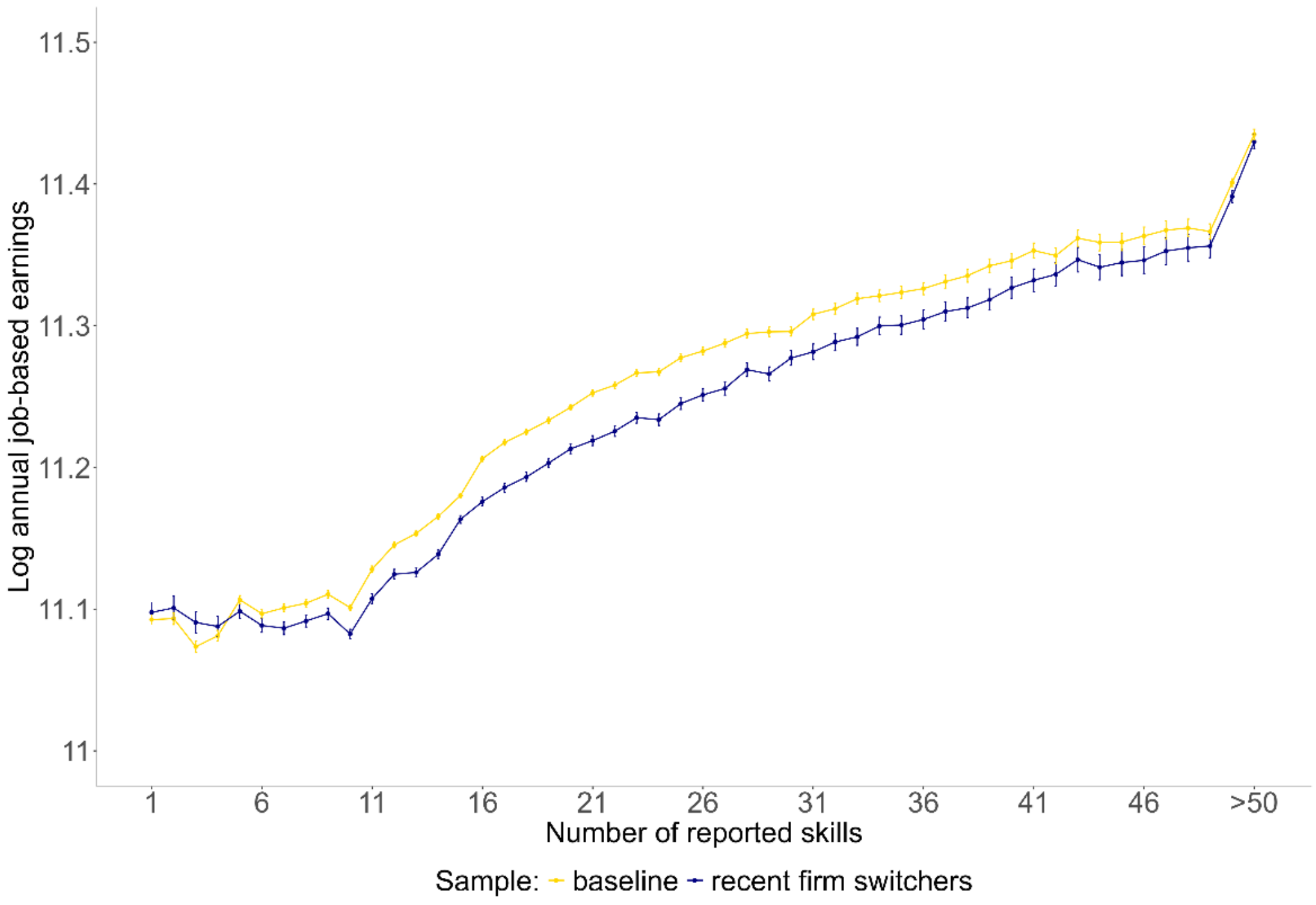

*Panel B: Job-based earnings by skill composition*

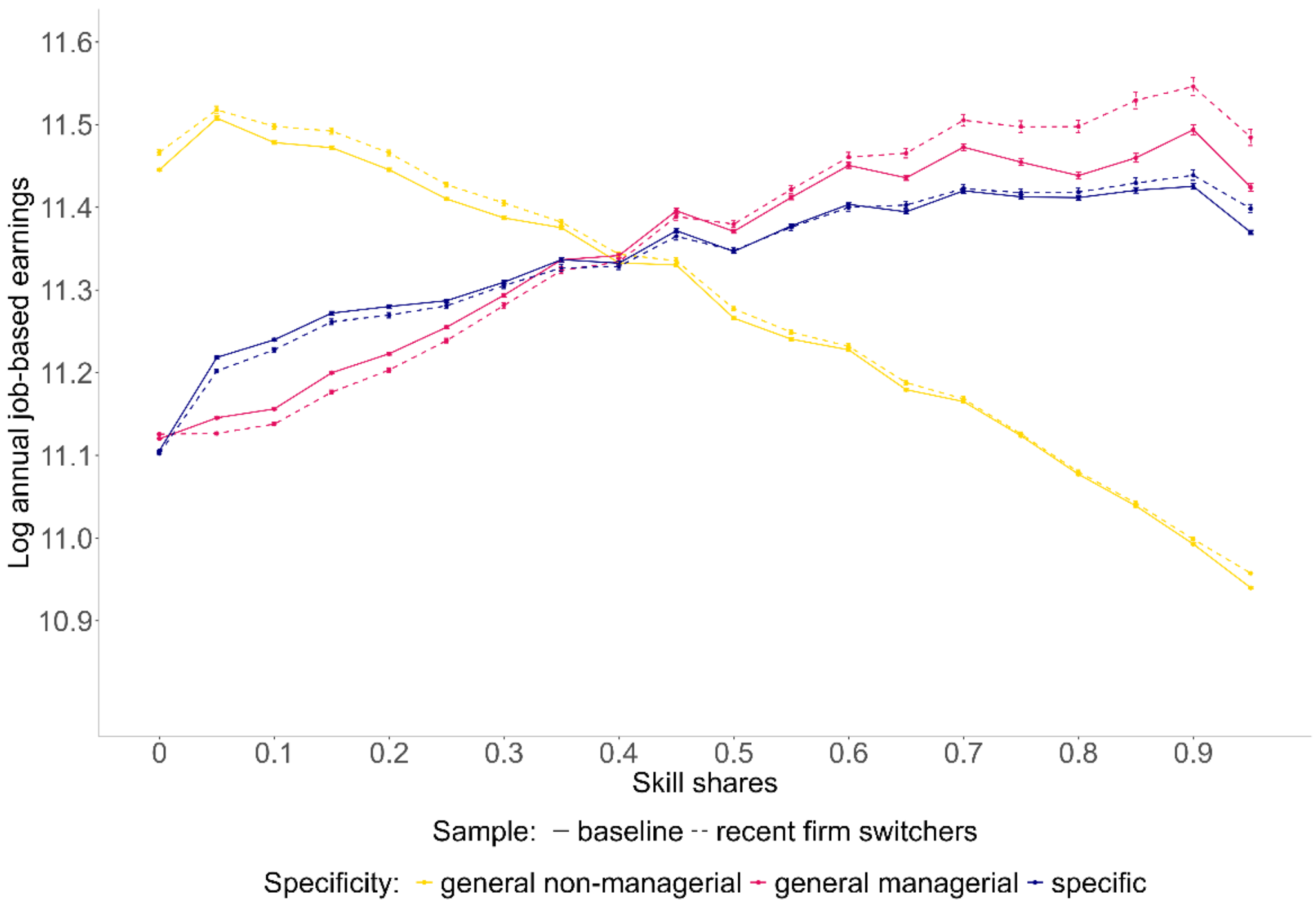

Notes: Average log annual earnings imputed by Revelio Labs' proprietary salary model by average number of reported skills and average fractions of general non-managerial, general managerial, and occupation-specific skills (five-percent bins), respectively. Error bars depict 99 percent confidence intervals.

**Figure B3: Number of reported skills by age and gender: Baseline sample vs. recent firm switchers**

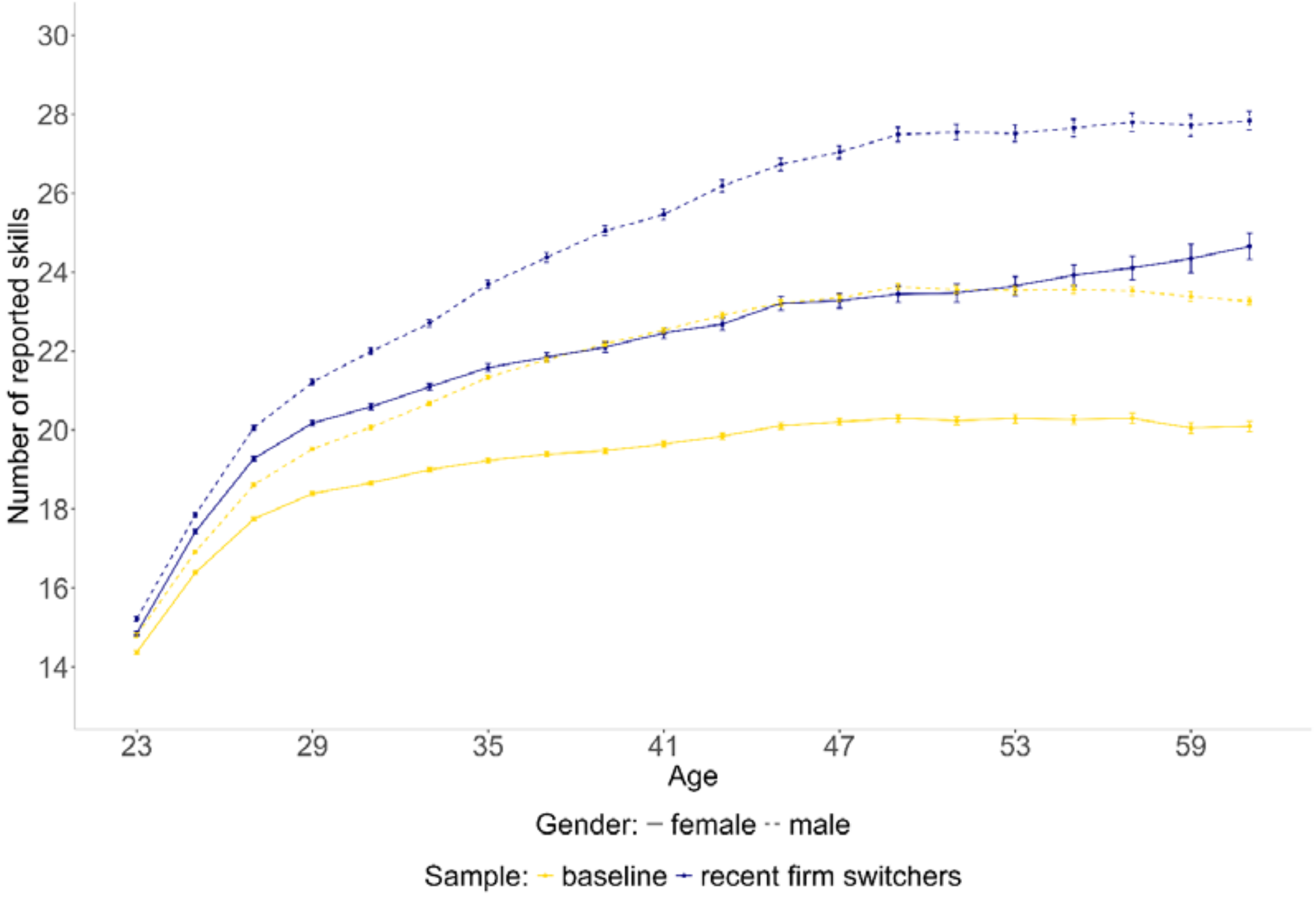

Notes: Average number of reported skills by age and gender. Two-year age bins; last bin combines ages 61-64. Error bars depict 99 percent confidence intervals.

## Appendix C: Results with ACS Population Weights

Our baseline results give equal weight to each observation in the analysis sample and are thus representative of U.S. college graduates with sufficiently complete LinkedIn profiles. As discussed in section 4.4, the composition of LinkedIn users is quite similar but not equal to the composition of the college-educated workforce in the US.

As an alternative to the baseline analysis, we re-estimate key results of our analysis using weights calculated from the pooled 2018-2019 American Community Survey (ACS) data that make the observations of our sample representative in terms of observables for the U.S. college-educated population at large. Both in the LinkedIn and in the ACS data, we compute the distribution of college graduates across 1,536 worker cells delineated by gender, eight five-year age bins (last age bin 58-64), two educational degrees (undergraduate, postgraduate), four U.S. census regions (Midwest, Northeast, South, West), and twelve broad occupational groups (following Deming and Noray 2020).[45] For the reweighted analysis, all observations of a given worker cell are weighted by the fraction of the cell's share in the ACS vs. in the LinkedIn data.

Panel A of Figure C1 indicates that the concave shape of the age-skill profile in Figure 1 similarly shows when reweighting the baseline sample to observationally look like the entire U.S. college-educated workforce. Weighting however shifts the graph downwards, indicating positive selection into the analysis sample in terms of number of reported skills. Panel B of Figure C1 shows that reweighting increases the fraction of general skills and reduces the fraction of managerial skills, though the changes are very modest in magnitude.

[45] The reweighting analysis omits farming/fishery occupations and military occupations for which cell sizes are very small.

**Figure C1: Number of reported skills and skill composition by age: Unweighted vs. weighted data**

*Panel A: Number of reported skills by age*

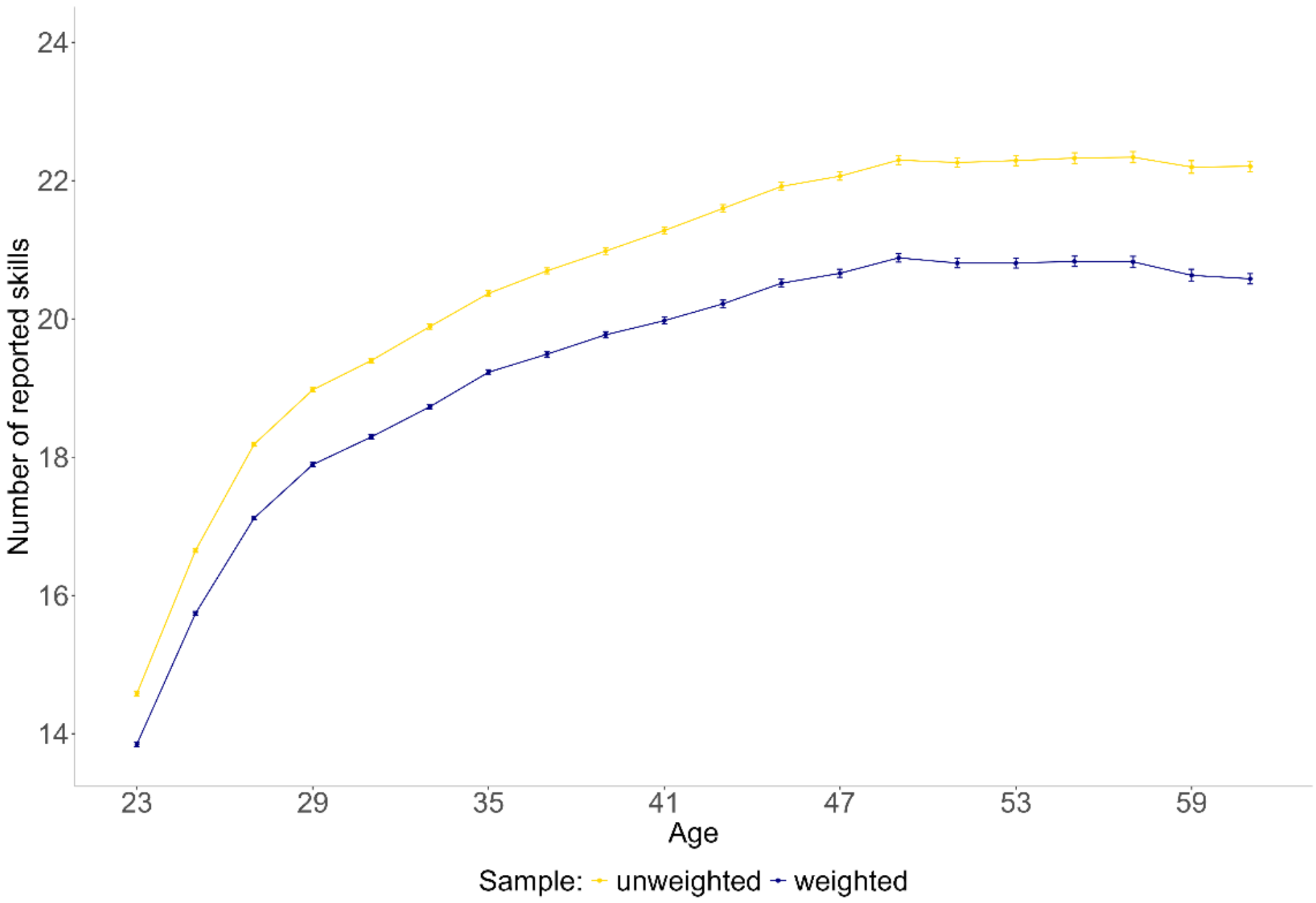

*Panel B: Share of general non-managerial, general managerial, and specific skills by age*

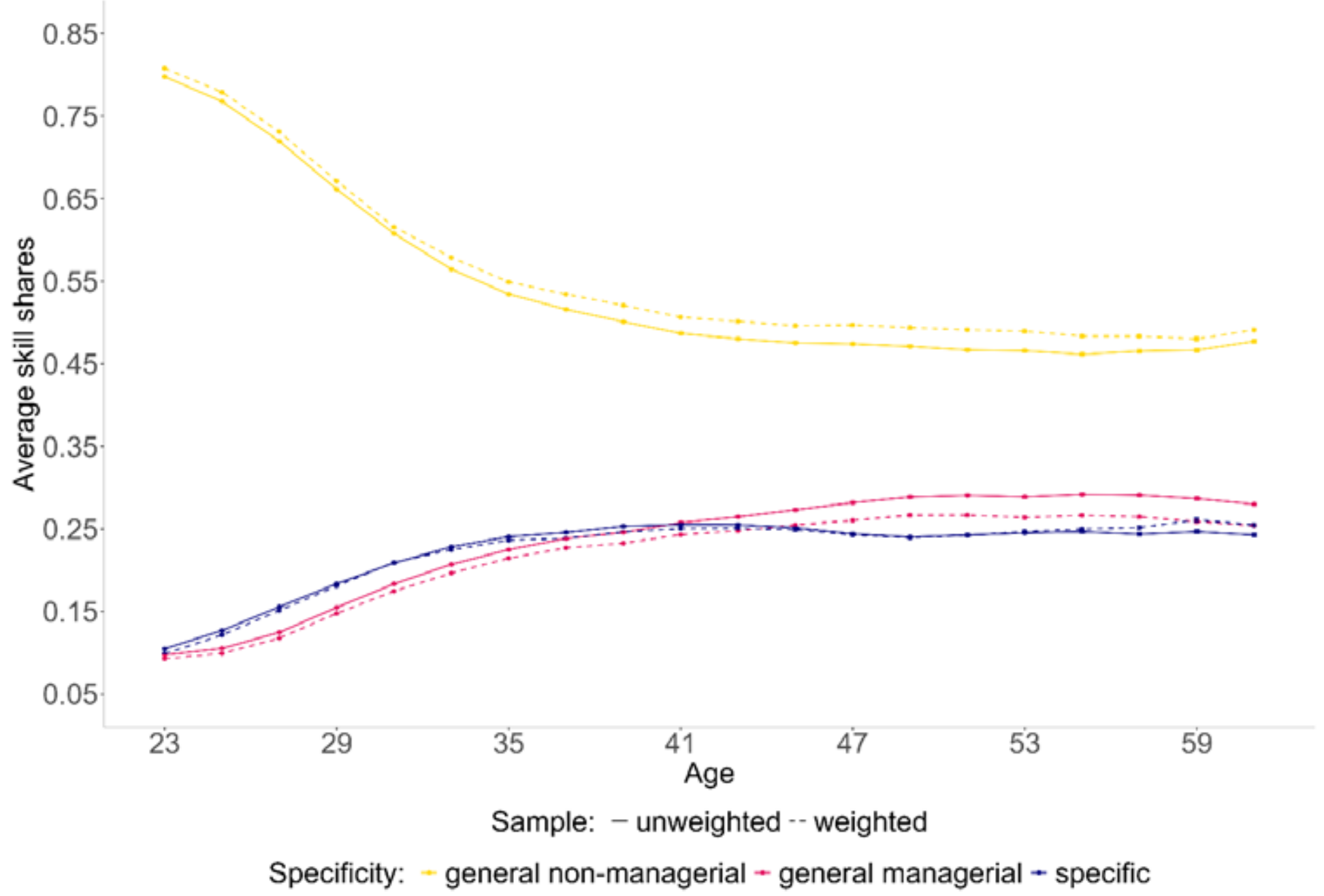

Notes: Average number of reported skills and average fractions of general non-managerial, general managerial, and occupation-specific skills, respectively, by age. Weighted graphs use sampling weights from the 2018-2019 American Community Survey (ACS) to make the estimation sample representative in terms of gender, five-year age bins, educational degrees, U.S. census regions, and occupation cells. Two-year age bins; last bin combines ages 61-64. Error bars depict 99 percent confidence intervals.

Panel A of Figure C2 indicates that the positive relationship between number of skills and job-based earnings of Panel A of Figure 5 is largely unchanged by reweighting, although there is a level shift to somewhat lower earnings at each level of the skill count. Panel B of Figure C2 confirms the finding of Panel B of Figure 5 that job-based earnings are increasing with the fractions of specific and managerial skills and decreasing with the fraction of general skills, and again indicates a lower level of earnings for all skill combinations.

Figure C3 shows that the gender-specific age-skill profiles of Figure 7 change little with reweighting, except that the level of reported skills is somewhat lower for both women and men at all ages.

Overall, reweighting the analysis sample to make it demographically representative of the college-educated workforce in the US reveals that our baseline sample is slightly biased towards individuals with higher skill and earnings levels. However, the basic relationships between skills, age, earnings, and gender remain qualitatively and often quantitatively unchanged. The reweighting analysis thus shows that our key findings are unlikely to be driven by patterns of differential participation on the LinkedIn platform for different demographic groups.

**Figure C2: Job-based earnings by number of reported skills and skill composition: Unweighted vs. weighted data**

*Panel A: Job-based earnings by number of reported skills*

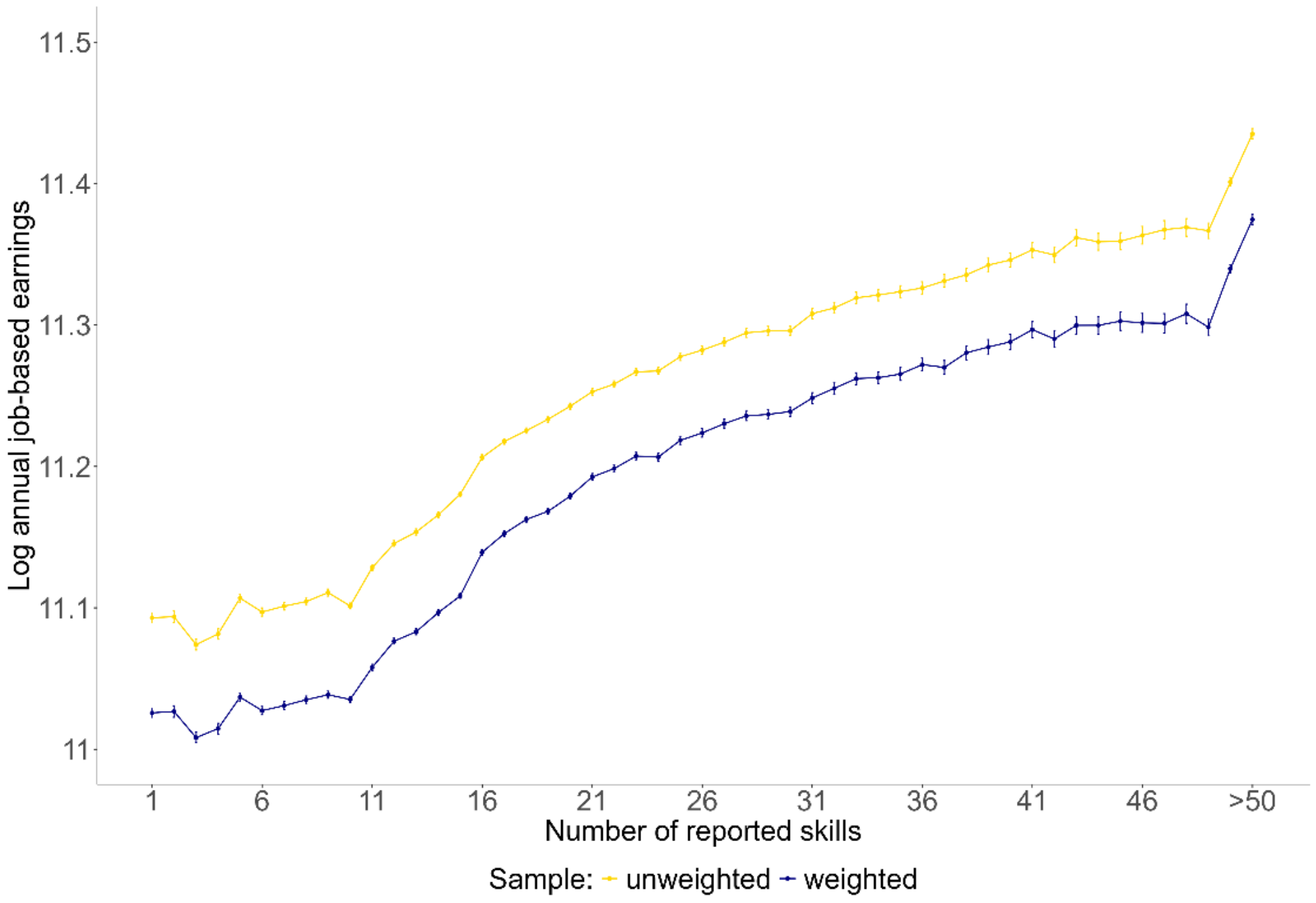

*Panel B: Job-based earnings by skill composition*

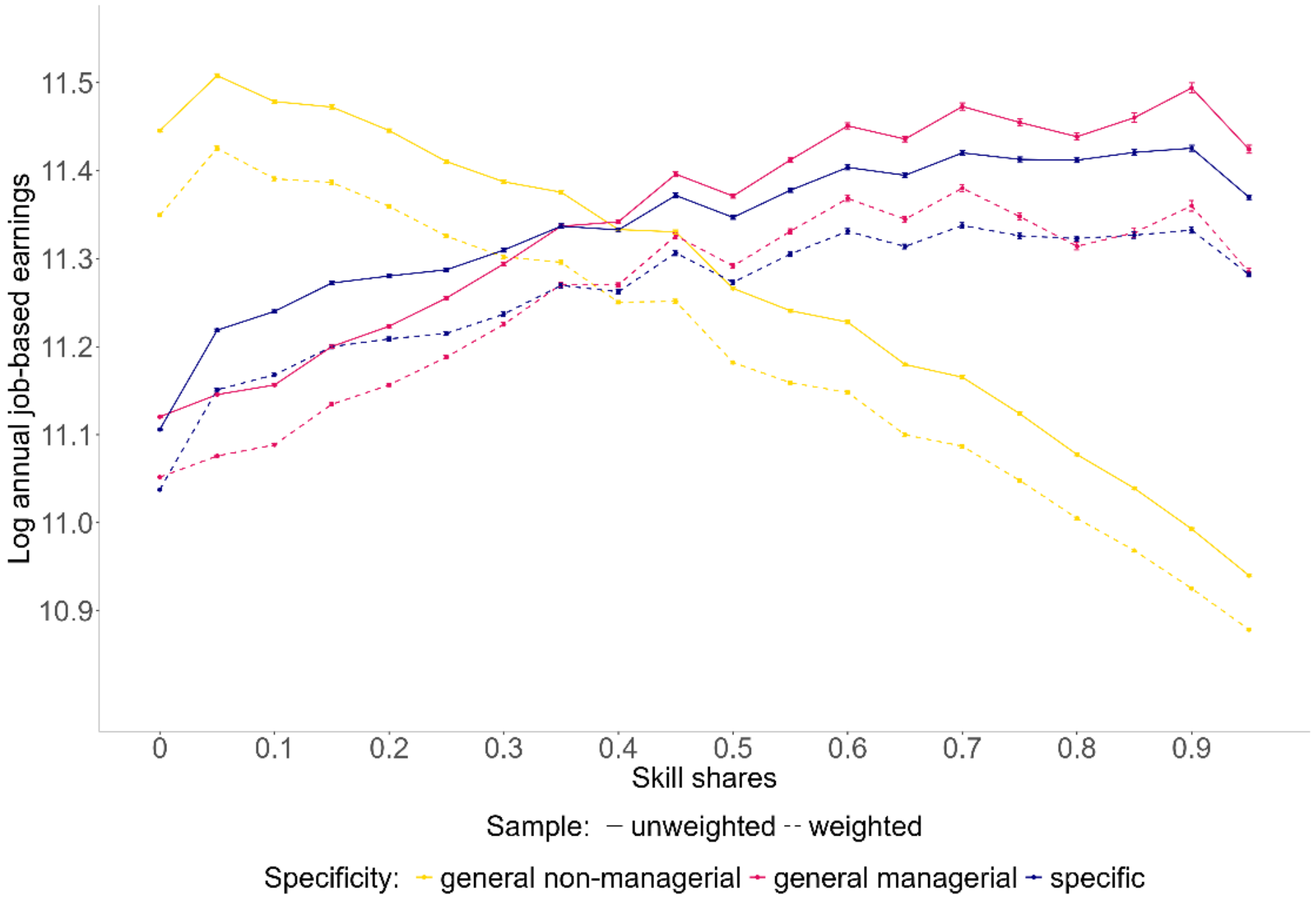

Notes: Average log annual earnings imputed by Revelio Labs' proprietary salary model by average number of reported skills and average fractions of general non-managerial, general managerial, and occupation-specific skills (five-percent bins), respectively. Weighted graphs use sampling weights from the 2018-2019 American Community Survey (ACS) to make the estimation sample representative in terms of gender, five-year age bins, educational degrees, U.S. census regions, and occupation cells. Error bars depict 99 percent confidence intervals.

**Figure C3: Number of reported skills by age and gender: Unweighted vs. weighted data**

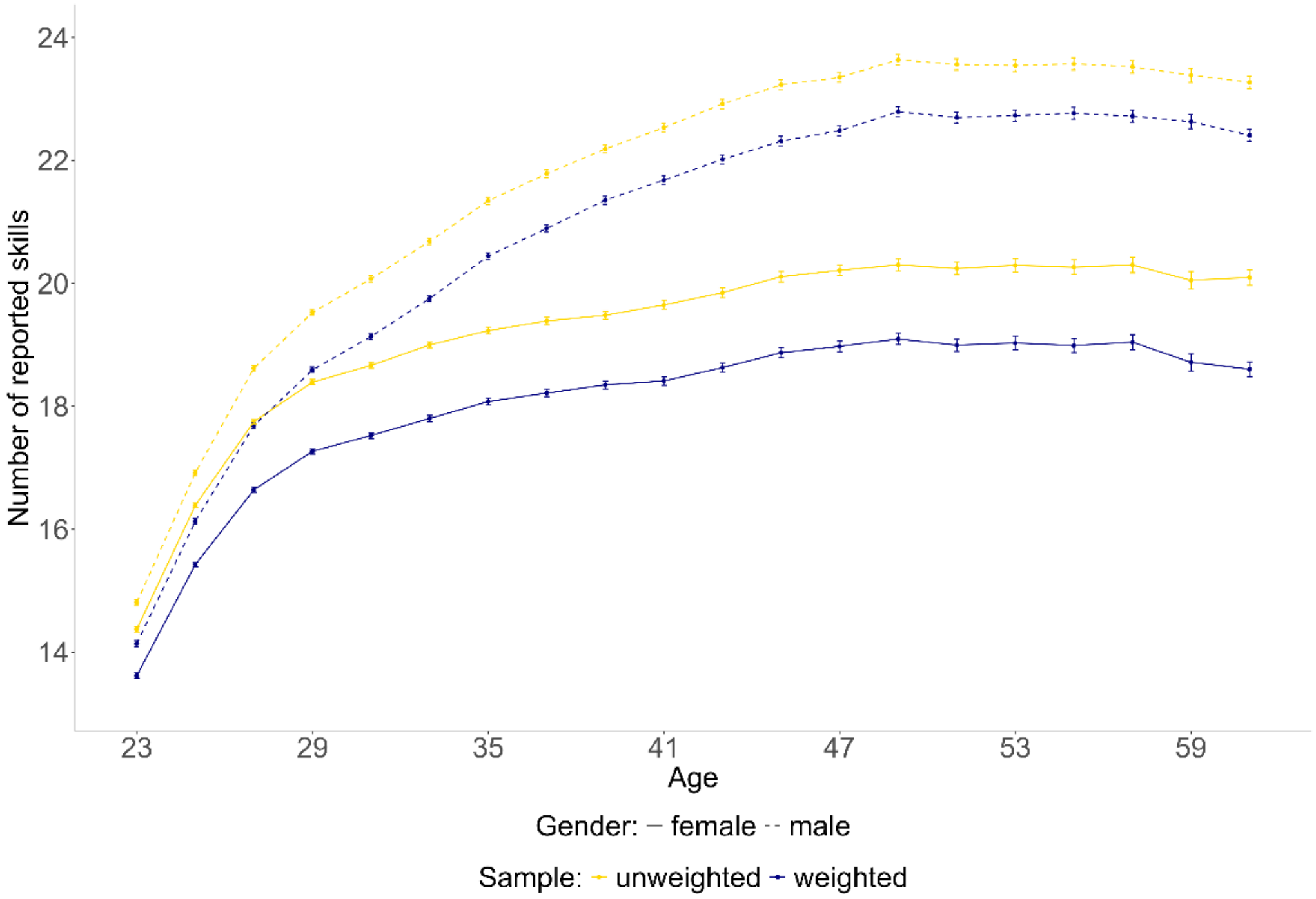

Notes: Average number of reported skills by age and gender. Weighted graphs use sampling weights from the 2018-2019 American Community Survey (ACS) to make the estimation sample representative in terms of gender, five-year age bins, educational degrees, U.S. census regions, and occupation cells. Two-year age bins; last bin combines ages 61-64. Error bars depict 99 percent confidence intervals.